\documentclass[runningheads]{llncs}
\pdfoutput=1
\usepackage{graphicx}
\usepackage{amsfonts}
\usepackage{amsmath}
\begin{document}
\title{Reduced order modeling of dynamical systems using artificial neural networks applied to water circulation\thanks{Supported by The Jefferson Project at Lake George, which is a collaboration of Rensselaer Polytechnic Institute, IBM, and The FUND for Lake George.}}
\titlerunning{Reduced Order Modeling applied to Water Circulation}
%
\author{Alberto Costa Nogueira Jr\inst{1} \and
Jo\~{a}o Lucas de Sousa Almeida\inst{1} \and
Guillaume Auger\inst{2} \and Campbell D Watson\inst{2}}
\authorrunning{A. Costa Nogueira Jr et al.}
%
\institute{IBM Research, Hortolandia SP 13186-900, Brazil \\
\email{albercn@br.ibm.com}\\
\url{https://www.research.ibm.com/labs/brazil/} \and
IBM Research, Yorktown Heights NY 10598, USA \\
\email{cwatson@us.ibm.com}}
\maketitle              
\begin{abstract}
General circulation models are essential tools in weather and hydrodynamic simulation. They solve discretized, complex physical equations in order to compute evolutionary states of dynamical systems, such as the hydrodynamics of a lake. However, high-resolution numerical solutions using such models are extremely computational and time consuming, often requiring a high performance computing architecture to be executed satisfactorily. Machine learning (ML)-based low-dimensional surrogate models are a promising alternative to speed up these simulations without undermining the quality of predictions. In this work, we develop two examples of fast, reliable, low-dimensional surrogate models to produce a 36 hour forecast of the depth-averaged hydrodynamics at Lake George NY, USA. Our ML approach uses two widespread artificial neural network (ANN) architectures: fully connected neural networks and long short-term memory. These ANN architectures are first validated in the deterministic and chaotic regimes of the Lorenz system and then combined with proper orthogonal decomposition (to reduce the dimensionality of the incoming input data) to emulate the depth-averaged hydrodynamics of a flow simulator called SUNTANS. Results show the ANN-based reduced order models have promising accuracy levels (within $6\%$ of the prediction range) and advocate for further investigation into hydrodynamic applications.

\end{abstract}
\section{Introduction}
Dynamical systems are mathematical descriptions for the evolution of many complex and sophisticated real-world processes. General circulation models simulate a class of dynamical systems that have well suited features for environmental applications, including weather and hydrodynamic prediction. These physics-based predictions are commonly used to make time-critical decisions in response to, for example, occurrences of extreme weather \cite{benjamin2016} or harmful algal blooms in water bodies \cite{wynne2013}. However, a drawback of general circulation models is the time duration for execution, often taking hours to complete. Reducing this execution time would be of benefit to many users.

Machine learning has emerged as a promising technique to significantly reduce the simulation time of dynamical systems. Artificial neural networks (ANNs) have been successfully developed as surrogate models of their more computationa\-lly-demanding counterparts. In essence, ANNs "learn" to perform tasks through examples without the need to be programmed with rigid rules to execute specific tasks. For example, \cite{navratil2019} replaced a reservoir simulator with ANNs trained on the input and the output of the model, making the surrogate model agnostic to the origin of the inputs and substantially decreasing the compute time. \cite{vlachas2018} used a method based on LSTM networks to simulate reduced order chaotic dynamical systems, including a barotropic climate model. Their method outperforms other techniques in short-term predictions, but long-term predictions experience a cumulative error, leading to erroneous long-term forecasts. 

Other machine learning methods used in fluid dynamics simulations include \cite{Wang2017}, who used a physics-informed machine learning approach to improve prediction of Reynolds stress in fluids from an estimation in a Reynolds averaged framework. They used the Random Forest method to build the regression functions and an available direct numerical simulation (DNS) data-set served as training for their framework. Their methodology improved the estimation of Reynolds stress, but generated spurious changes in some cases. \cite{Byungsoo2019} used a convolutional neural network to accurately reconstructed the velocity field in a fluid, with a 700 times speedup, including the divergence-free condition for the fluid in their loss function.


A recent review of machine learning for fluid mechanics \cite{brunton2020} highlighted that combining data driven methods (e.g., ANNs) with reduced order models (ROMs) is a compelling technique that can outperform each of its components. \cite{brunton2020} highlighted encouraging results by \cite{wan2018}, who used a type of recurrent neural network with long-short term memory (LSTM) to create data-driven predictions of extreme events in a complex dynamical system. 

With this in mind, we hereby develop two low-dimensional ANN-based ROMs to simulate the hydrodynamics of a freshwater lake in New York, USA. ROMs represent a technique for reducing the computational complexity of mathematical models in numerical simulations. A common technique for order reduction is proper orthogonal decomposition (POD), which is a numerical scheme that compresses data and preserves the essence of the original information in the form of an orthonormal basis matrix (which is optimal in the least-squares sense) \cite{wang2019}. The POD method can decouple the approximate solution of a dynamical system (e.g., water circulation in a lake) into spatial and temporal components; the spatial components can then be computed offline while an ANN can predict the temporal coefficients of the ROM. 

Hence, the objective of this work is to investigate the feasibility of using ANN-based ROMs to simulate the hydrodynamics of a freshwater lake. Specifically, we task ourselves with generating a 36 hour hydrodynamic forecast and compare its skill to the full-order approximate solution given by a high-resolution, hydrostatic model.

The paper is arranged as follows. First, the dimensionality reduction to produce the ROM is described, followed by the philosophy and workflow for two ANN-based ROMs. We then demonstrate the effectiveness of an ANN at predicting the deterministic and chaotic regimes of a discrete system called the Lorenz system, followed by results from the ANN-based ROMs predicting the depth-averaged hydrodynamics of a freshwater lake.

\section{Dimensionality Reduction}\label{DR_section}
Consider the full order well-posed dynamical system given by
\begin{equation}
\frac{\partial Q(\mathbf{x},t)}{\partial t}+\mathcal{N}(Q(\mathbf{x},t)) = S(\mathbf{x},t), \ \  (\mathbf{x},t) \in \Omega \times \mathcal{T}
\end{equation}
with suitable initial and boundary conditions. Above, $\Omega \subset \mathbb{R}^{d} (d=1,2,3)$ and $\mathcal{T} \subset [0,T]$ refer to the space domain and time, respectively. ${Q,S}: \Omega \times \mathcal{T}\longrightarrow \mathbb{R}^{n}$ denote the space-time solution and  source term, respectively, with $n$ being the number of dependent variables in the system. $\mathcal{N}$ is a general nonlinear operator associated to the dynamical system of interest.

After spatial discretization by any suitable numerical technique, the full order system is reduced to a system of ordinary differential equations (ODE) as
\begin{equation}
\frac{d Q_{h}(t)}{d t}+\mathcal{N}_{h}(Q_{h}(t)) = S_{h}(t), \ \  t \in \mathcal{T}
\label{ODE-sys}
\end{equation}
where $Q_{h} : \mathcal{T} \longrightarrow \mathbb{R}^{M}$ is the discrete solution, $M$ is the number of degrees of freedom (DOFs) of the discrete dynamical system, $\mathcal{N}_{h}$ and $S_{h} : \mathcal{T} \longrightarrow \mathbb{R}^{M}$ are the corresponding discrete nonlinear operator and source term, respectively.

Repeatedly solving the discrete system of ODEs, which is the case in many forecasting applications, is a hard working and time consuming task. It motivates us to seek an approximate solution of the full-order problem based on a linear combination of reduced basis functions $\{{\psi}_{l}\}_{1 \leq l \leq L} \subset \mathbb{R}^{L}$ with $L \ll M$. 
Thus, the reduced space spanned by the reduced basis functions is
\begin{equation}
\mathbb{V}_{rb} = span\{\psi_{1},\ldots,\psi_{L}\} \subset \mathbb{V}_{h}
\end{equation}
where $\mathbb{V}_h$ is a finite dimensional subspace of a Hilbert space $\mathbb{V}$ defined over the domain $\Omega$.

An approximation of the full-order system can be expressed by the ansatz
\begin{equation}
Q(\mathbf{x},t) \approx Q_{rb}(\mathbf{x},t) = \overline{q}(\mathbf{x})+q'(\mathbf{x},t) = \overline{q}(\mathbf{x}) + \sum_{i=1}^{L}a_{i}(t)\psi_{i}(\mathbf{x})
\label{rb_approx}
\end{equation}
where $\mathbf{a}(t) = [a_{1}(t),\ldots,a_{L}(t)]^{T} \in \mathbb{R}^{L}$ is defined as the vector of coefficients of the approximate solution and, $\overline{q}$ and $q'$ are mean and fluctuating quantities that split the approximation $Q_{rb}$.  According to \cite{wang2019}, such splitting prevents the first reduced coefficient from containing most of the energy of the original system and therefore adds stability to the reduced system. It is worth noting that we only expand the term $q'$ since it depends on both time and space.

For the sake of conciseness, we denote $\Psi_{i} = \psi_{i}(\mathbf{x})$, and define the following reduced basis matrix $\Phi = [\Psi_{1}, \ldots, \Psi_{L}] \in \mathbb{R}^{M \times L}$ which will be used later on.

\subsection{Proper Orthogonal Decomposition}
A sharp definition of the POD method can be found in \cite{wang2019} where the authors state that it is a numerical technique that compresses data preserving the essence of the original information through an orthonormal basis matrix. Such matrix is built to be the optimal solution of a least-squares problem.

To construct a reduced basis using the POD, we start by arranging a collection of $N$ snapshots of the fluctuating quantities introduced in equation (\ref{rb_approx}) column-wise as $\mathcal{Q'} = [q'(\mathbf{x},t_{1}),\ldots, q'(\mathbf{x},t_{N})] \in \mathbb{R}^{M \times N}$, which is called the snapshot matrix. At this stage, we consider that $q'$ is evaluated as a uniform lattice or a set of randomly distributed points. However, the fluctuation values still depend on $\mathbf{x}$ since these variables must be integrated with respect to the space coordinates to form a correlation matrix (as we shall see later in the solution of the maximization step of the POD method).

Based on the expansion defined in equation (\ref{rb_approx}) and using the reduced basis matrix definition, we can write $\mathcal{Q'}$ as
\begin{equation}
\mathcal{Q'} = \Phi\mathcal{A} \ \ \textnormal{with} \ \  \Phi \in \mathbb{R}^{M \times L}, \ \mathcal{A} \in \mathbb{R}^{L\times N},
\label{full_space-time_decomp}
\end{equation}
where we used the matrix of temporal coefficients defined as $\mathcal{A} = \left[ \mathbf{a}(t_{1}),\ldots,\mathbf{a}(t_{N}) \right]$.

It is straightforward to show that each $\Psi_{i}$ basis function satisfies the following equation
\begin{equation}
\Psi_{i}=\lambda_{i}^{-1/2}\mathcal{Q'}v_{i}
\end{equation}
with $\lambda_{1}\geq\lambda_{2}\geq\ldots\geq\lambda_{L}$, where $\lambda_{i}^{-1/2}$ is the normalization factor and $v_{i}$ are the eigenvectors of the following eigen problem
\begin{equation}
\mathbf{C}v_{i}=\lambda_{i}v_{i},
\end{equation}
where $\mathbf{C} = \int_{\Omega} \mathcal{Q'}^{T}\mathcal{Q'} \ d\Omega$ is the correlation matrix\footnote{Writing the correlation matrix as an integral over the domain $\Omega$ means that we are weighting each variable in the vector $q'$ associated to a lattice point $\mathbf{x}_{i}$ with the corresponding volume $V_{i}$ of a fictitious control volume enclosing that lattice point. For fluctuation data $q'$ collected from unstructured finite element-like grids, the enclosing volume of a point $\mathbf{x}_{i}$ taken at the centroid of a mesh cell is the cell's own volume.}

A standard way to estimate the dimension $L$ of the reduced basis $\Phi$ is the following criterion
\begin{equation}
\frac{\sum_{i=1}^{L} \lambda_{i}}{\sum_{i=1}^{M} \lambda_{i}} \geq \gamma, 
\end{equation}
where $0 < \gamma < 1$ is a threshold that indicates the amount of energy of the original full order system being preserved by the reduced order system.

A simple projection operation is sufficient to recover the temporal coefficients associated to the spatial basis functions
\begin{equation}
a_{j}\left(t_{i}\right) = \left( \mathcal{Q'}\left(\mathbf{x},t_{i}\right) \right)^{T}\Psi_{j}.
\label{time_coeff}
\end{equation}
It provides the necessary inputs for training the artificial neural networks which compose the reduced order models proposed in this work.

\section{Artificial Neural Networks}
As the name suggests, an ANN is a computational system inspired by brain functioning \cite{nielsen2015}. An ANN consists of a collection of units called artificial neurons and virtual wired connections between neurons called edges that mimic synapses in a biological brain. Information is transmitted from input neurons towards output neurons according to the paths imposed by the connections forming a directed graph. Neurons are typically arranged into layers and a sequence of such interconnected layers creates the ANN itself. Usually, weights and biases are associated with edges and neurons, respectively, to adjust the learning process of the network for any specific application. In this work, we consider two types of ANNs: the traditional fully connected neural network (FCNN) \cite{schmidhuber2015} and a specific recurrent neural network (RNN) \cite{miljanovic2012} called the LSTM neural network \cite{hochreiter1997}. The code implementation was written in Python using Numpy and Keras together with the TensorFlow AI library.

\section{ANN-based Reduced Order Models}
The POD method produces a separation of variables of the original dynamical system (cf. equation \ref{rb_approx}). The spatial basis can be computed offline from the snapshot matrix and remains fixed for each simulation data set. The projection of the fluctuating quantities onto the spatial basis provides the input for training the ANN-based ROM that we seek. Such an approach gives us an approximation of the temporal coefficients such that
\begin{equation}
\mathcal{C} \approx \mathcal{A} = \Phi^{T}\mathcal{Q'},
\end{equation}
where $\mathcal{C} = \left[ \mathbf{c}(t_{1}),\ldots,\mathbf{c}(t_{N}) \right], \ \mathcal{C} \in \mathbb{R}^{L\times N}$, analogous to the definition of matrix $\mathcal{A}$.
It is woth noting from equation (\ref{full_space-time_decomp}) that we can easily recover $\mathcal{A}$ since $\Phi$ is an orthonormal matrix. Once the approximation of the temporal coefficients is obtained, we can write the desired reduced basis approximate solution of the original dynamical system as
\begin{equation}
Q_{rb} = \overline{q} + \Phi \ \mathcal{C}.
\label{Q_approx}
\end{equation}
This result accomplishes our goal of having a ROM that works as a proxy model of the original full order system. In the following, we show how to build two types of ANN-based ROMs.

\subsection{FCNN ROM}
After performing the POD method, we can rewrite equation (\ref{rb_approx}) as an exclusive time dependent set of equations
\begin{equation}
Q_{rb}(t) = \overline{q}+q'(t) = \overline{q} + \sum_{i=1}^{L}a_{i}(t)\Psi_{i},
\end{equation}
and we can differentiate them with respect to time to obtain
\begin{equation}
\frac{d Q_{rb}(t)}{dt} = \sum_{i=1}^{L}\Psi_{i}\frac{da_{i}(t)}{dt} = \Phi\mathbf{\dot{a}}(t).
\end{equation}
Noting that equation (\ref{ODE-sys}) can be recast as
\begin{equation}
\frac{d Q_{h}(t)}{d t} = S_{h}(t) - \mathcal{N}_{h}(Q_{h}(t)),
\end{equation}
we can combine both equations in such a way that we isolate the time derivative of the reduced basis coefficients on the left and keep all the nonlinearities and sources of the dynamical system on the right; that is,
\begin{equation}
\mathbf{\dot{a}}(t) \approx \Phi^{T}\left( S_{h}(t) - \mathcal{N}_{h}(Q_{h}(t))\right).
\end{equation}
It suggests that all the complexity of the dynamical system under analysis can be described by the evolution of the temporal coefficients $\mathbf{a}(t)$. Given that, we set up a FCNN as a regression tool to estimate $\mathbf{\dot{a}}(t)$ as follows
\begin{equation}
\mathbf{\dot{a}}(t) \approx \frac{d \mathbf{c}(t)}{d t} = \mathbf{W}^{[l]} \mathbf{G}^{[l-1]}+\mathbf{b}^{[l]},
\end{equation}
where $\mathbf{W}^{[l]}$ is the weight matrix of the last layer of the FCNN, $\mathbf{G}^{[l-1]}$ is the activation matrix of the penultimate layer and $\mathbf{b}^{[l]}$ is the bias vector of the last layer. 

\begin{figure}[h!]
\centering
\vspace{-0.3cm}
\hspace{-0.3cm}
\includegraphics[scale=.158]{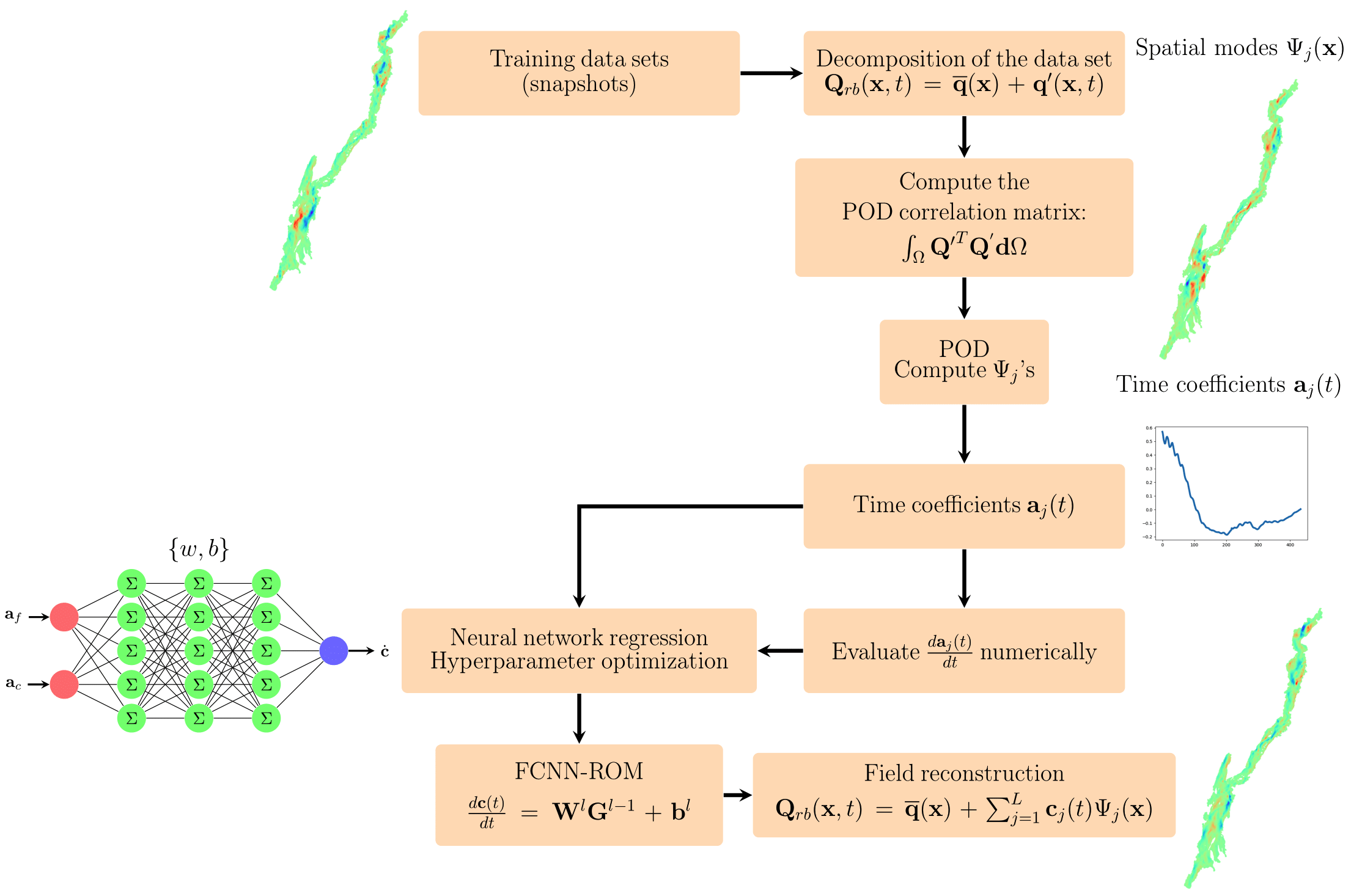}
\caption{Schematic of the FCNN ROM workflow. On the bottom left, the inputs of the ANN $\mathbf{a}_{c}$ and $\mathbf{a}_{f}$ are the time coefficients associated with the lake circulation flow fields and the atmospheric forcing variables, respectively. The output $\mathbf{\dot{c}}$ is the approximation of the time derivative of the time coefficients associated to the water circulation.}
\label{FCNN_workflow}
\end{figure}

Figure \ref{FCNN_workflow} shows a schematic representation of the operations performed to construct the FCNN ROM. The FCNN is designed to receive the temporal coefficients $\mathbf{a}(t)$ as input from the (lake circulation) flow field variables and the atmospheric forcing terms (cf. input vectors $\mathbf{a}_{c}$ and $\mathbf{a}_{f}$ on the bottom left of Figure \ref{FCNN_workflow}, respectively) and to return an approximation of the time derivative of the flow field coefficients as output. Such an approach is inspired by the work of \cite{lui2019} which focused on computational fluid dynamics applications, though it did not depend on external atmospheric forcing terms.

The FCNN's loss function is minimized with respect to the time derivative of the temporal coefficients $\mathbf{a}(t)$, computed with a high resolution finite difference scheme. To recover the approximate time coefficients $\mathbf{c}(t)$ computed by the FCNN, we use a 5-stage 4th order strong stability preserving Runge-Kutta (RK) scheme. Such an approach allows us to predict the field variables of the dynamical system beyond the training window because $\overline{q}(\mathbf{x})$ and $\Psi_{i}(\mathbf{x})$ depend only on the spatial coordinates. Furthermore, RK time-steps used to reconstruct field variables don't need to be the same size as those used for training the FCNN, ensuring great flexibility for the FCNN ROM.

\vspace{-0.8cm}

\subsection{LSTM ROM}
The LSTM neural network is a type of Recurrent Neural Network (RNN) which has been primarily utilized in speech modeling \cite{sak2014} and language translation \cite{sutskever2014}. However, some authors have applied it successfully in forecasting problems including turbulent flows \cite{mohan2018}, weather prediction \cite{karevan2020}, and runoff applications \cite{xiang2020}.

For a given time series $a_i(t)$, the LSTM uses a double-leaf moving window with input and output time lengths $\Delta t_{in}$ and $\Delta t_{out}$, respectively, to create a set of training sequences. Upon minimization of the loss function, given a section of the time series with length $\Delta t_{in}$ on the left leaf of the moving window, the LSTM learns how to predict the adjacent section of the time series with length $\Delta t_{out}$ on the right leaf of the moving window. Once trained, the LSTM can perform a forecast of length $\Delta t_{out}$ using the last $\Delta t_{in}$ section of the time series. If we want to stretch the forecast length for multiple time lengths $\Delta t_{out}$, we feedback the trained LSTM with each predicted extrapolation.

Looking at the time coefficients $a_{i}(t)$ in equation (\ref{rb_approx}), we can interpret them as a set of $L$ time series of length $N$ associated with each spatial mode $\Psi_{i}$. Given that, we use an LSTM neural network with a suitable moving window to estimate the temporal coefficients $\mathbf{a}(t)$ directly so that we can forecast the dynamical system under analysis by extrapolating those time series. Equation (\ref{time_coeff}) provides the necessary input for training the LSTM ROM.

We should remark that the LSTM ROM workflow is nearly identical to that of the FCNN ROM except that the LSTM neural network does not rely on the derivative of temporal coefficients ($\mathbf{\dot{a}}(t)$) to minimize the loss function. The FCNN box in Fig. \ref{FCNN_workflow}, which computes weights and biases to estimate $\mathbf{\dot{a}}(t)$, can therefore be replaced by an LSTM neural network that outputs the temporal coefficients $\mathbf{a}(t)$ themselves.

\section{Numerical Results}
This section demonstrates the effectiveness of the proposed ANN-based ROMs via two examples: predicting the state of a discrete system called a Lorenz attractor, and the movement of water in a freshwater lake. Note that while the Lorenz attractor is a discrete system and does not require dimensionality reduction, it is, from a theoretical standpoint, a suitable case study for evaluating the ability of ANNs to forecast deterministic and chaotic physical behavior.

\subsection{Lorenz Attractor}
The Lorenz attractor is a set of chaotic solutions of the Lorenz system which consists of the following set of three ODEs:
\begin{align}
\frac{\mathrm{d}x}{\mathrm{d}t} &= \sigma (y - x), \\[6pt]
\frac{\mathrm{d}y}{\mathrm{d}t} &= x (\rho - z) - y, \\[6pt]
\frac{\mathrm{d}z}{\mathrm{d}t} &= x y - \beta z.
\end{align}
This system, originally developed by Edward Lorenz in the early 1960s, was an attempt to represent atmospheric convection through a two dimensional fluid flow of uniform depth with an imposed temperature difference in the vertical direction. Thus, in the system above, $x$ is a quantity describing the rate of convection, $y$ is the horizontal temperature variation, and $z$ is the vertical temperature variation. Parameters $\sigma$, $\rho$ and $\beta$ are the Prandtl number, a number proportional to the Rayleigh number and a geometric factor, respectively.

The time derivatives of the coordinates $x$, $y$ and $z$ can be interpreted as three separate time series. Once this system of three ODEs is numerically integrated along a period $T$ using a time step $dt$, we have the necessary dataset to train and test the ANN which models the Lorenz system behavior. As noted above, a dimensionality reduction is not required.

In this numerical example, we trained a FCNN to predict the deterministic and chaotic regimes of the Lorenz system. As the chaotic regime was supposed to be much harder to predict than the deterministic one, we performed a hyperparameter optimization using the library Optuna (https://optuna.org/) to find the best neural network architecture for predicting the chaotic case. Then we used the same FCNN to predict both system behaviors. The best hyperparameters configuration is shown in table \ref{table:param_setup}.

\begin{table}[h!]
\vspace{-0.3cm}
\caption{Best FCNN hyperparameters configuration.}
\centering
\begin{tabular}{cccccc}
\hline
Neurons & Learning & $L^2$ & Activation & Loss & Optimizer \\
per layer & rate & Regularization & function & function &   \\
\hline
$\lbrace3, 24, 25, 33, 22, 26, 3\rbrace$ & 5.2e-06 & 2.3e-05 & ELU & Mean Squared & ADAM /  \\
  &  &  &  & Error & L-BFGS-B \\
\hline
\end{tabular} 
\label{table:param_setup}
\end{table}

The ANN's weights and biases were optimized by two algorithms applied successively. First, the ADAM algorithm iterates 2000 cycles and then the L-BFGS-B scheme completes the optimization process until the convergence is reached. ADAM is a variant of the classical Stochastic Gradient Descent (SGD) method which is an iterative method for optimizing an objective function. Similarly, L-BFGS-B is a second order quasi-Newton algorithm that performs the same task with the advantage of showing greater accuracy. As the former method is computationally more efficient, it gives a quick enhanced initial guess to the latter. The physical parameters of the deterministic Lorenz system were set as: $\rho = 14.0, \ \sigma = 10.0, \ \beta = 8/3, \ T = 12.0s, \ dt = 0.001s$.

\begin{figure}[h!]
\vspace{-0.5cm}
\centering
\begin{tabular}{cccc}
\hspace{-0.3cm}
\includegraphics[scale=0.21]{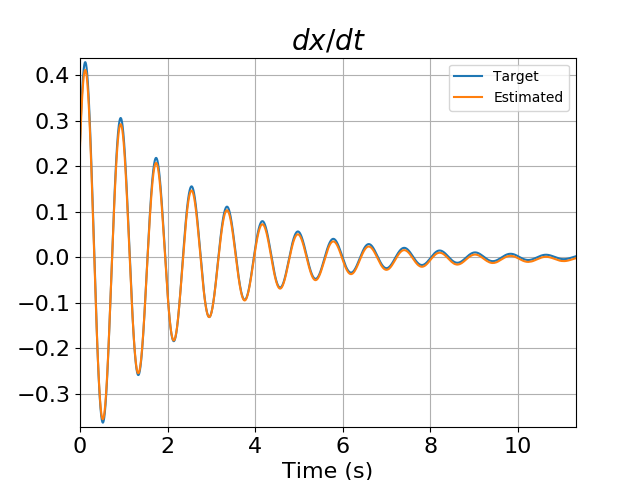} &
\hspace{-0.5cm}
\includegraphics[scale=0.21]{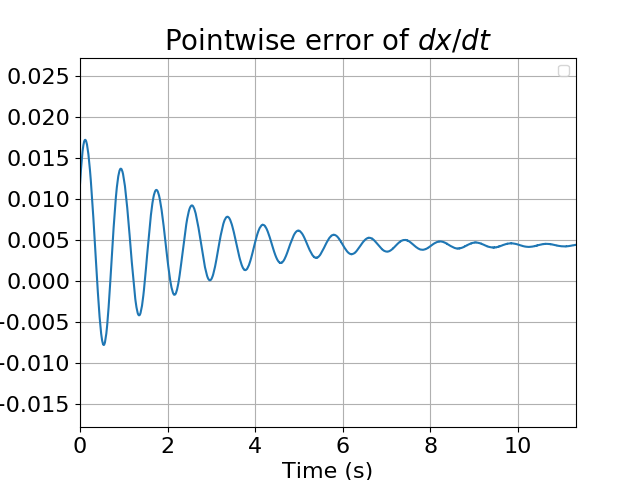} &
\hspace{-0.5cm}
\includegraphics[scale=0.21]{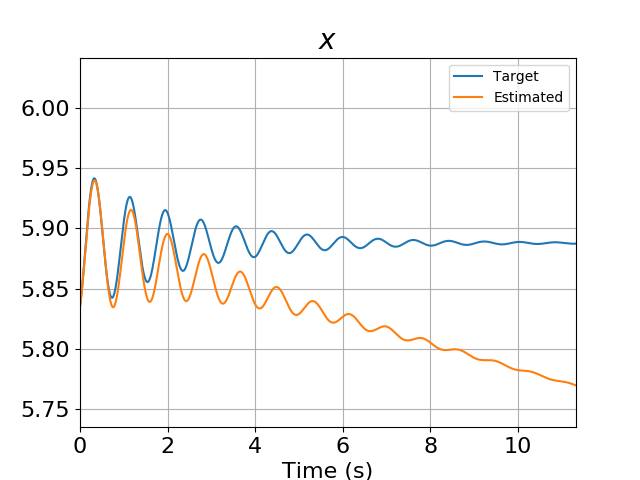} &
\hspace{-0.5cm}
\includegraphics[scale=0.21]{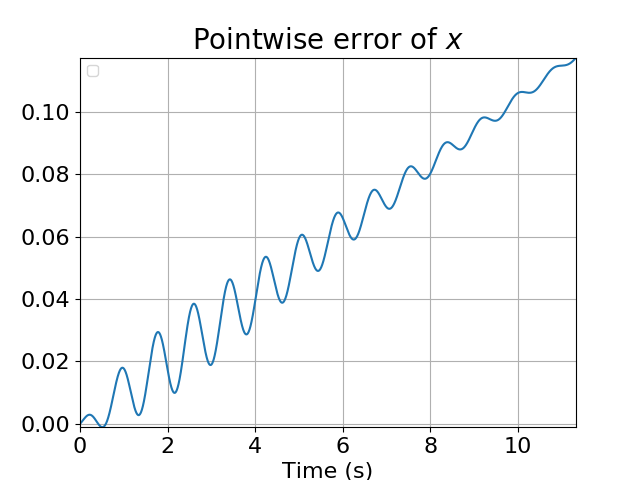} \\
\hspace{-0.3cm}
\includegraphics[scale=0.21]{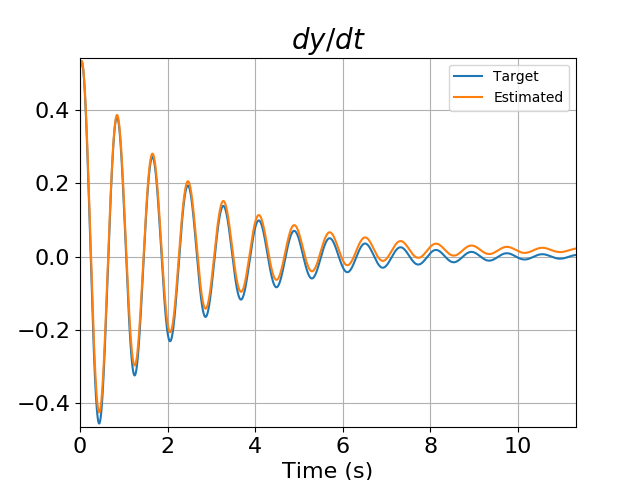} &
\hspace{-0.5cm}
\includegraphics[scale=0.21]{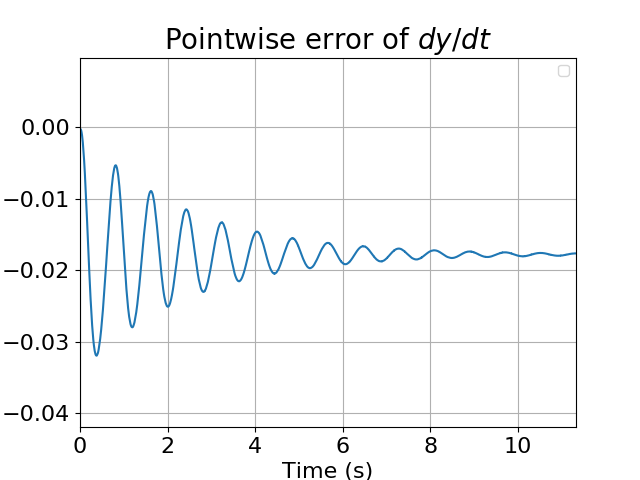} &
\hspace{-0.5cm}
\includegraphics[scale=0.21]{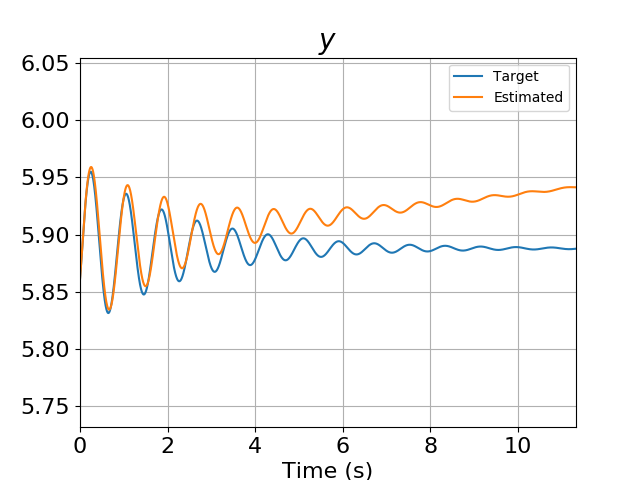} &
\hspace{-0.5cm}
\includegraphics[scale=0.21]{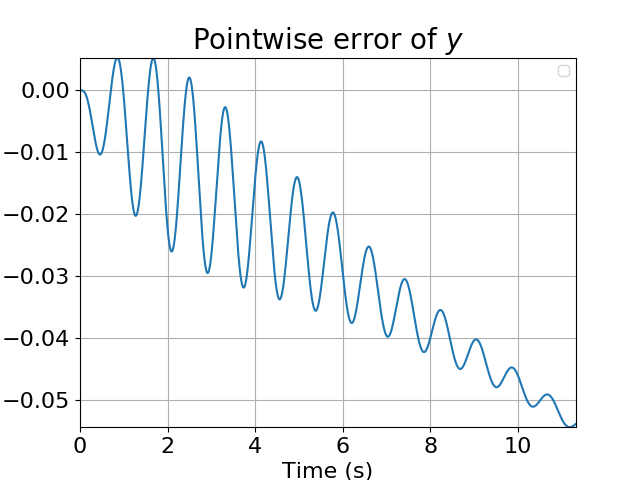} \\
\hspace{-0.3cm}
\includegraphics[scale=0.21]{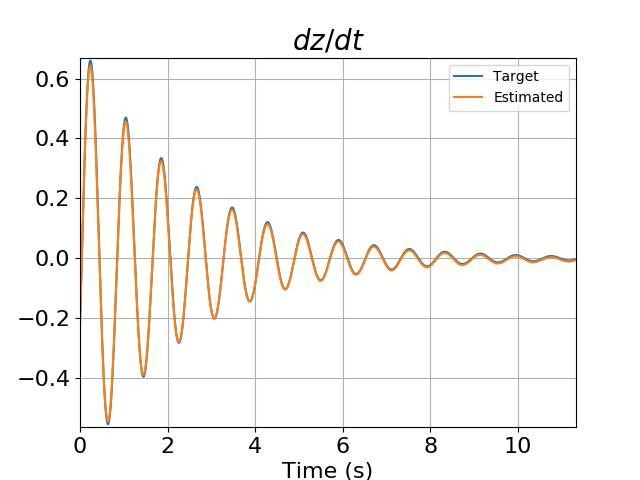} &
\hspace{-0.5cm}
\includegraphics[scale=0.21]{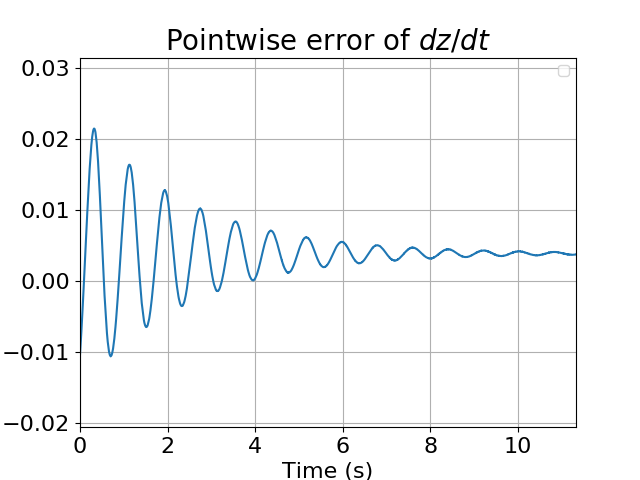} &
\hspace{-0.5cm}
\includegraphics[scale=0.21]{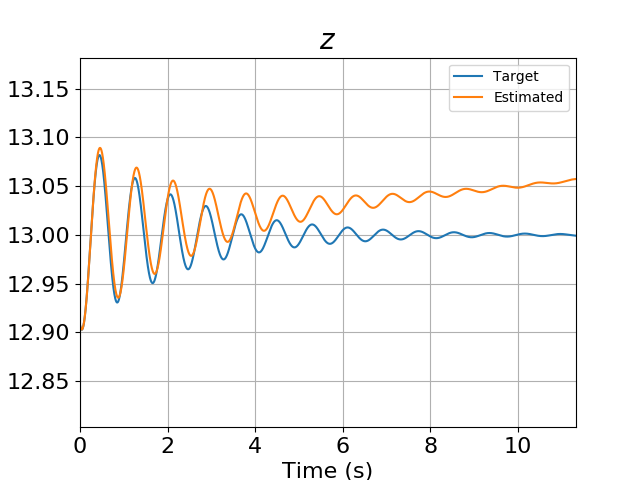} &
\hspace{-0.5cm}
\includegraphics[scale=0.21]{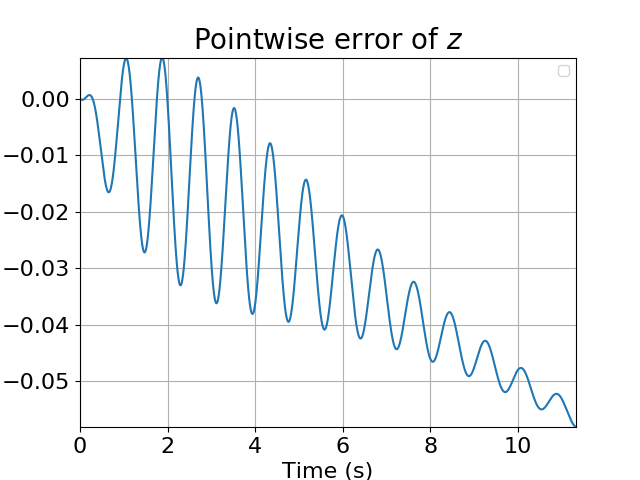} \\
\hspace{-0.3cm}
a) & b) & c) & d)
\end{tabular}
\caption{Deterministic Lorenz system: a) $\dot{x}, \dot{y}, \dot{z}$ true and predicted; b) $error_{\dot{x}} = \dot{x}_{true}-\dot{x}_{pred}, \ error_{\dot{y}} = \dot{y}_{true}-\dot{y}_{pred}, \ error_{\dot{z}} = \dot{z}_{true}-\dot{z}_{pred}$; c) $x, y, z$ true and predicted; d) $error_{x} = x_{true}-x_{pred}, \ error_{y} = y_{true}-y_{pred}, \ error_{z} = z_{true}-z_{pred}$.}
\label{lorenz_determ}
\end{figure}

Figure \ref{lorenz_determ} shows plots of the discrete variables and their time derivatives compared to the true solutions over time, along with the absolute errors\footnote{Relative errors are not suited for the analysis because the time derivative of the reference solution has many values close to zero.}. We observe that the FCNN can predict variables $x$, $y$ and $z$ with moderate accuracy only for a short period (about two seconds) since the errors of the time derivatives of the discrete variables are one order of magnitude smaller than the derivative values themselves. However, all time derivatives reach the system steady state with some positive or negative bias resulting in a clear discrepancy between predicted and reference solutions for $x$, $y$ and $z$. As the architecture of the FCNN was optimized to predict persistent unsteady behavior, typical of chaotic regimes, it was expected that only the oscillatory patterns would be captured by the ANN in the deterministic case.

\begin{figure}[h!]
\vspace{-0.5cm}
\centering
\begin{tabular}{cccc}
\hspace{-0.3cm}
\includegraphics[scale=0.21]{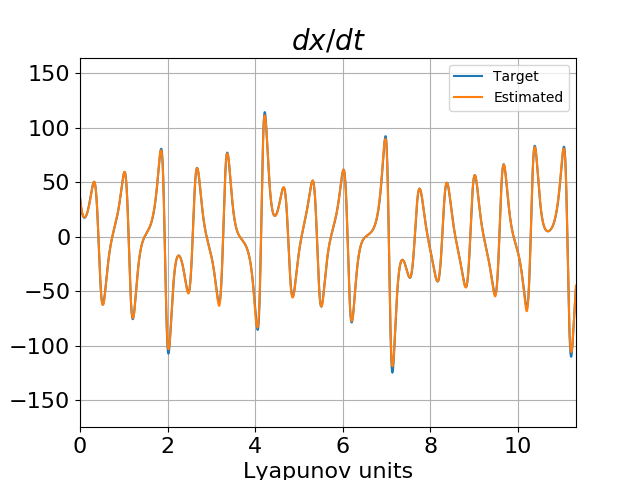} &
\hspace{-0.5cm}
\includegraphics[scale=0.21]{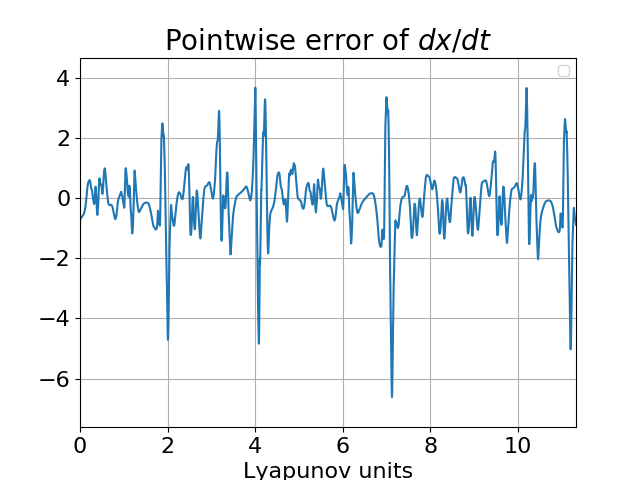} &
\hspace{-0.5cm}
\includegraphics[scale=0.21]{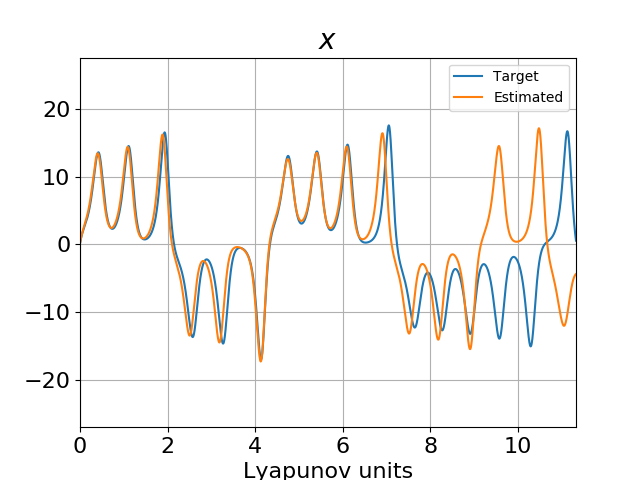} &
\hspace{-0.5cm}
\includegraphics[scale=0.21]{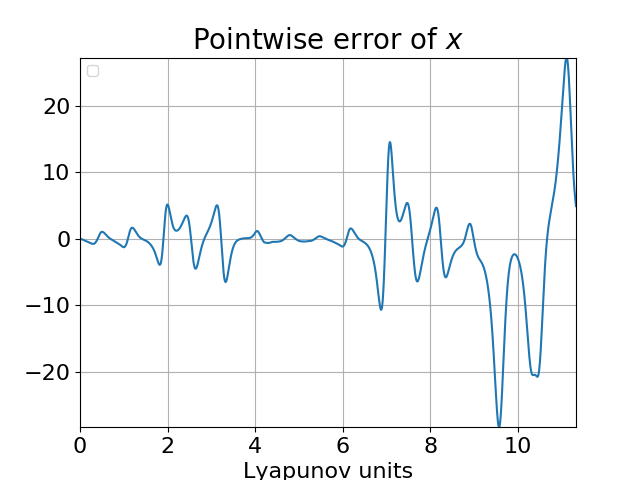} \\
\hspace{-0.3cm}
\includegraphics[scale=0.21]{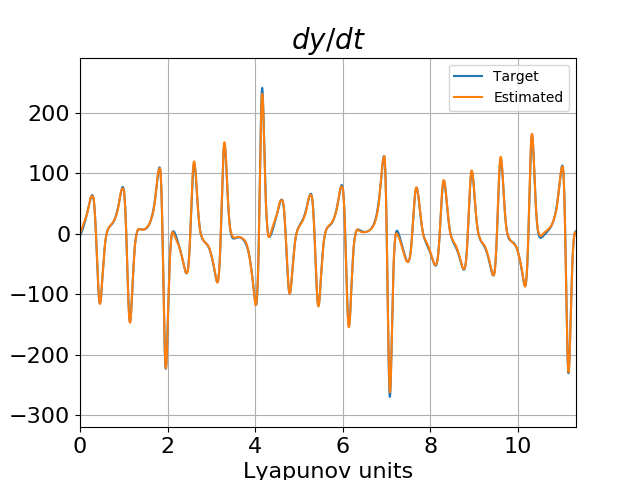} &
\hspace{-0.5cm}
\includegraphics[scale=0.21]{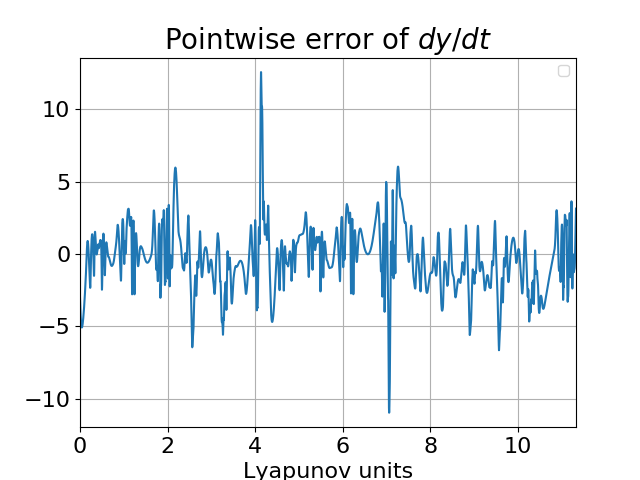} &
\hspace{-0.5cm}
\includegraphics[scale=0.21]{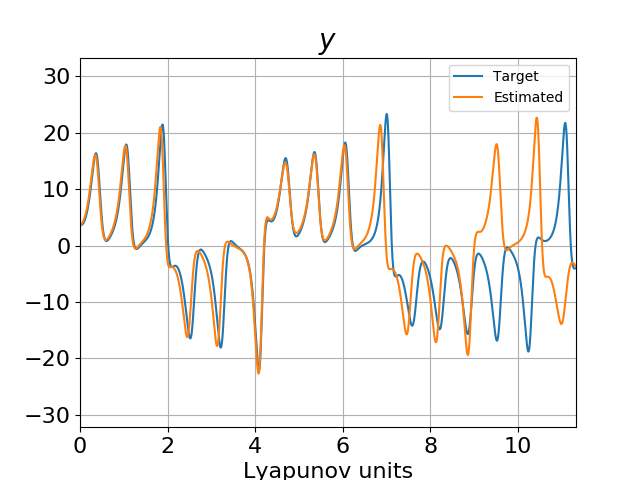} &
\hspace{-0.5cm}
\includegraphics[scale=0.21]{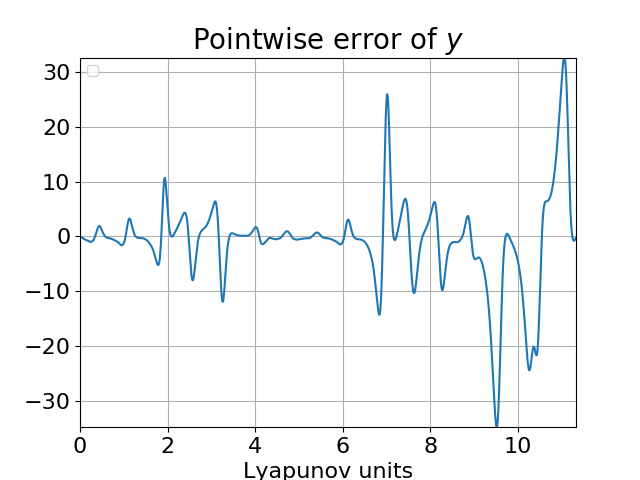} \\
\hspace{-0.3cm}
\includegraphics[scale=0.21]{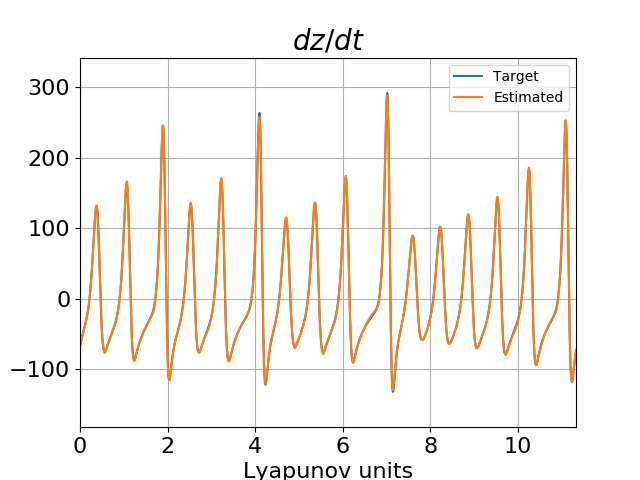} &
\hspace{-0.5cm}
\includegraphics[scale=0.21]{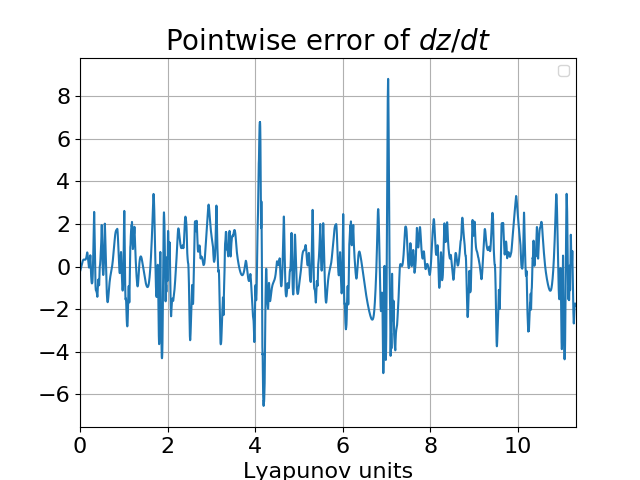} &
\hspace{-0.5cm}
\includegraphics[scale=0.21]{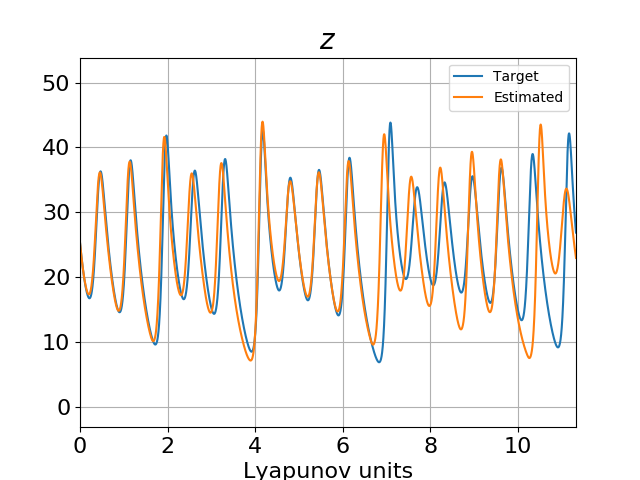} &
\hspace{-0.5cm}
\includegraphics[scale=0.21]{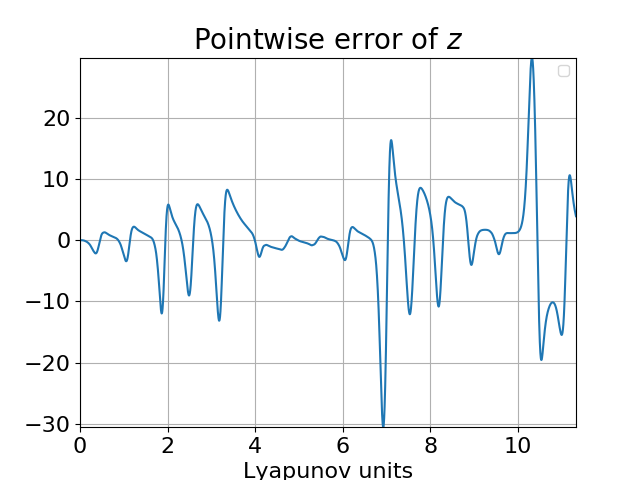} \\
\hspace{-0.3cm}
a) & b) & c) & d)
\end{tabular}
\caption{Chaotic Lorenz system: a) $\dot{x}, \dot{y}, \dot{z}$ true and predicted; b) $error_{\dot{x}} = \dot{x}_{true}-\dot{x}_{pred}, \ error_{\dot{y}} = \dot{y}_{true}-\dot{y}_{pred}, \ error_{\dot{z}} = \dot{z}_{true}-\dot{z}_{pred}$; c) $x, y, z$ true and predicted; d) $error_{x} = x_{true}-x_{pred}, \ error_{y} = y_{true}-y_{pred}, \ error_{z} = z_{true}-z_{pred}$.}
\label{lorenz_chaotic}
\end{figure}

Figure \ref{lorenz_chaotic} shows the same distribution of plots for the chaotic regime, which has the same physical parameters as in the deterministic system except for $\rho = 28.0$. In this case, we report time in terms of Lyapunov timescale which mirrors the limits of the predictability of the system: counting the number of Lyapunov units over which the ANN matches the dynamical system's patterns provides a threshold beyond which the ANN will fail to predict the system's behavior. In the numerical experiments, we computed the Lyapunov unit as $\approx 1/\lambda_{max} = 1.06 s.$, where $\lambda_{max}$ is the system's largest positive Lyapunov exponent (LLE). We observe that the errors of the predicted time derivatives of the discrete variables are two orders of magnitude smaller than the maximum values reached by the derivatives themselves (Figs. \ref{lorenz_chaotic}a and \ref{lorenz_chaotic}b). Such performance is sufficient to provide a remarkable correspondence between predicted and reference time series for all discrete variables $x$, $y$ and $z$ for nearly 9 Lyapunov units, as shown in Fig. \ref{lorenz_chaotic}c. It is clear that hyperparameter optimization tuned for the chaotic regime increased the quality of the FCNN prediction substantially. However, it is not obvious how to extend such predictability limits to realistic hydrodynamics systems since the LLE are not readily available for such complex systems.

For the sake of brevity, we omitted the results of the LSTM approach for the Lorenz system as they are quite similar to the FCNN ones.

\vspace{-0.8cm}

\subsection{Hydrodynamics at Lake George}
Lake George is a freshwater lake in upstate New York, USA. It is considered of medium size ($51.5 \ km$ x $2.15 \ km$) with a total surface area of $117.4 \ km^{2}$. It is the subject of a multiyear research effort called The Jefferson Project that has a goal of understanding the impact of human activity on fresh water, and how to mitigate those effects. A operational prediction system has developed as part of The Jefferson Project to perform short-term (1-7 day) forecasts of the weather, hydrology and hydrodynamics (\emph{viz.} water circulation) at Lake George \cite{watson2018}.

Currently, hydrodynamic predictions are performed daily using the Stanford unstructured-grid, nonhydrostatic, parallel coastal ocean model (SUNTANS). SUNTANS solves the three dimensional Reynolds-averaged Navier-Stokes (RANS) equations on an unstructured, horizontal grid and fixed z-level vertical domain using a finite-volume (FV) discretization. SUNTANS has been extensively used in a variety of hydrodynamic applications with good results \cite{fring06}. For Lake George, the SUNTANS grid has a varying horizontal resolution of $10-50 \ m$ and a vertical resolution of $0.5\ m$ close to the surface and $1.6 \ m$ at the deepest point of the lake. Its configuration for daily forecasts at Lake George contains over 40,000 grid cells and takes around $1 \ 1/2$ hours to complete a 36 hour forecast using 30 high-end processing cores.

To simulate realistic conditions, SUNTANS requires a realistic atmospheric forcing. For Lake George, this is provided by the Weather Research and Forecasting (WRF) model v3.9.1 \cite{Skamarock2008}, which has been configured to generate daily, 36 hour forecasts for Lake George at 0.33 km horizontal resolution. More details about the model setup can be seen in \cite{watson2018}. WRF is said to be one-way coupled to SUNTANS (i.e., the forecasted state of SUNTANS does not feedback into WRF).

The time duration and obvious energy requirements demanded by SUNTANS motivates us to develop a hydrodynamic surrogate model using ANN-based ROMs. In the following, we discuss the numerical experiments made with the FCNN and LSTM ROMs. Note that we are not attempting to reconstruct the full 3D hydrodynamic state of Lake George; rather, we focus on reconstructing the 2D depth-averaged values of four variables: temperature, density, and northward and eastward water velocity. With such a simplified reconstruction, a ML-based surrogate takes only a couple of seconds to perform a 36 hour water circulation forecast. Compared with the $1 \ 1/2$ hours of a regular forecast made by SUNTANS, it demonstrates a significant prediction speed up. While a time complexity analysis for the forward propagation of feed forward neural networks results in $O(N^{4})$, such an analysis seems to be virtually unattainable given the huge complexity of a PDE solver based on a FV discretization method like SUNTANS.

Training data was provided by daily WRF and SUNTANS forecasts with 10 minute output resolution for the period from April $1^{st}$ to $20^{th}$, 2019. Each forecast overlaps by 12 hours and was included in the training. Data from a single 36 hour forecast by WRF and SUNTANS on April $21^{st}$, 2019 was used for testing. The data contained four atmospheric variables from WRF: surface air temperature and pressure, and northward and eastward surface wind velocity; and four depth-averaged hydrodynamic variables from SUNTANS: density, temperature and northward and eastward water velocity. All data was collected from the centroid of each SUNTANS cell (including the WRF data which was projected onto the SUNTANS grid).

The aim is to generate a 36 hour depth-averaged hydrodynamics forecast of Lake George starting from April $21^{st}$, with a $5\%$ error in the L2-norm with respect to the full order approximate solution (given by SUNTANS across the same forecast interval).

\subsubsection{FCNN ROM for hydrodynamics}
\begin{figure}[h!]
\vspace{-0.8cm}
\centering
\begin{tabular}{ccccc}
\hspace{-0.3cm}
\includegraphics[scale=0.064]{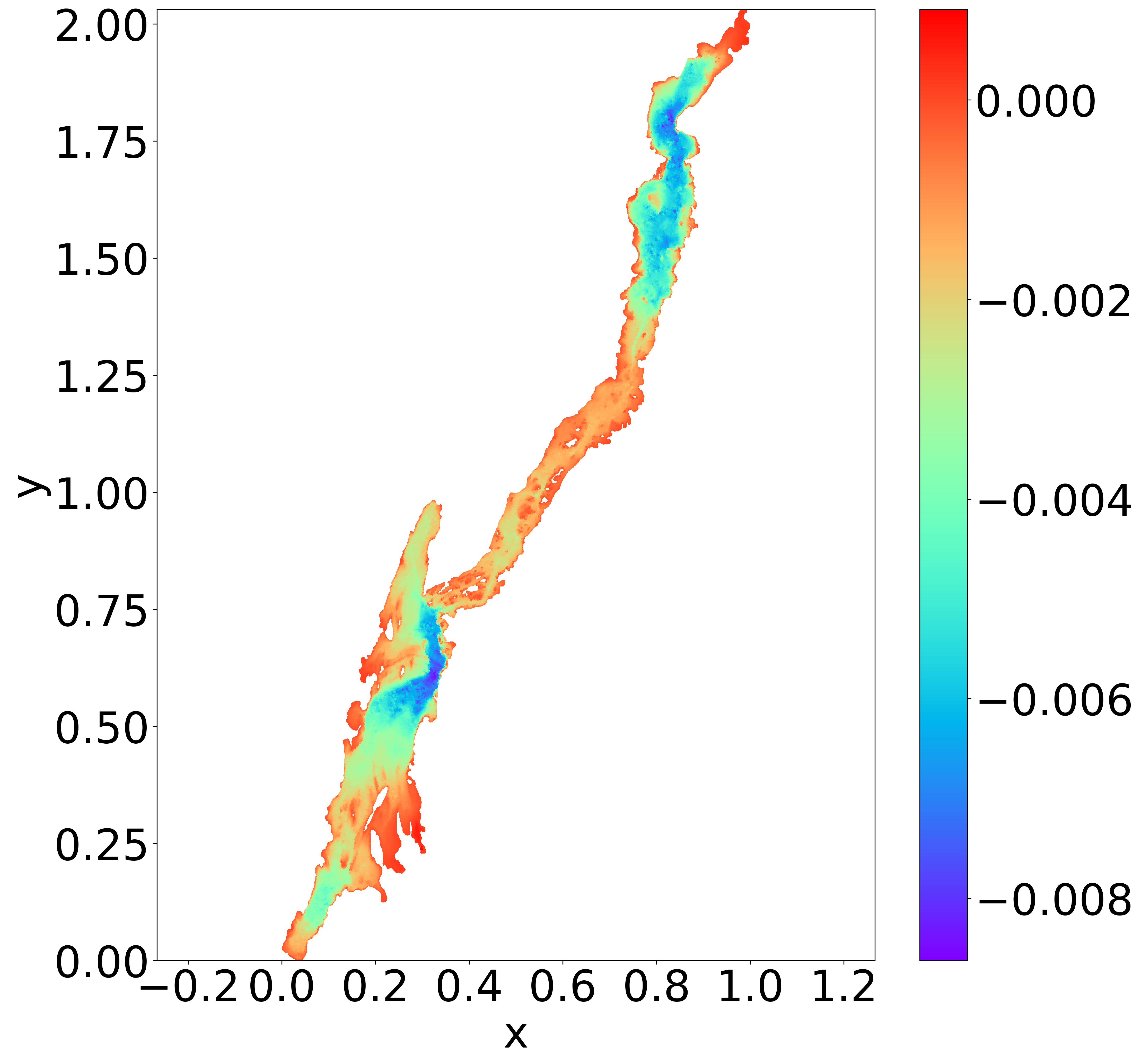} &
\includegraphics[scale=0.064]{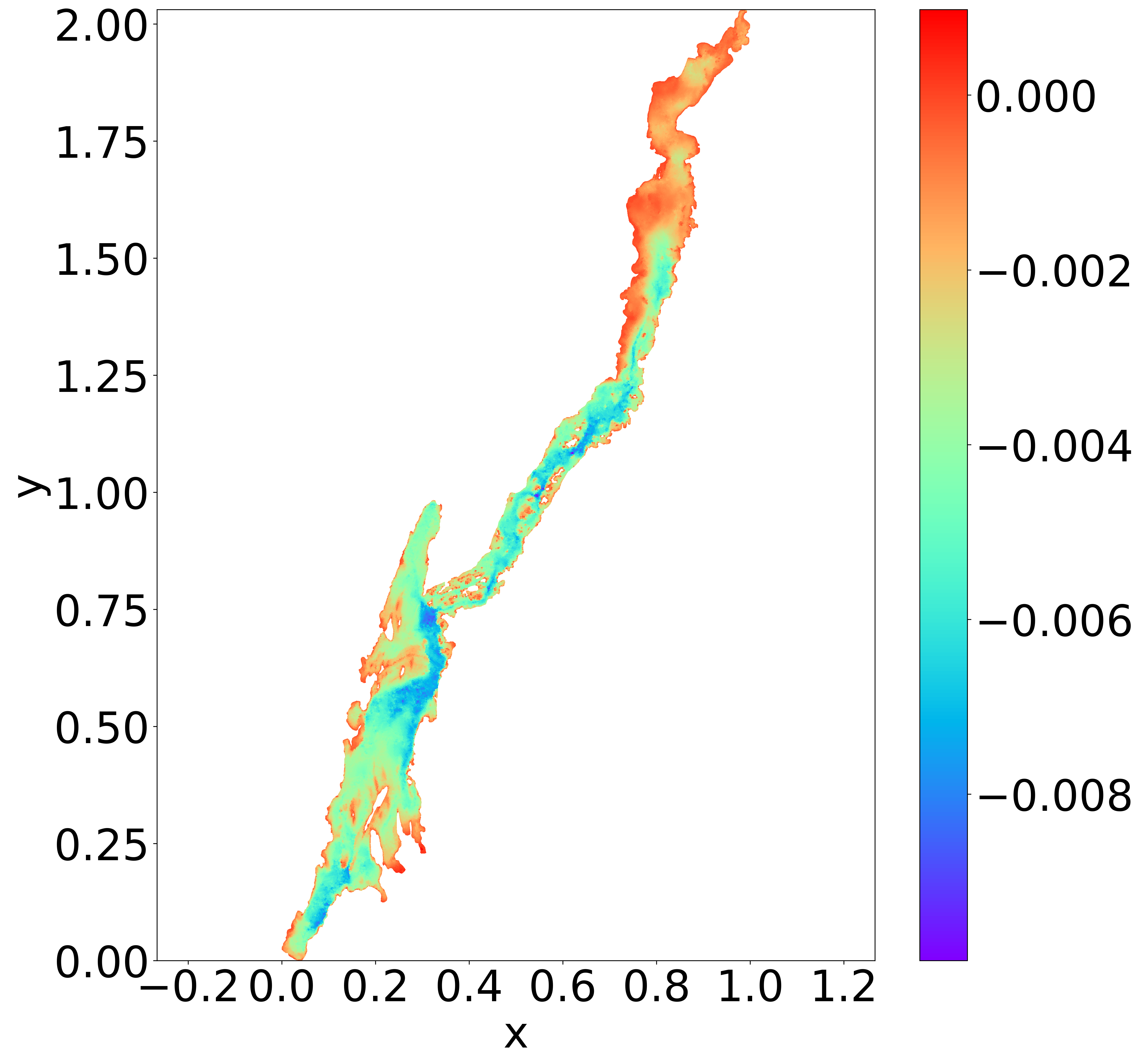} &
\includegraphics[scale=0.064]{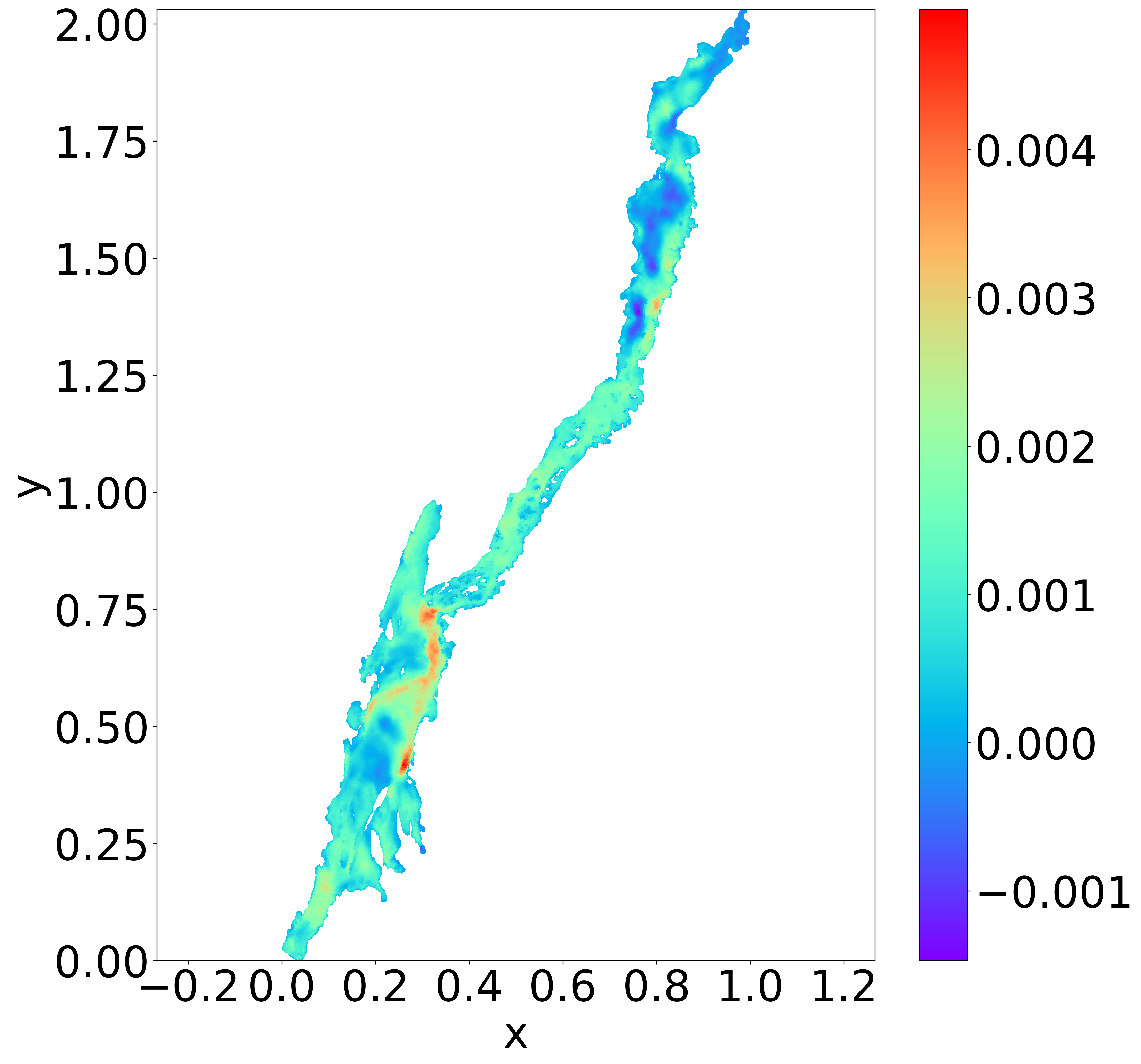} &
\includegraphics[scale=0.064]{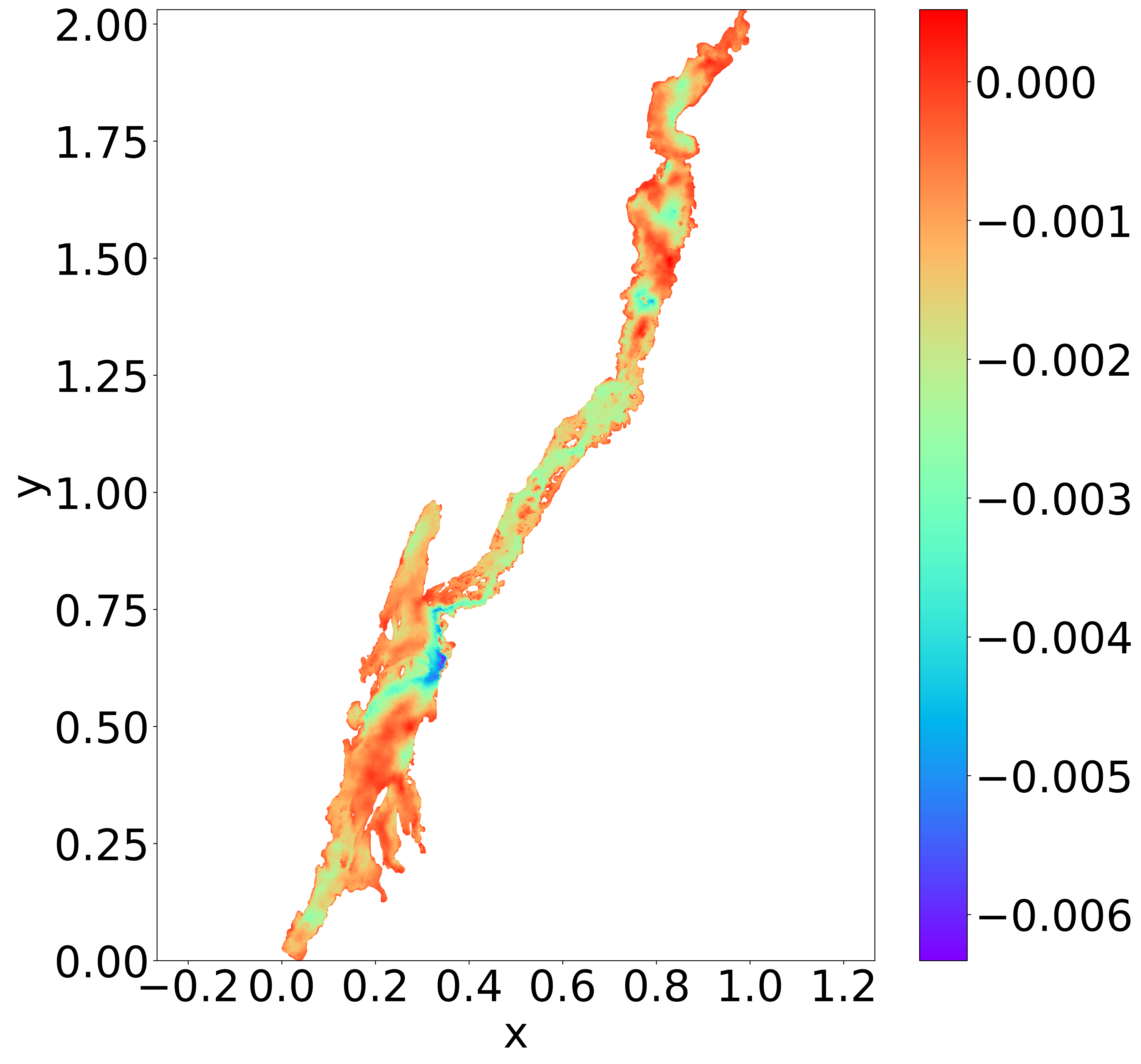} &
\includegraphics[scale=0.064]{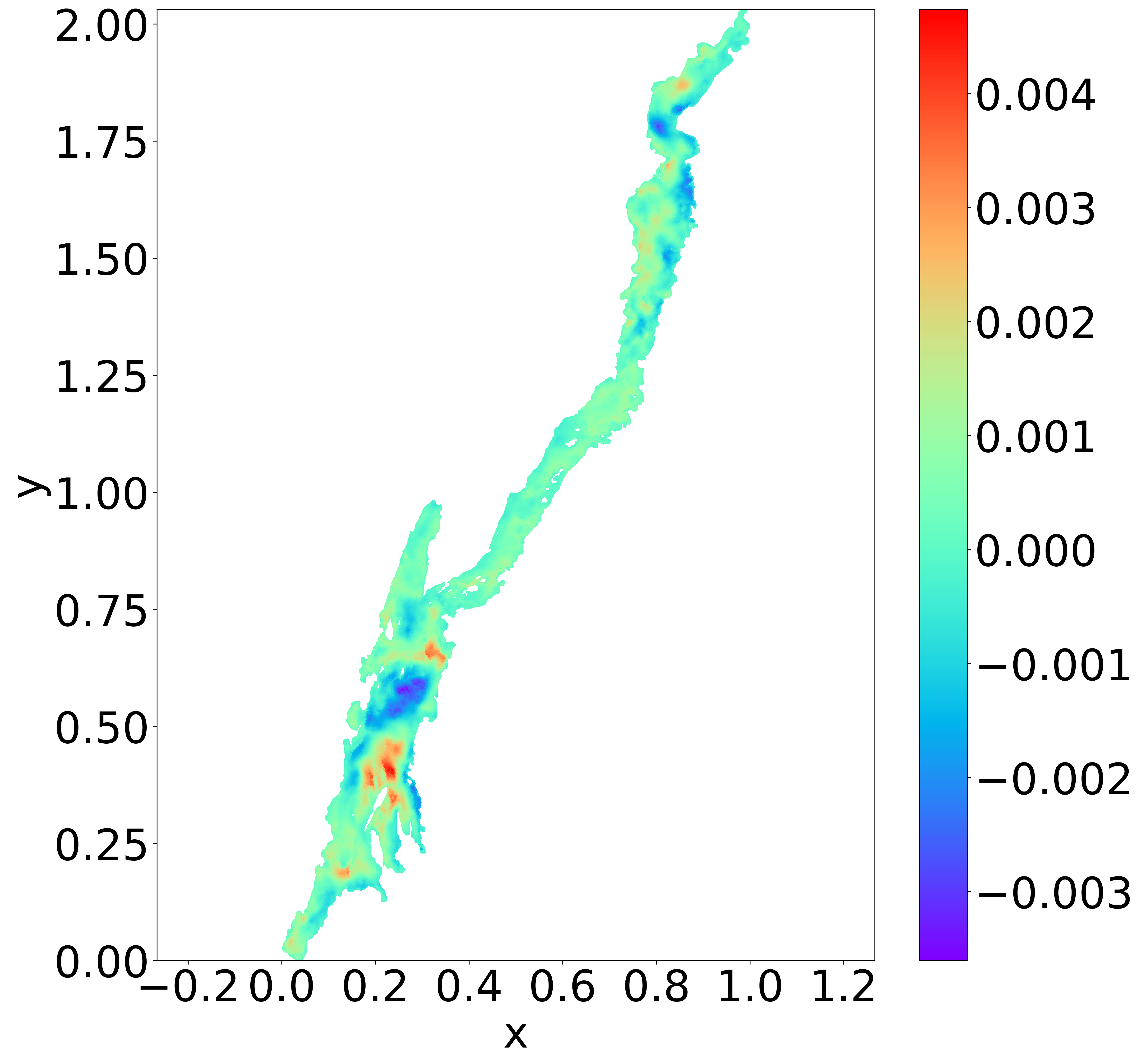} \\
\hspace{-0.3cm}
\includegraphics[scale=0.064]{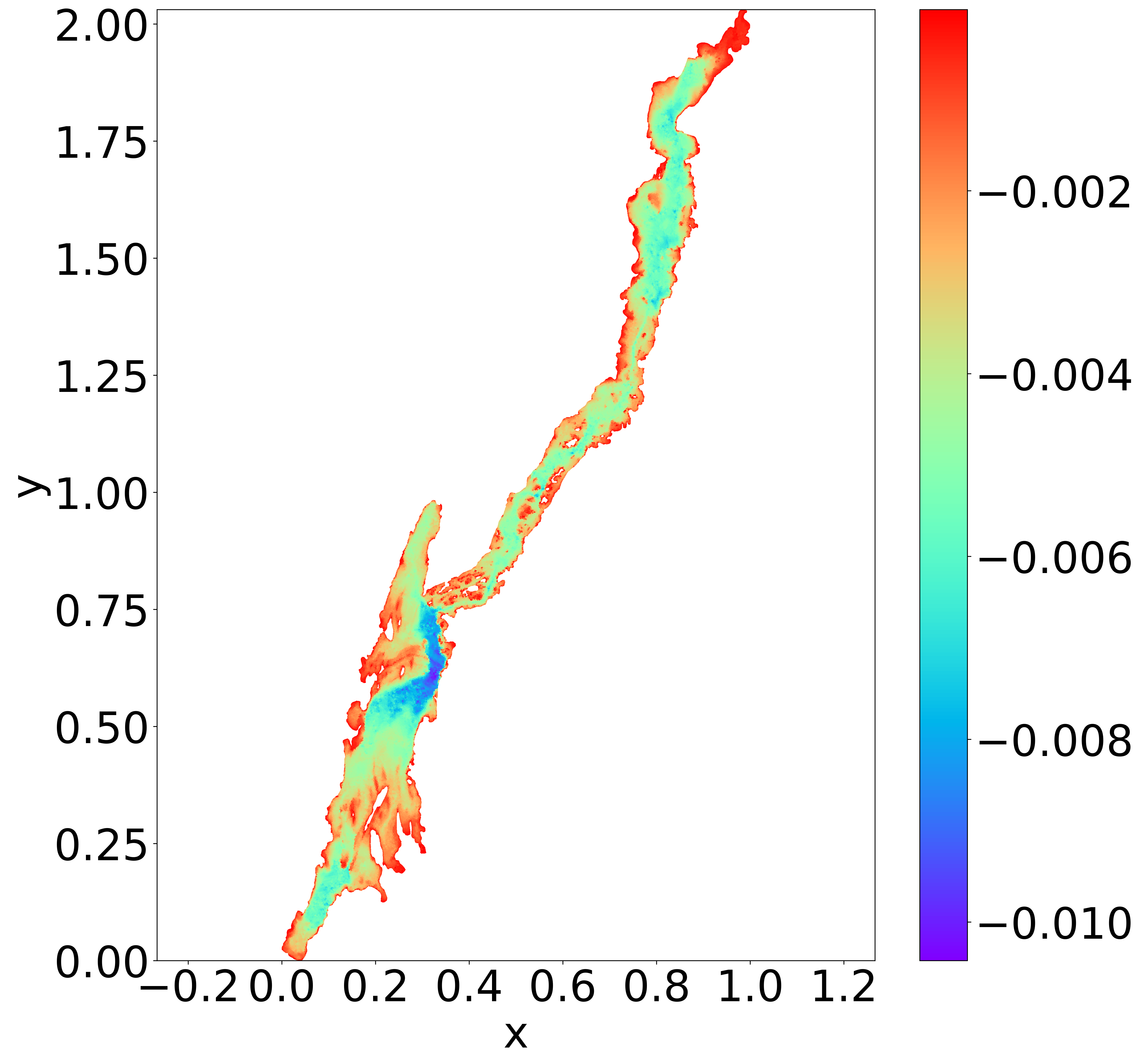} &
\includegraphics[scale=0.064]{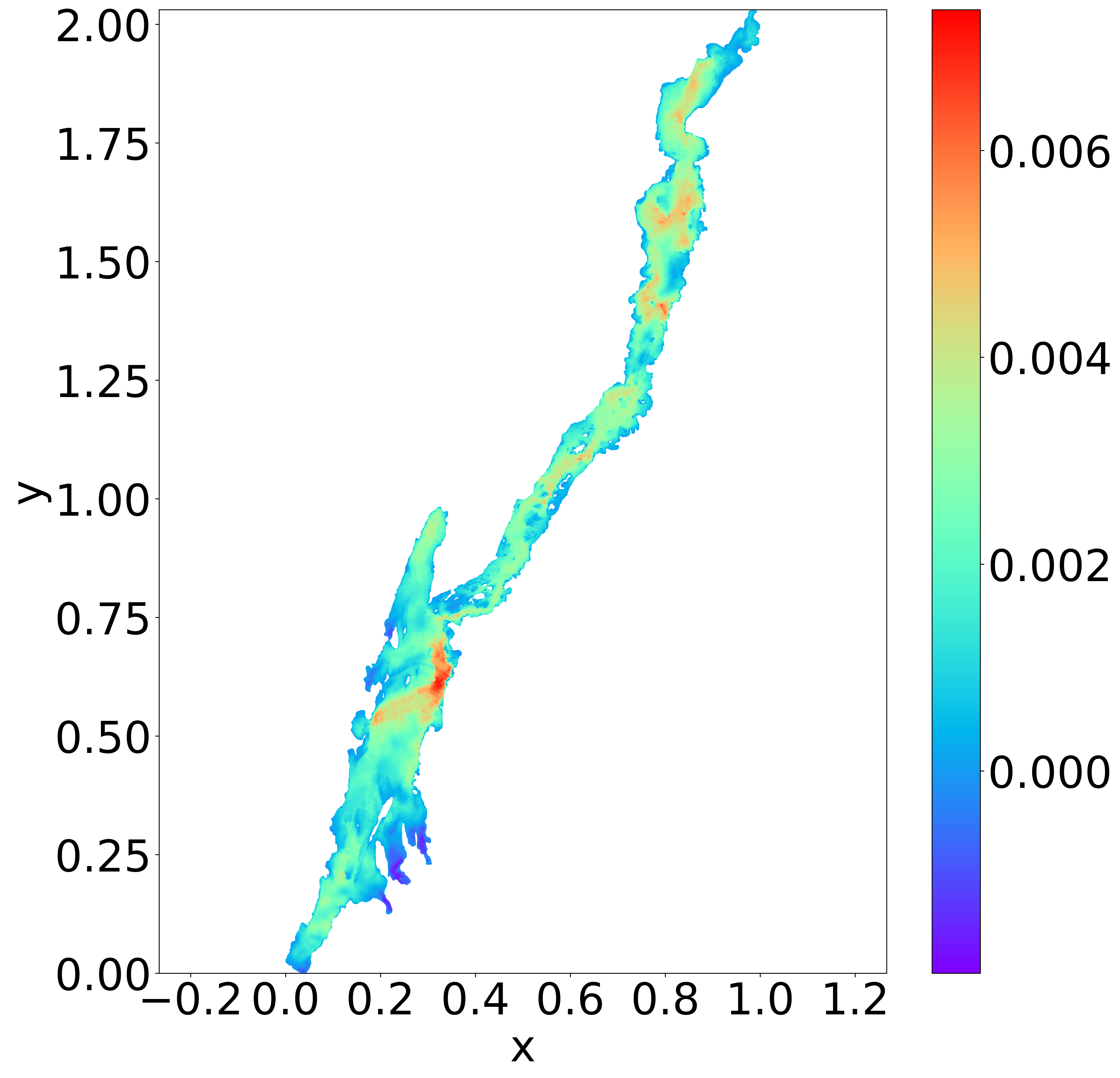} &
\includegraphics[scale=0.064]{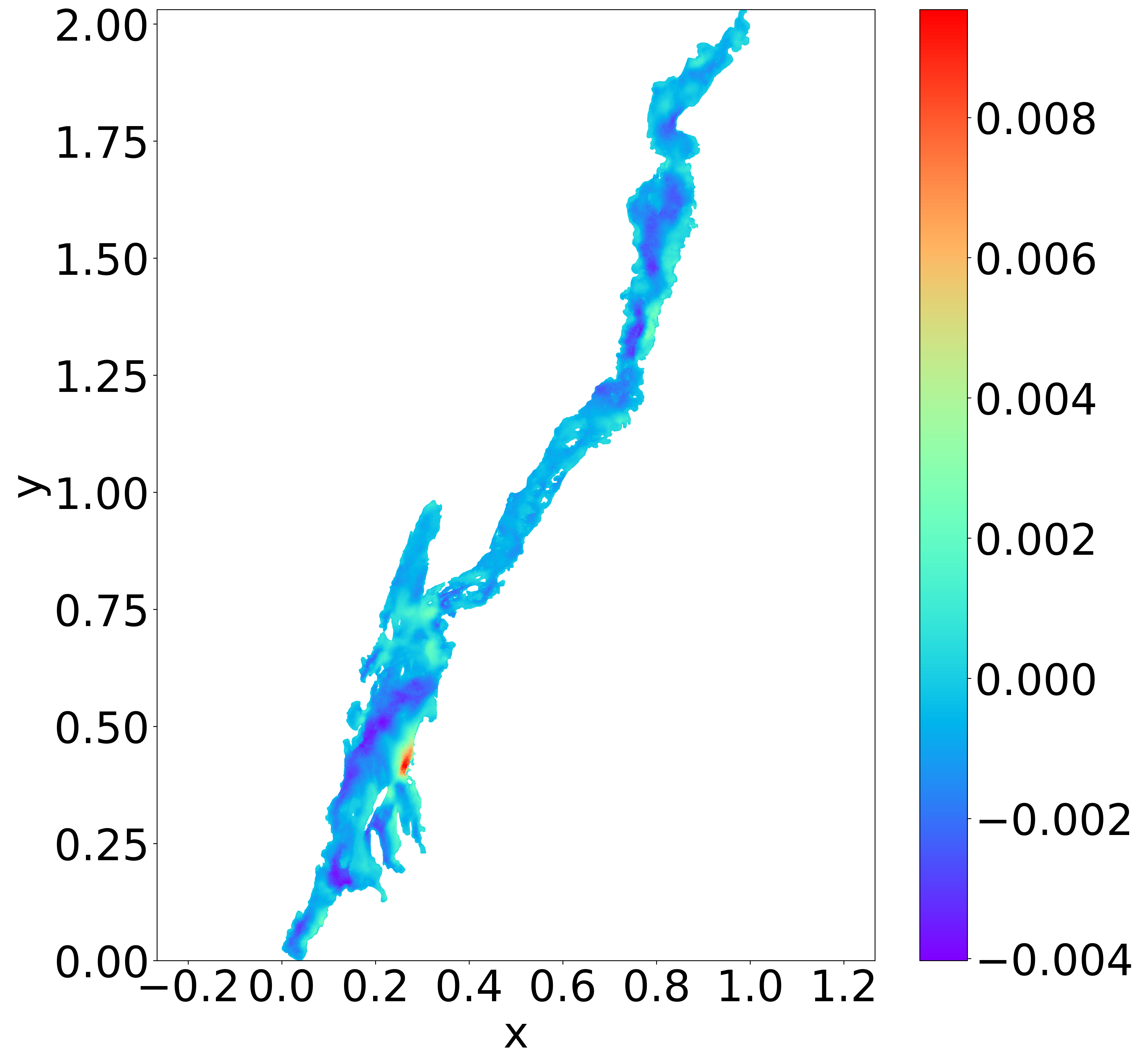} &
\includegraphics[scale=0.064]{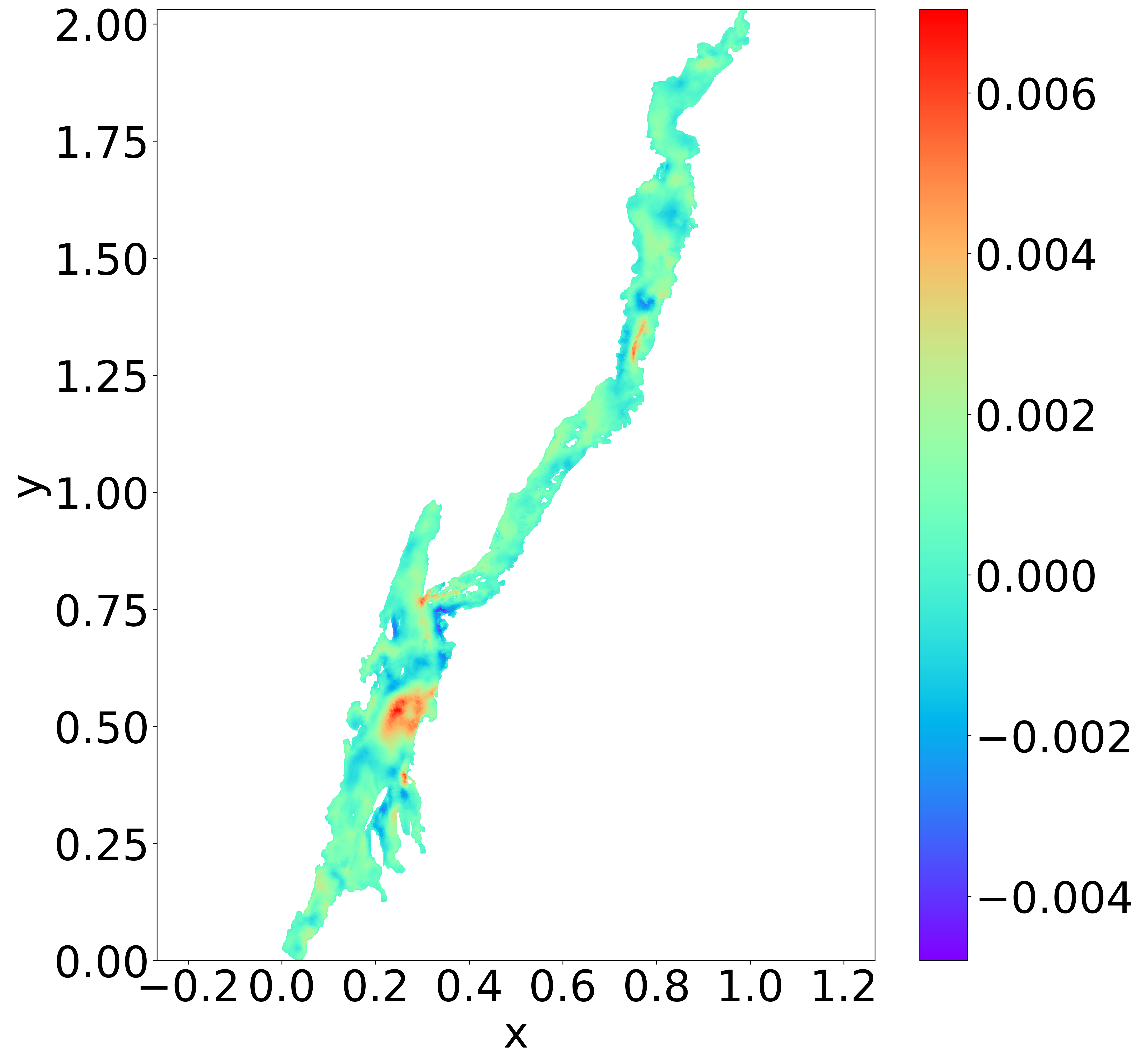} &
\includegraphics[scale=0.064]{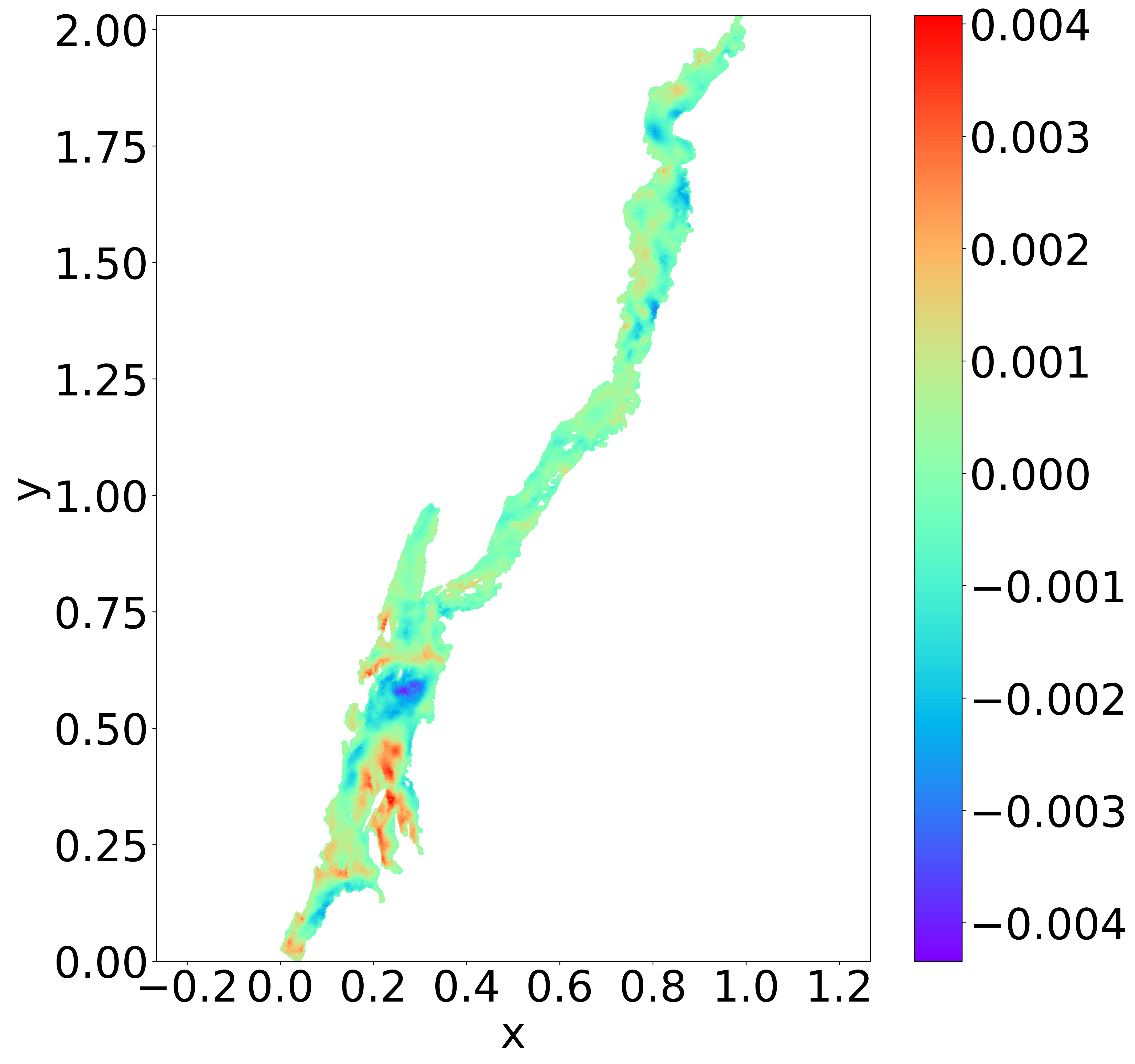} \\
\hspace{-0.3cm}
\includegraphics[scale=0.064]{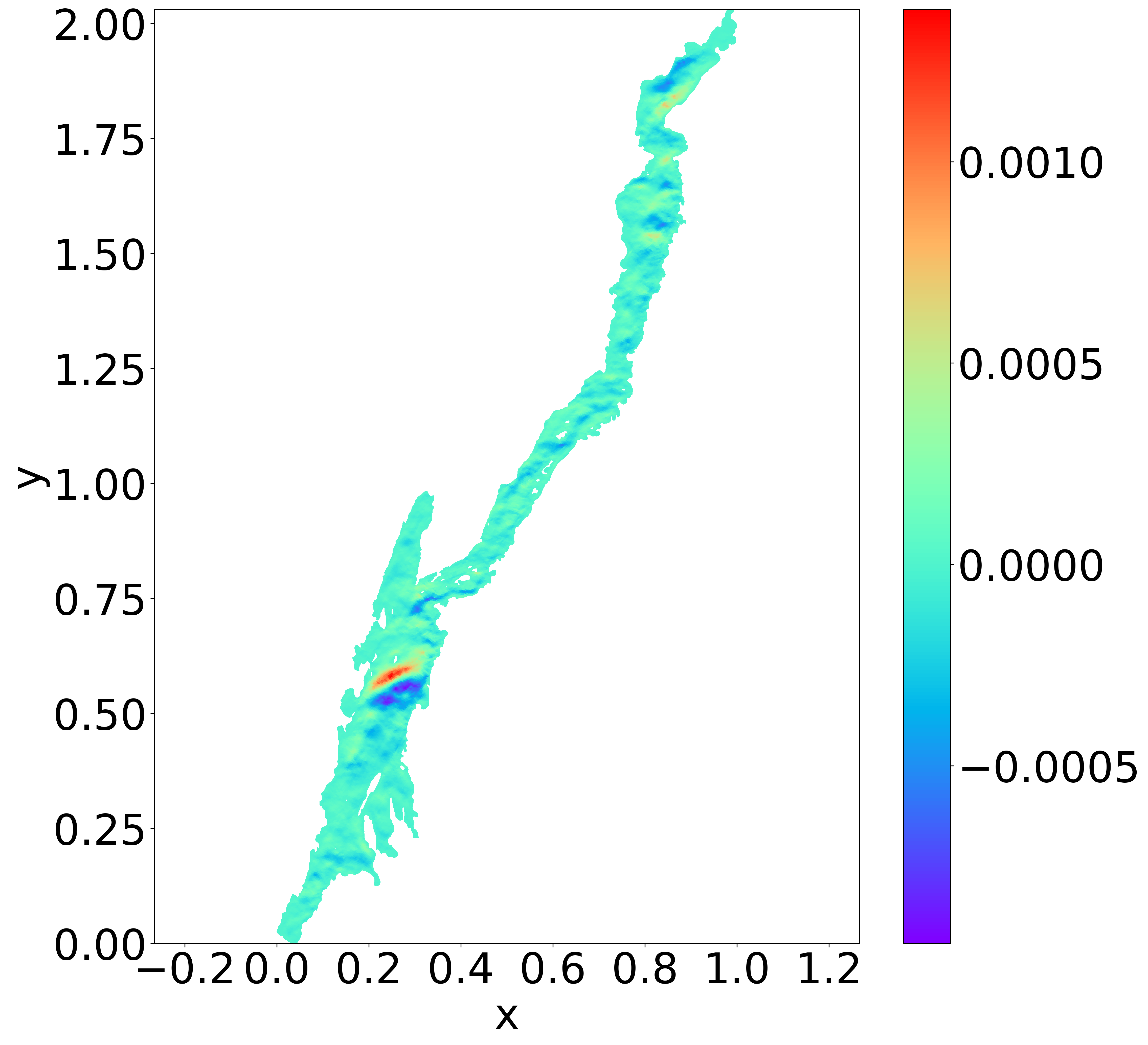} &
\includegraphics[scale=0.064]{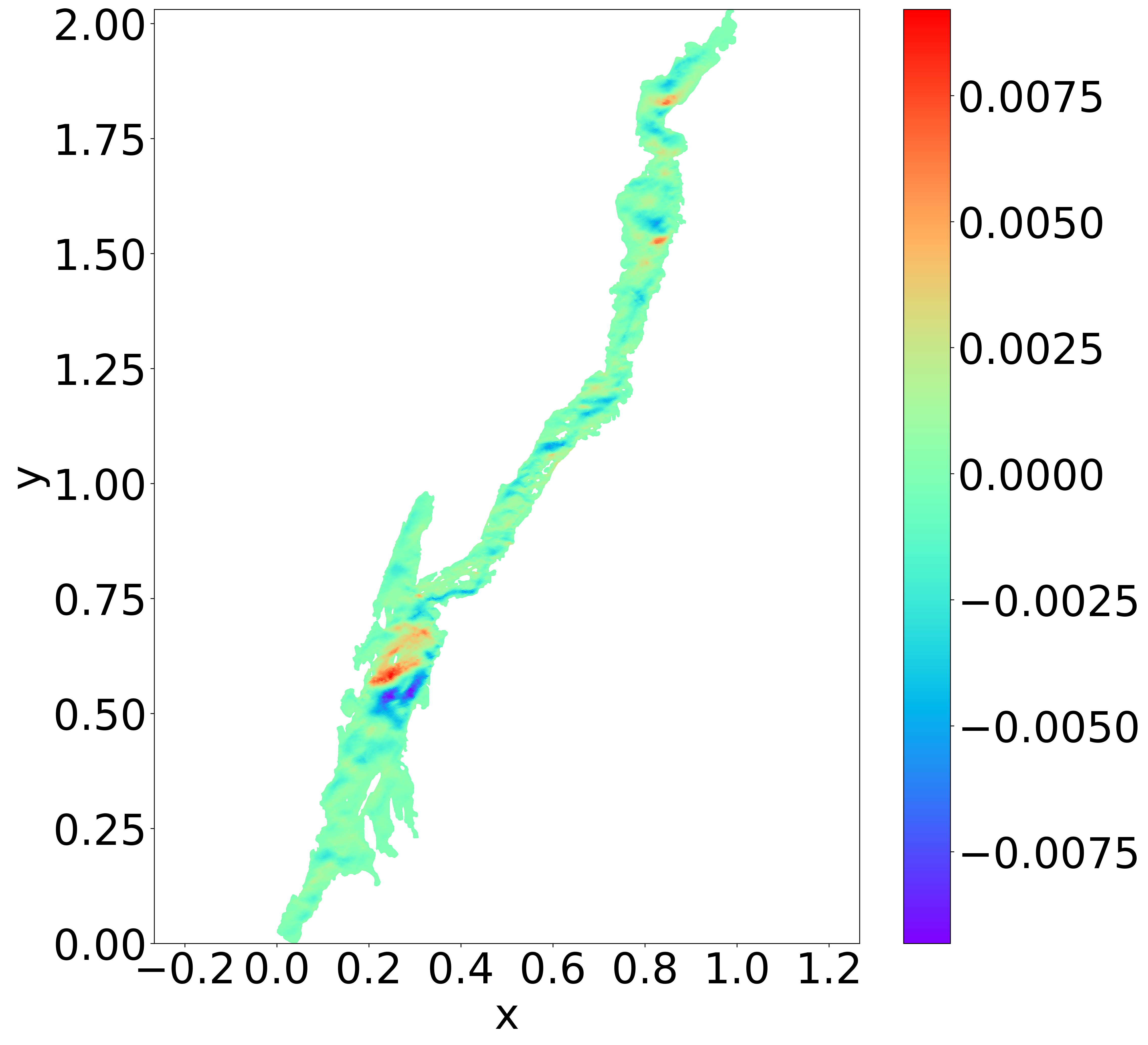} &
\includegraphics[scale=0.064]{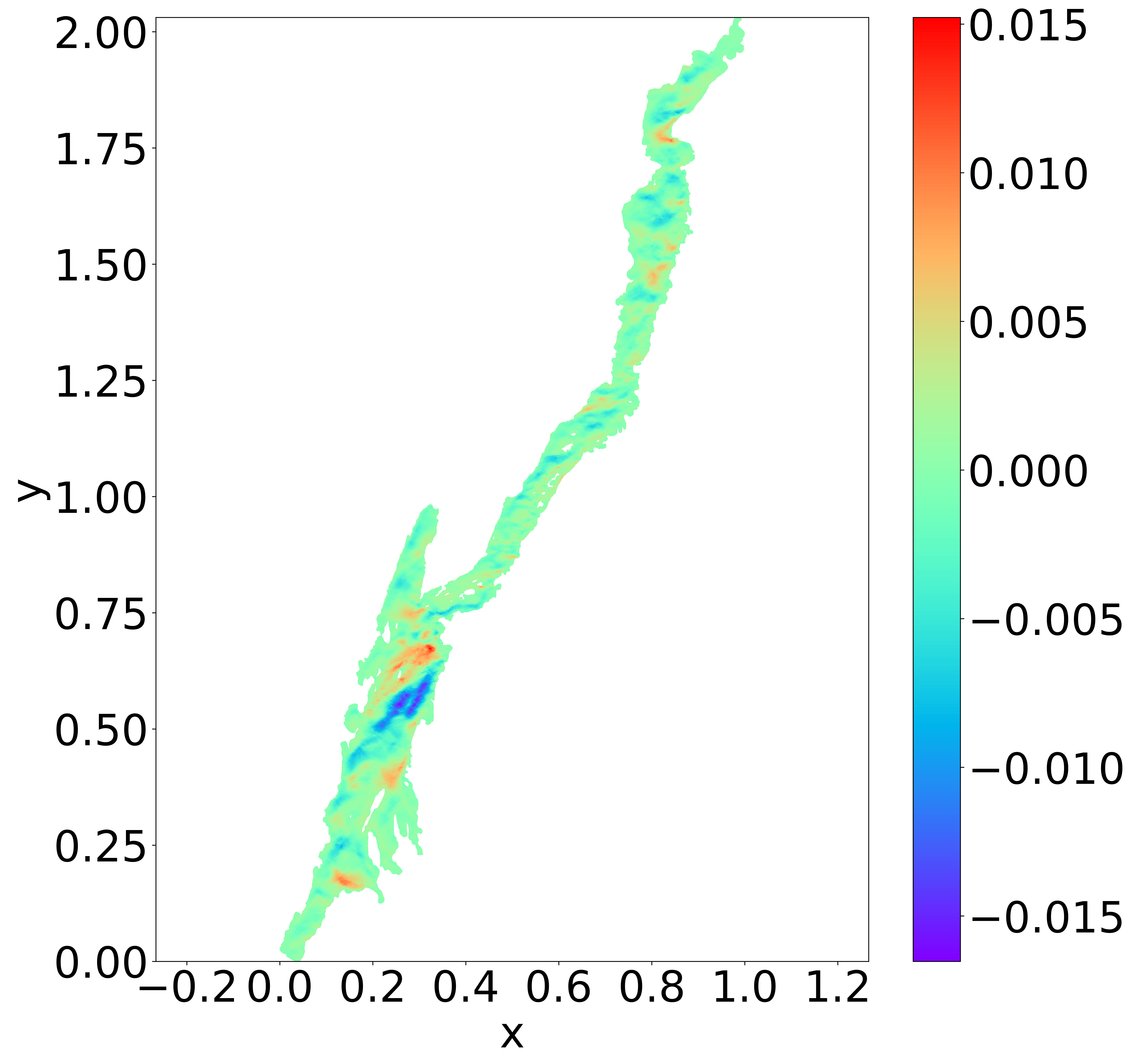} &
\includegraphics[scale=0.064]{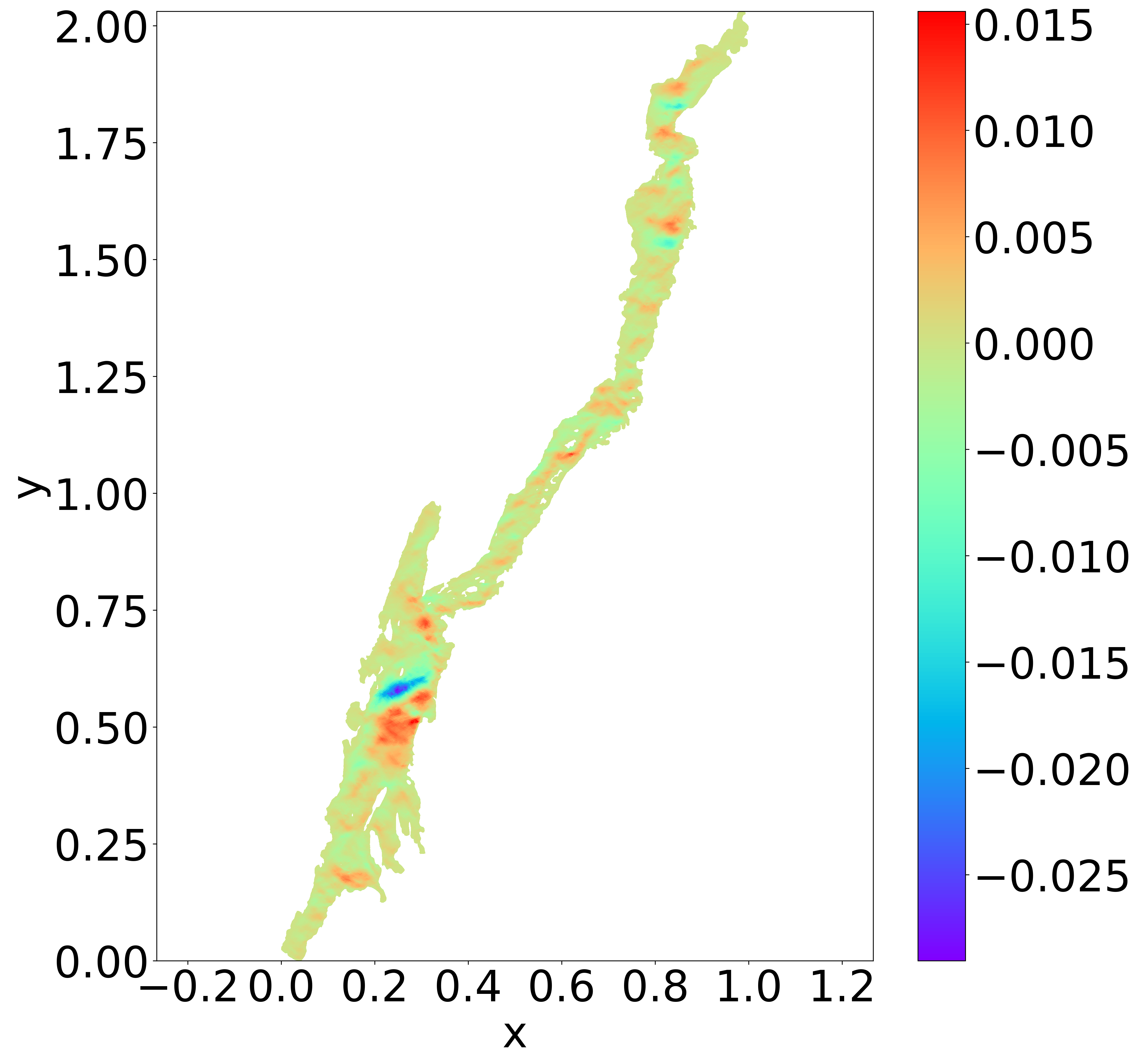} &
\includegraphics[scale=0.064]{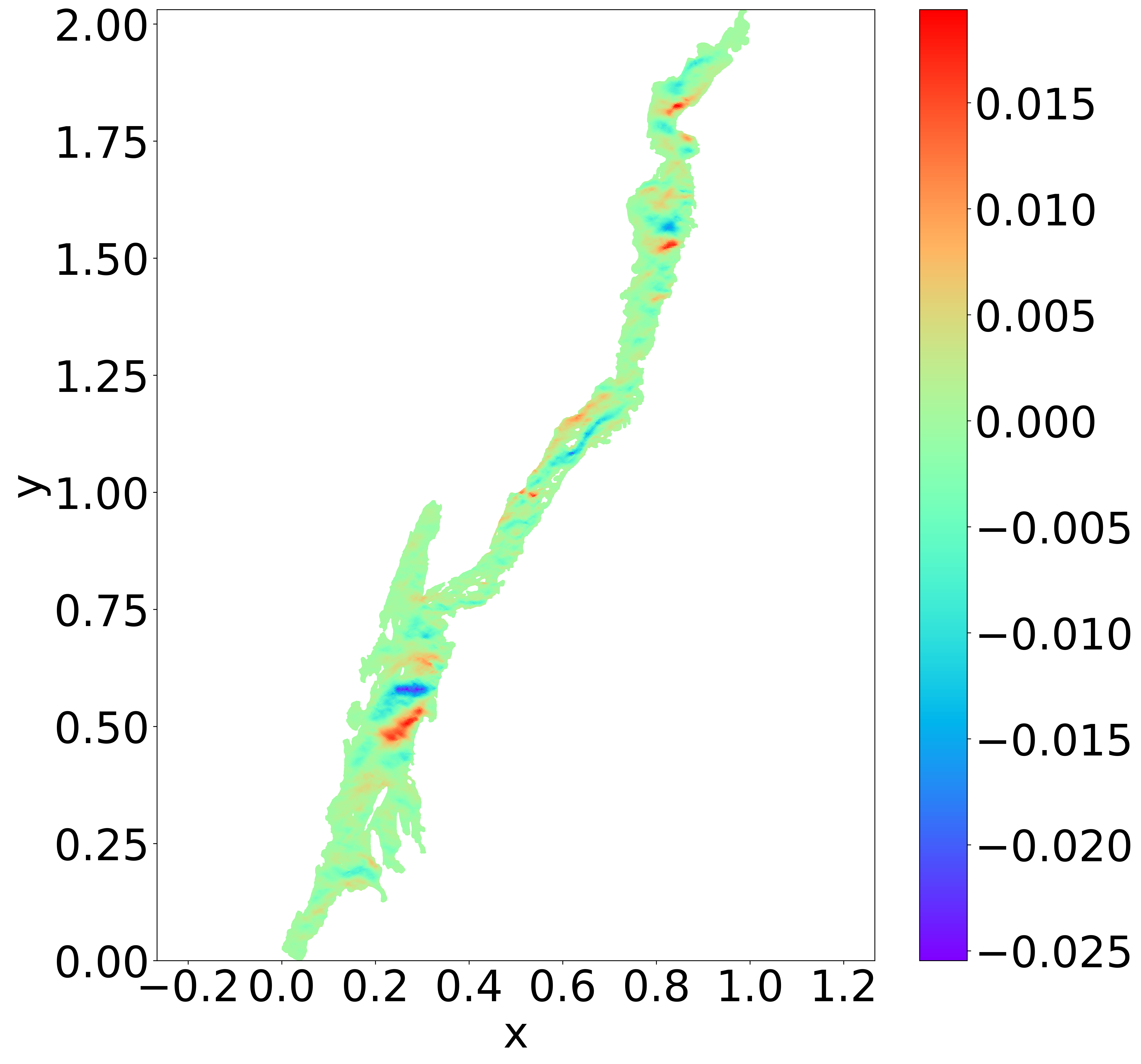} \\
\hspace{-0.3cm}
\includegraphics[scale=0.064]{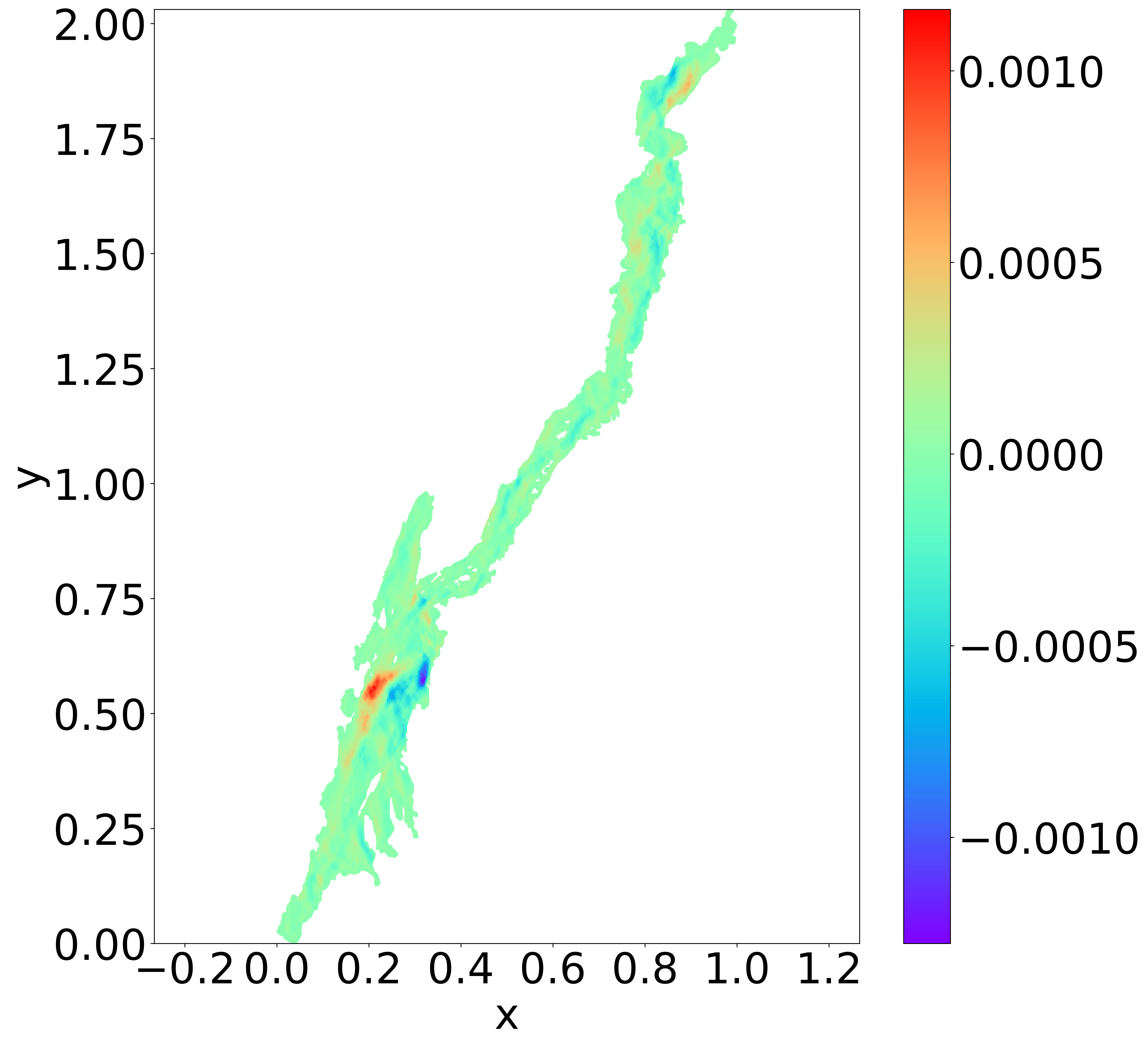} &
\includegraphics[scale=0.064]{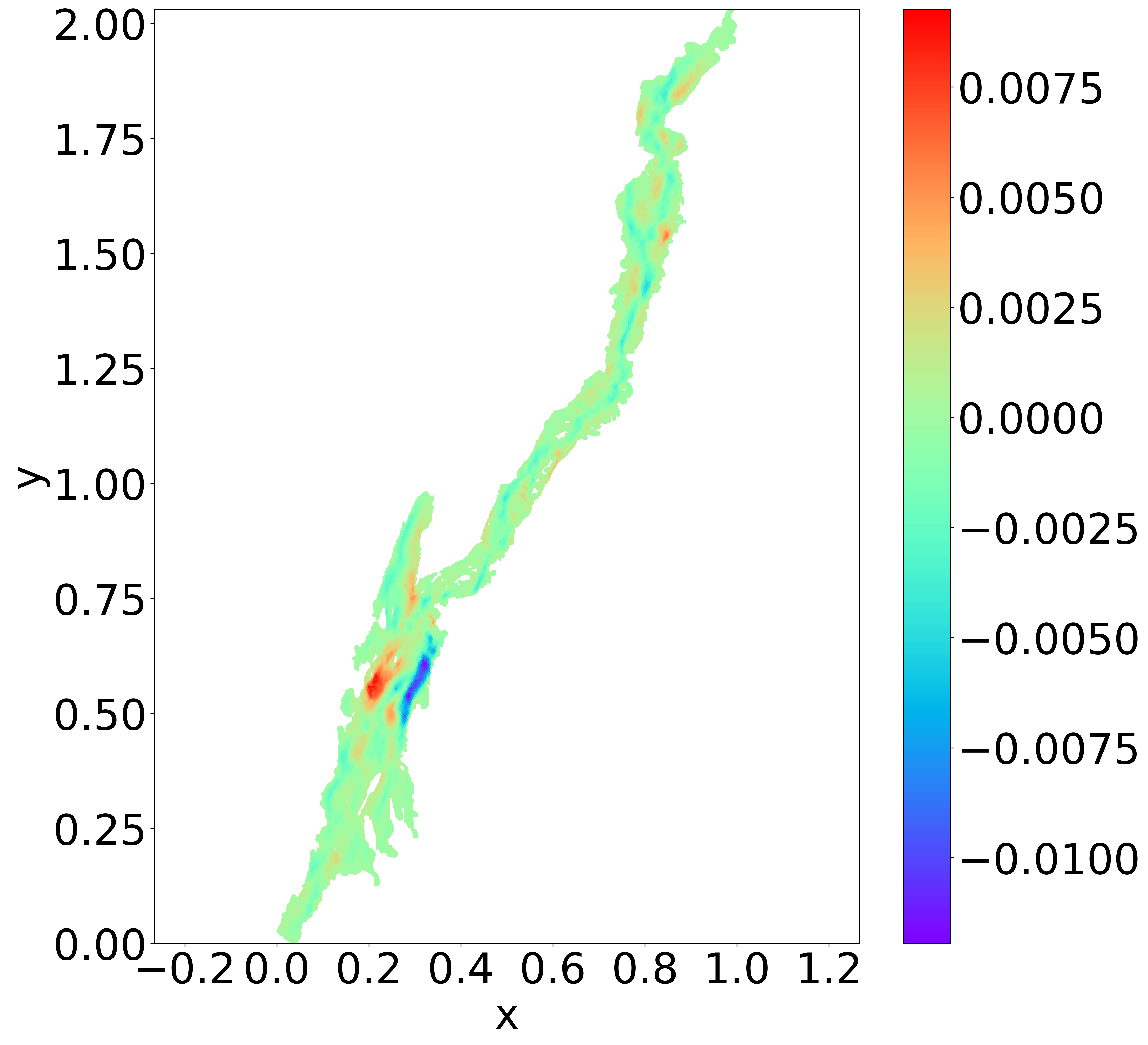} &
\includegraphics[scale=0.064]{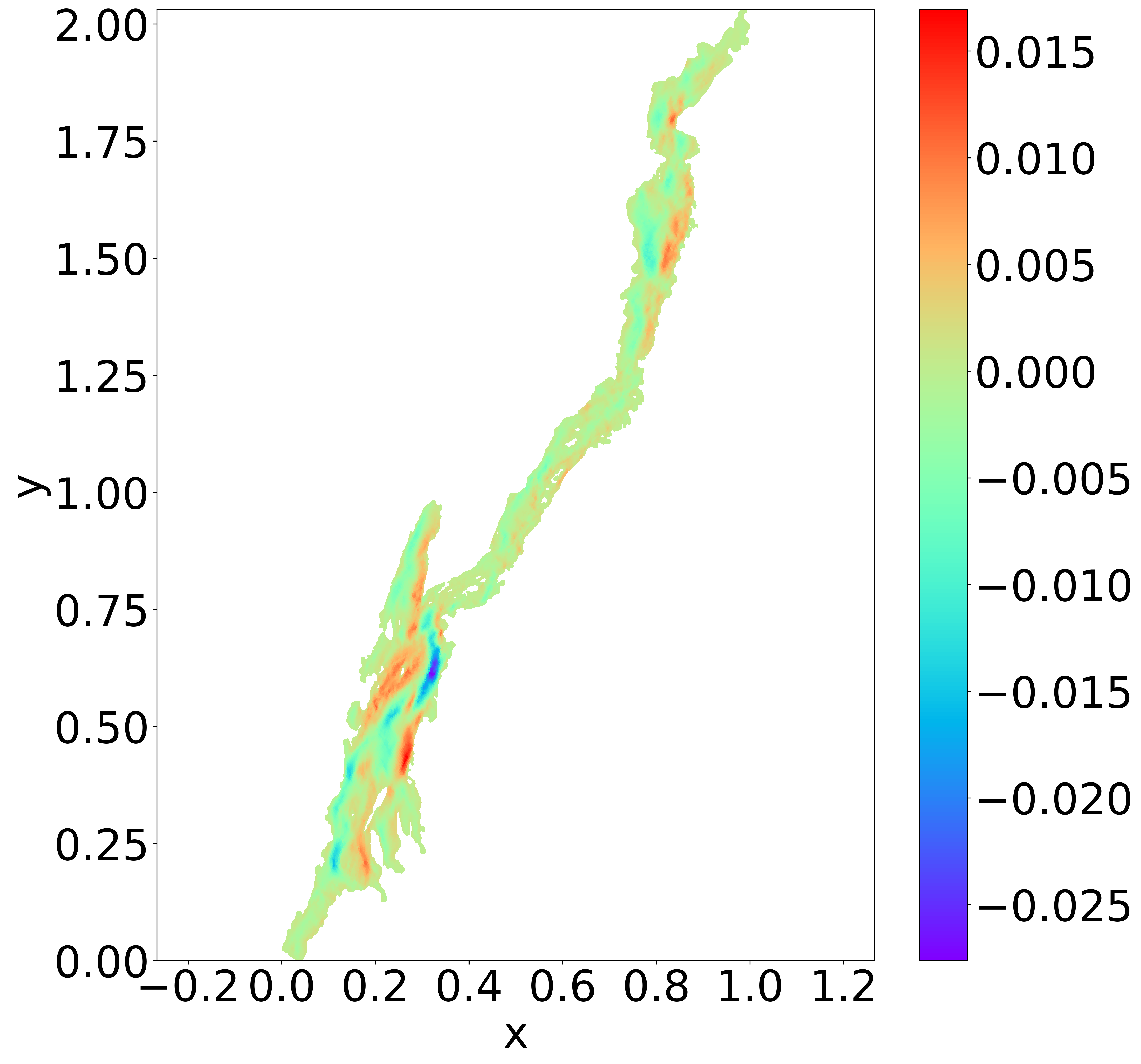} &
\includegraphics[scale=0.064]{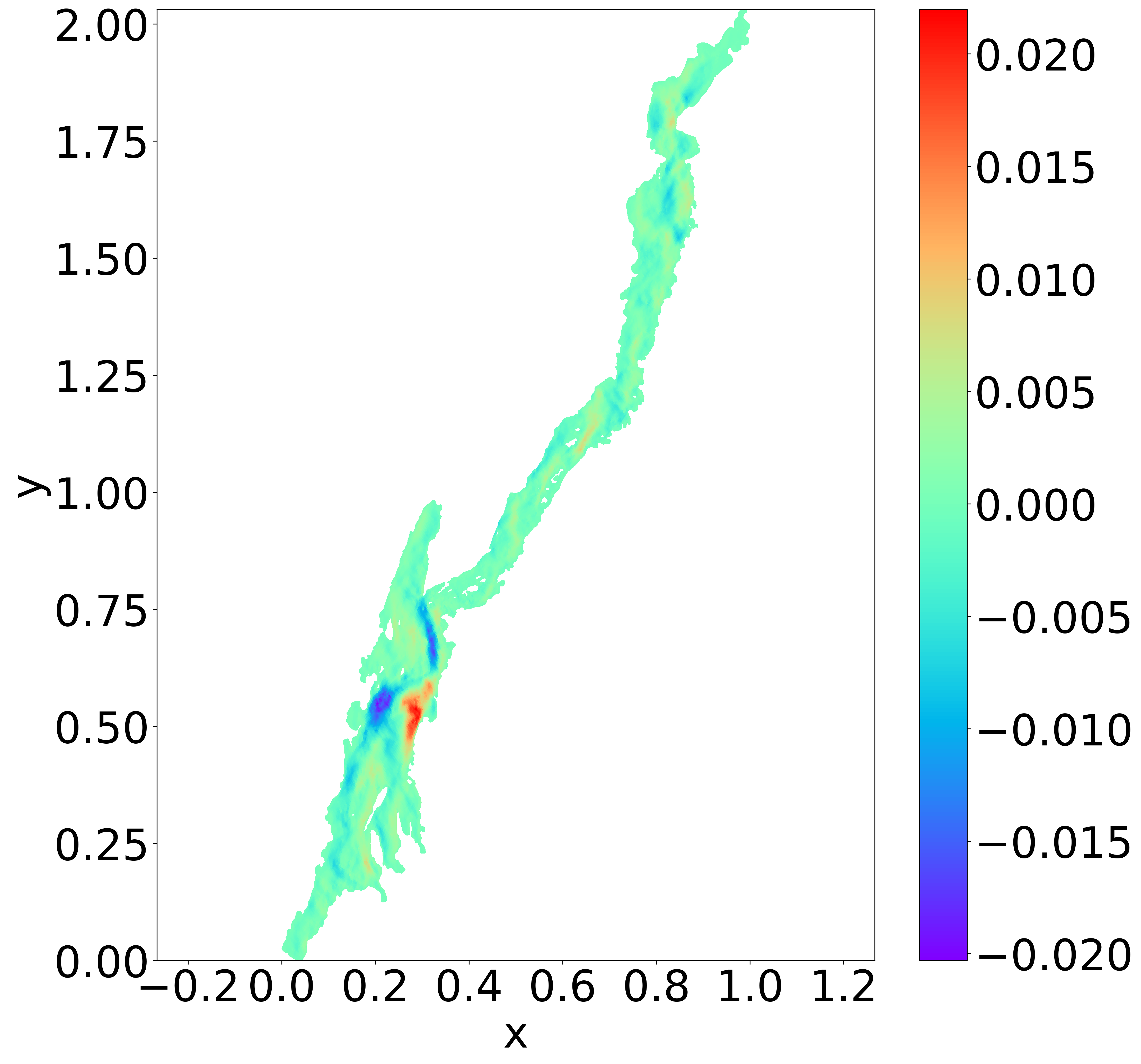} &
\includegraphics[scale=0.064]{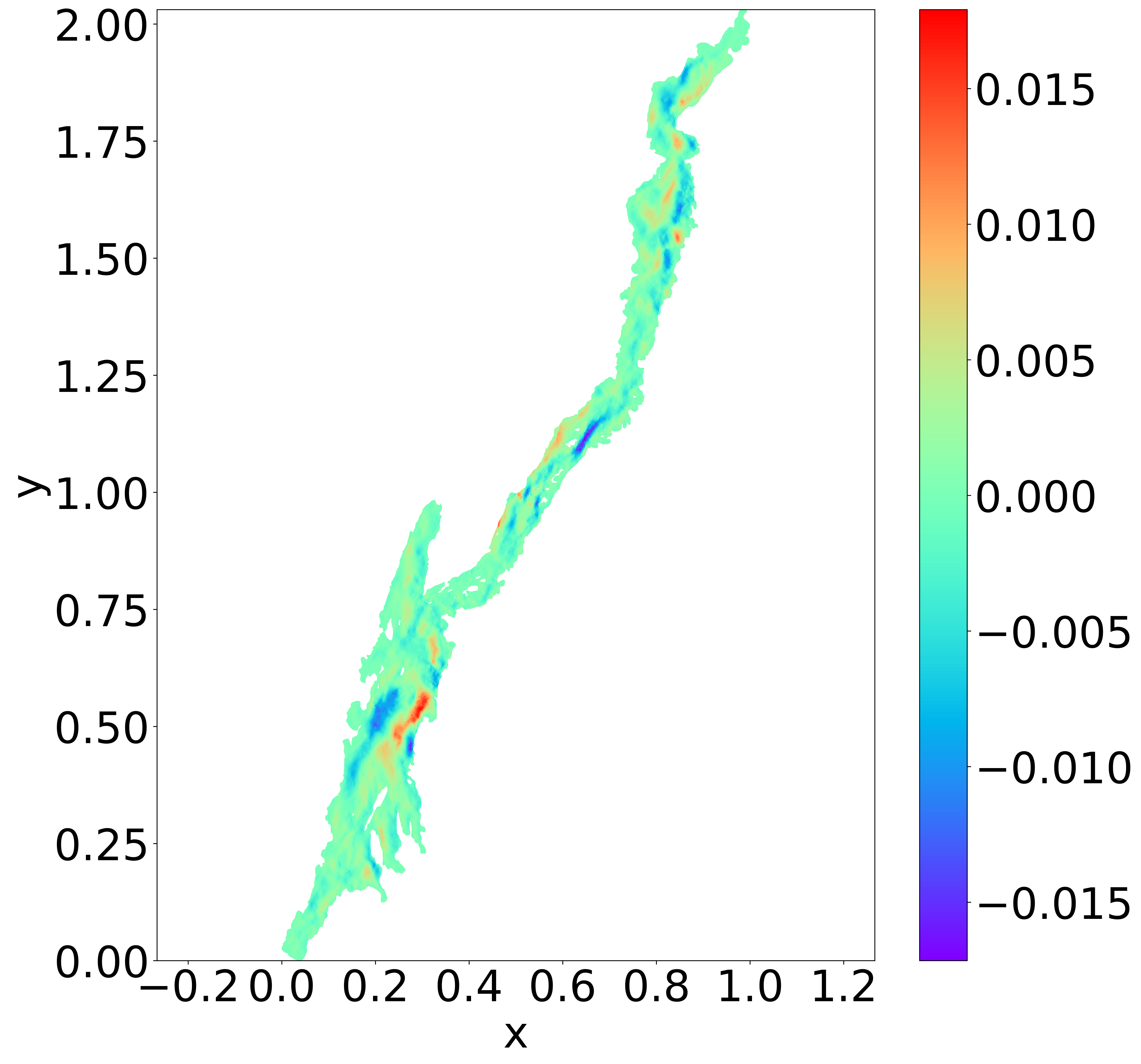} \\
\hspace{-0.3cm}
 a) $\Psi_{1}(\mathbf{x})$ & b) $\Psi_{2}(\mathbf{x})$ & c) $\Psi_{3}(\mathbf{x})$ & d) $\Psi_{4}(\mathbf{x})$ & e) $\Psi_{5}(\mathbf{x})$
\end{tabular}
\caption{From top to bottom: Spatial modes relative to Density, Temperature, East and North velocity components.}
\label{space_modes_n-vel}
\end{figure}

To find the best performance models, we applied a hyperparameter optimization on the FCNN architecture of $600$ configurations randomly chosen with different numbers of hidden layers, neurons per layer and optimizer iterations, and dropout values and learning rates. Empirical tests revealed that 5 spatial modes $\Psi_{j}$ represent the main circulation features over the entire lake (see Fig. \ref{space_modes_n-vel}), preserving $96\%$ of the total energy of the full order system. Table \ref{preserved_energy_modes} shows the percentage preservation by each spatial mode, indicating how representative each mode is at explaining the variance in each circulation variable (e.g, density, temperature and velocity components). It is worth noting that the preserved energy of each spatial mode is the same regardless of the circulation variable as the eigenvalues of the correlation matrix are the same for each circulation variable.
\begin{table}[h!]
\caption{Percentage of preserved energy in each spatial mode.}
\label{preserved_energy_modes}
\centering
\begin{tabular}{cccccc}
\hline
$\Psi_{1}(\mathbf{x})$ & $\Psi_{2}(\mathbf{x})$ & $\Psi_{3}(\mathbf{x})$ &  $\Psi_{4}(\mathbf{x})$ & $\Psi_{5}(\mathbf{x})$ & Total preserved energy\\
\hline
$91.4\%$ & $2.4\%$ & $1.1\%$ & $0.6\%$ & $0.5\%$ & $96.0\%$ \\
\hline
\end{tabular}
\end{table}

\begin{figure}[h!]
\vspace{-0.5cm}
\centering
\begin{tabular}{ccc}
\includegraphics[scale=0.21]{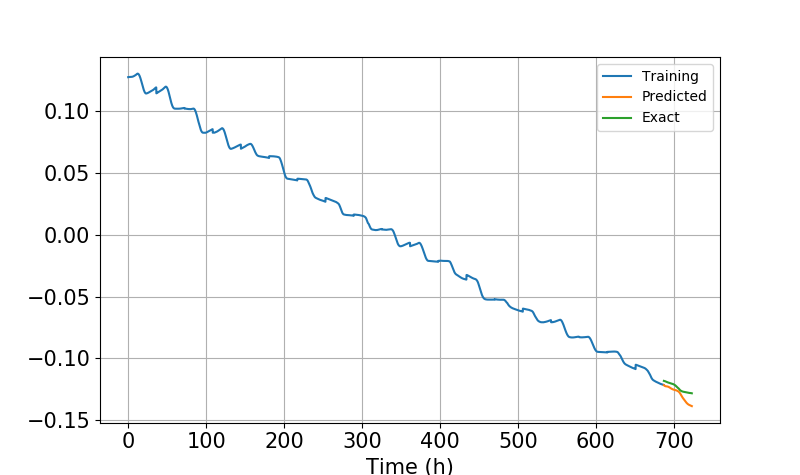} & 
\hspace{-0.4cm}
\includegraphics[scale=0.21]{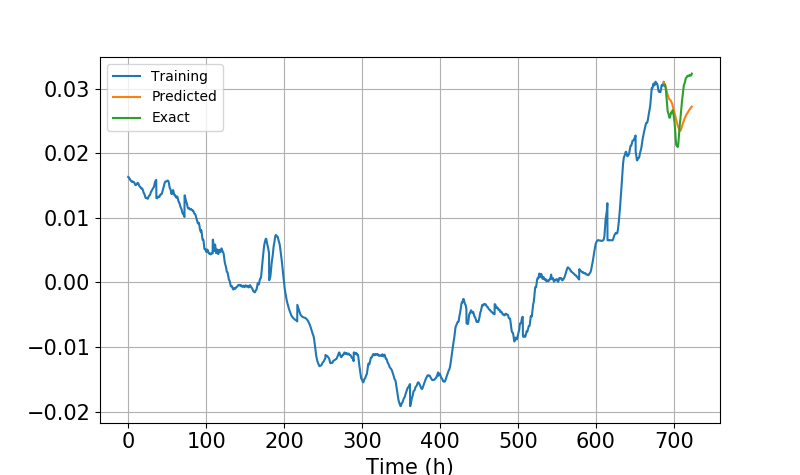} &
\hspace{-0.4cm}
\includegraphics[scale=0.21]{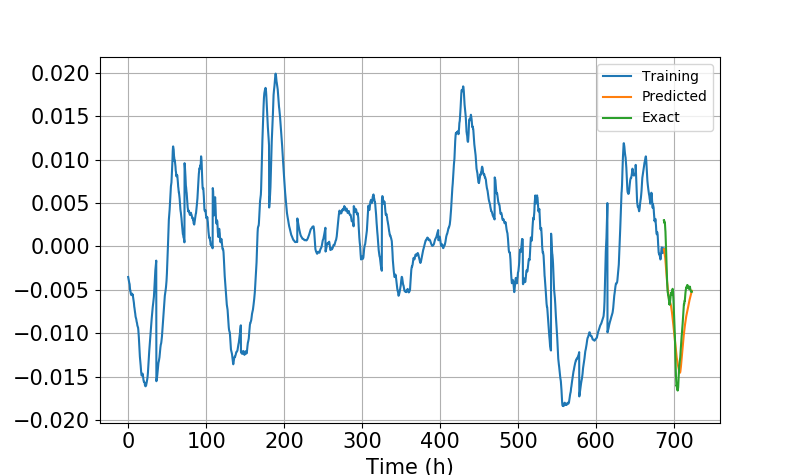} \\
\includegraphics[scale=0.21]{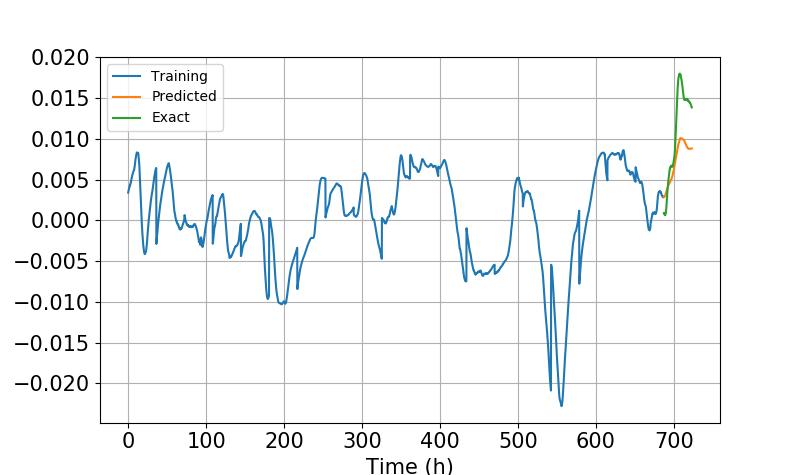} &
\hspace{-0.4cm}
\includegraphics[scale=0.21]{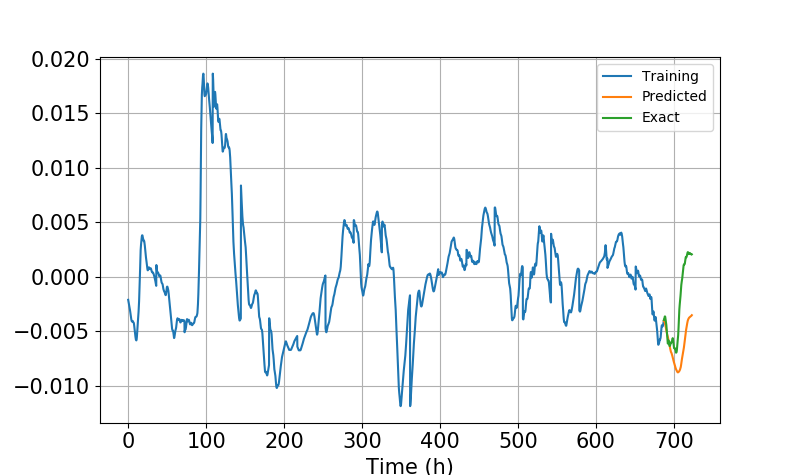} & 
\end{tabular}
\caption{\small{Temporal coefficients of the FCNN ROM corresponding to the 5 spatial modes. From left to right, first row: $\mathbf{a}_{1}(t), \mathbf{a}_{2}(t)$ and  $\mathbf{a}_{3}(t)$; second row: $\mathbf{a}_{4}(t)$ and $\mathbf{a}_{5}(t)$.}}
\label{fcnn_time_coeff}
\end{figure}

\vspace{-0.5cm}
Figure \ref{fcnn_time_coeff} shows the time series of the 5 temporal coefficients for each spatial mode. In these plots, blue lines correspond to input used for training the machine learning model, green lines represent the "true" values simulated by SUNTANS for testing, and the orange lines represent predicted values by the FCNN ROM. The time series corresponding to the FCNN ROM temporal coefficients are predicted fairly well. It suggests that the numerical differentiation of the temporal coefficients recovered from the full order model are quite consistent and reliable, ensuring an accurate reconstruction of the ROM coefficients through the RK time integration procedure. Although the error of the predicted time series corresponding to the fourth and fifth modes had been relatively large, their contributions to the approximating capacity of the surrogate model is quite insignificant as shown in table \ref{preserved_energy_modes}.

\begin{figure}[h!]
\centering
\begin{tabular}{cccc}
\hspace{-0.2cm}
\textbf{Density} $\rho$ & \textbf{Temperature} $T$ & \textbf{East velocity} $u$ & \textbf{North velocity} $v$\\
\hspace{-0.2cm}
\includegraphics[scale=.080]{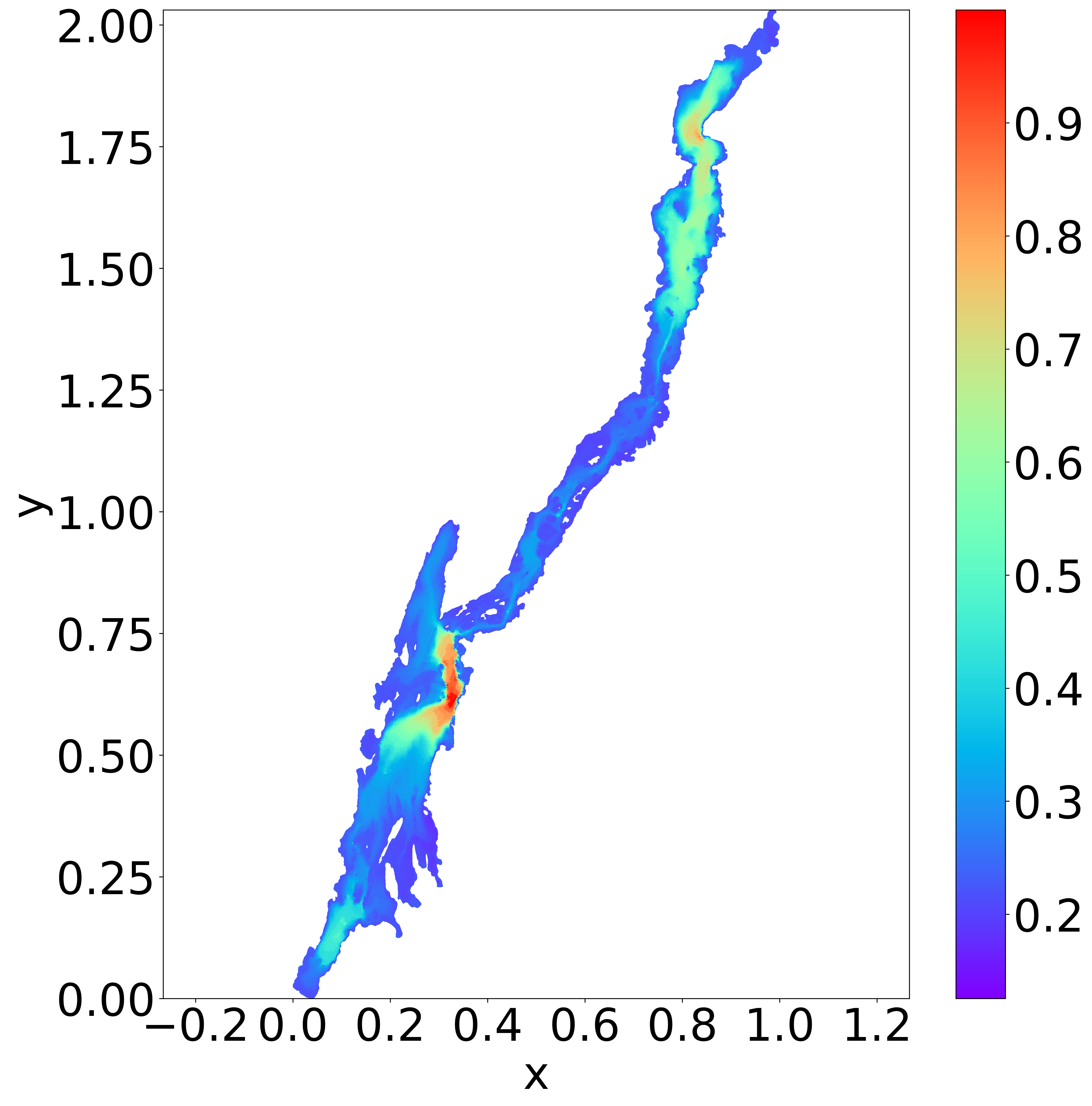} &
\includegraphics[scale=.080]{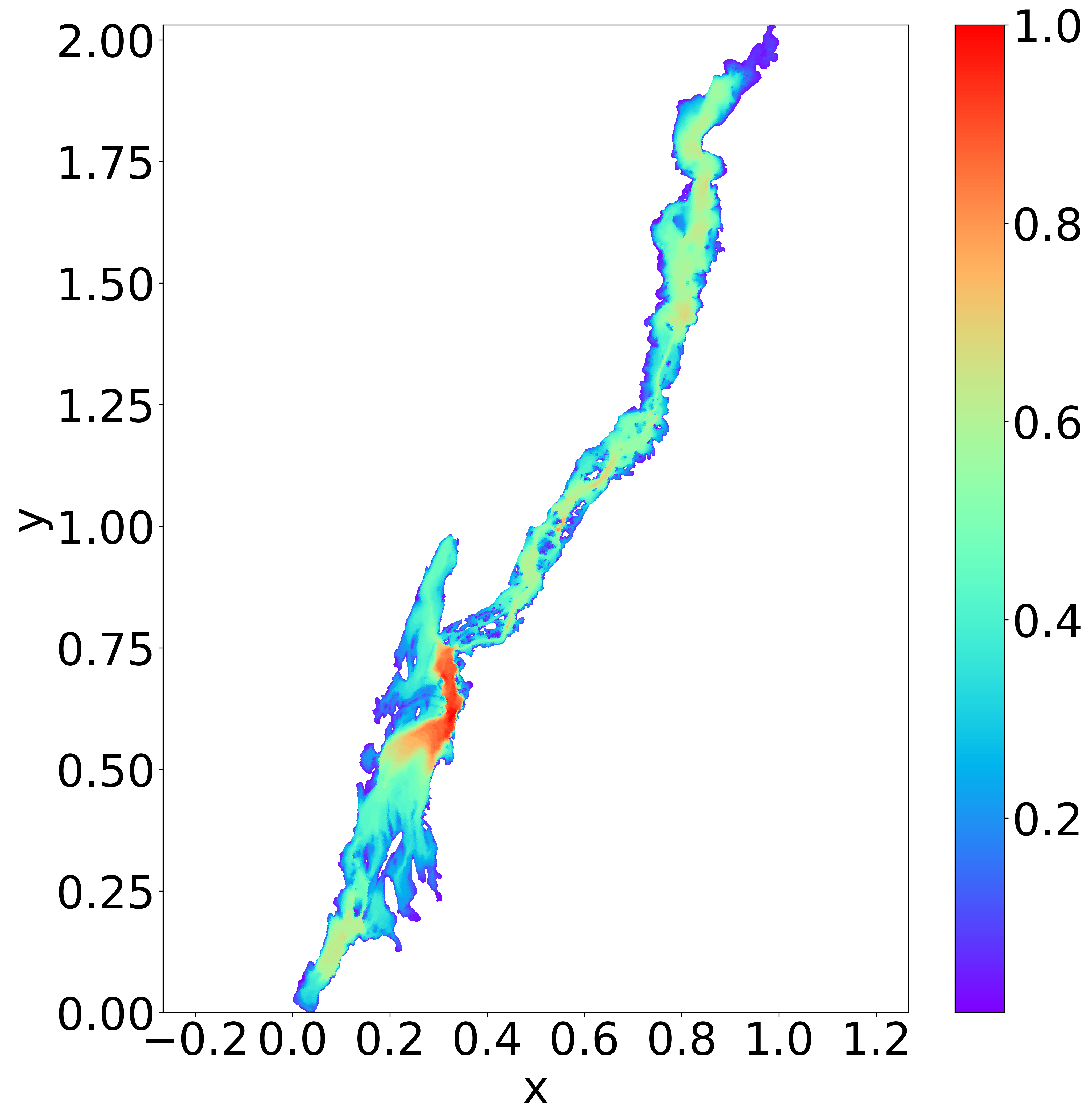} &
\includegraphics[scale=.080]{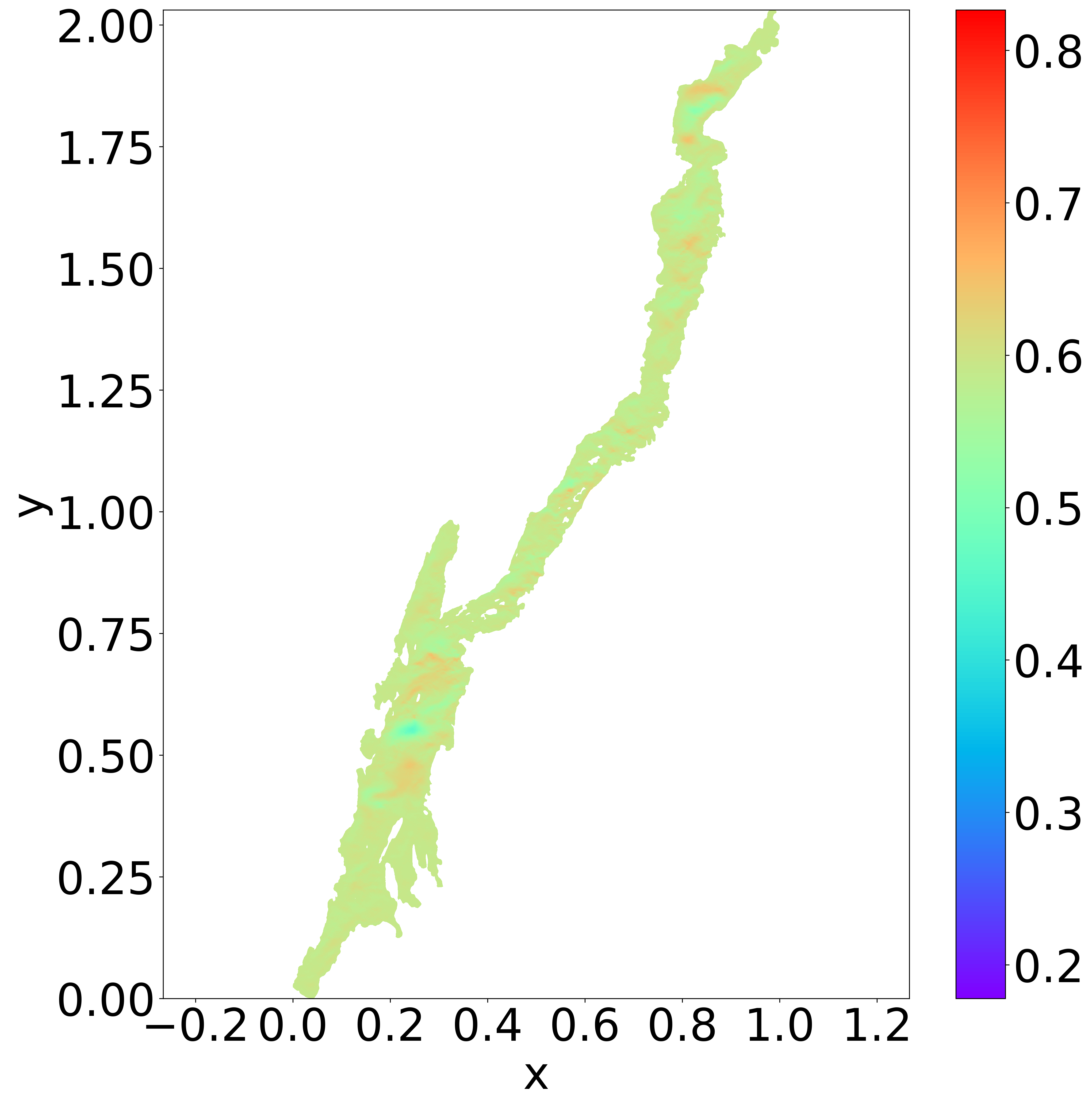} &
\includegraphics[scale=.080]{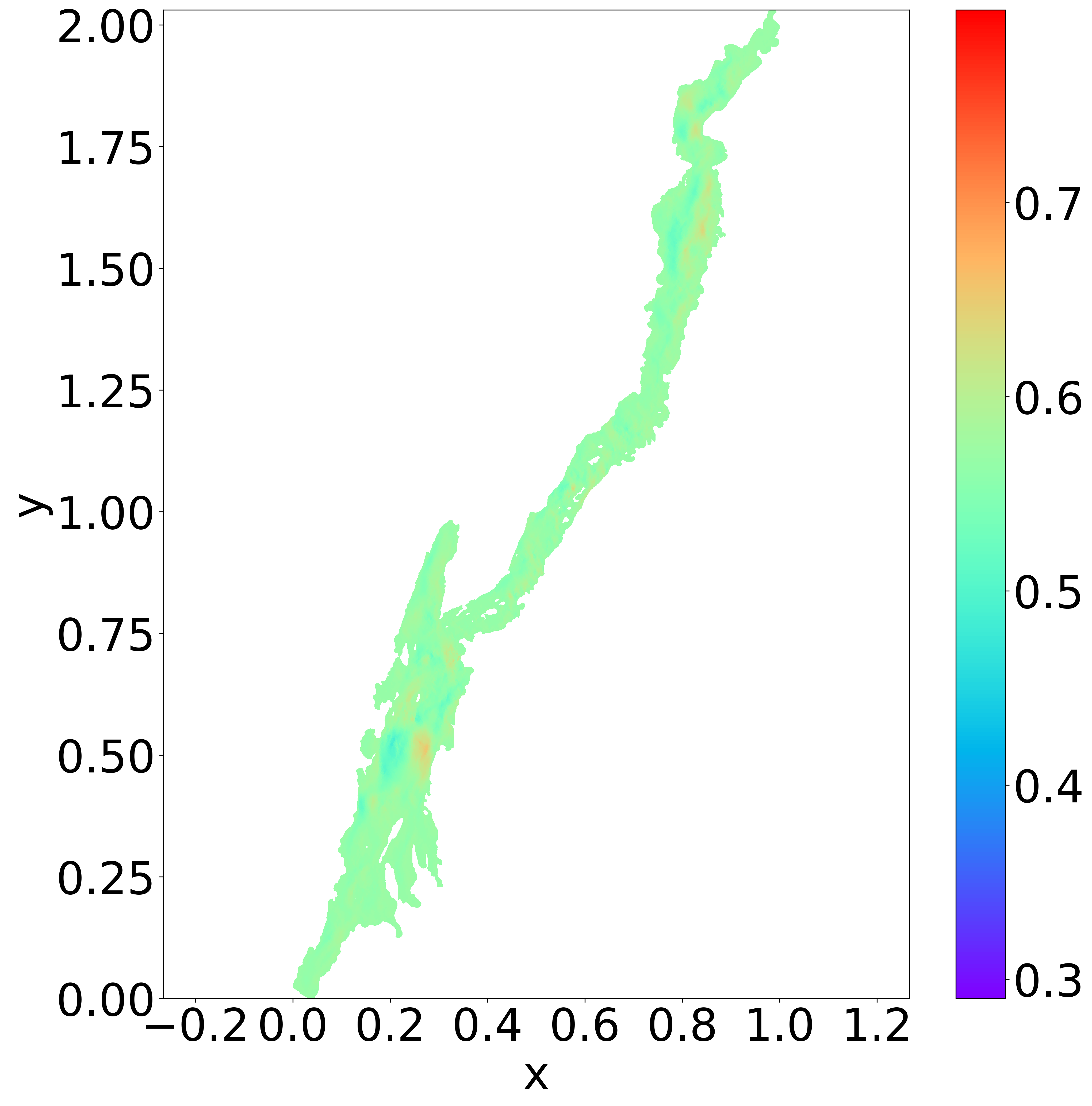}\\
\multicolumn{4}{c}{\textbf{True}} \\
\hspace{-0.2cm}
\includegraphics[scale=.080]{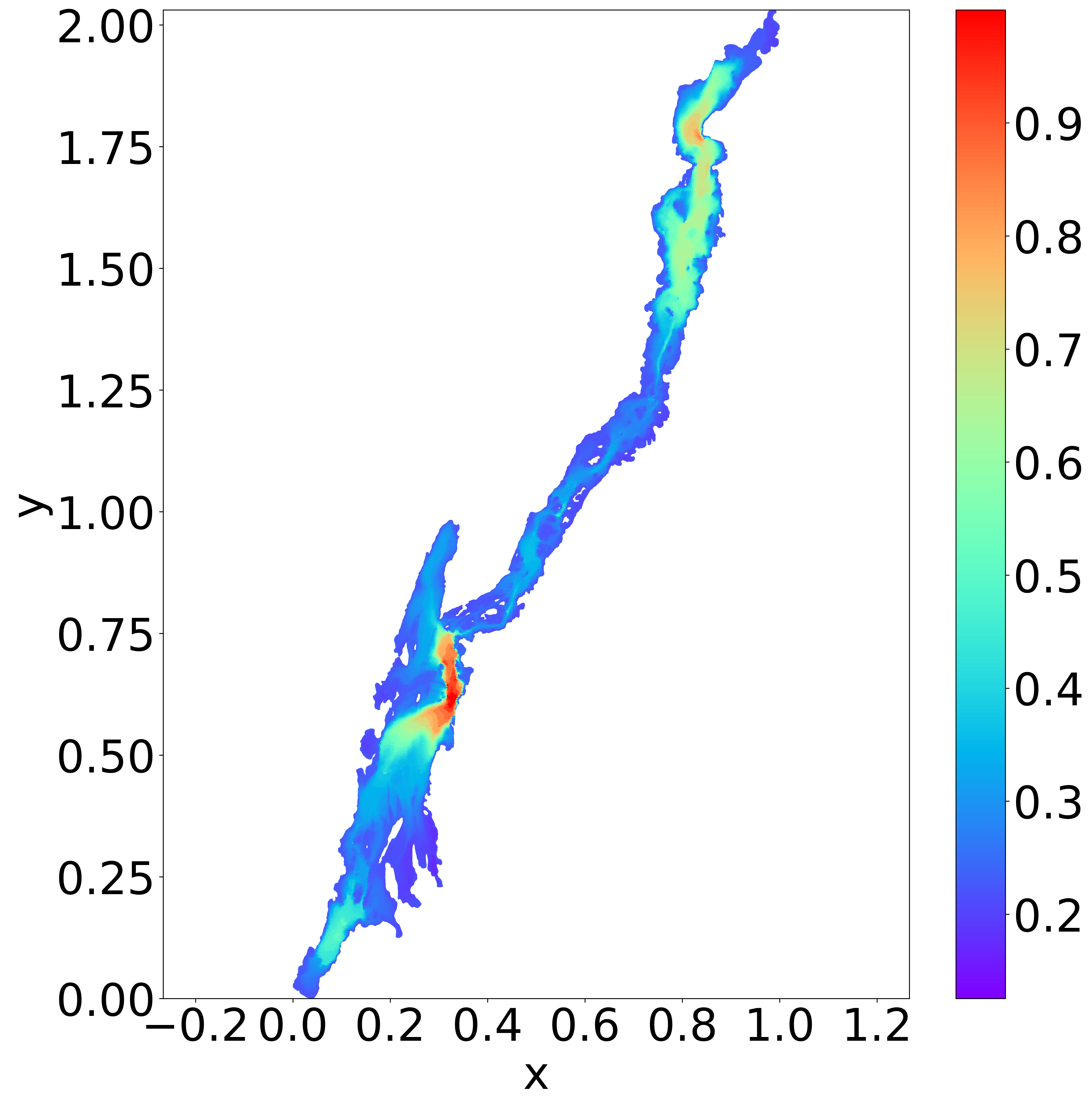} &
\includegraphics[scale=.080]{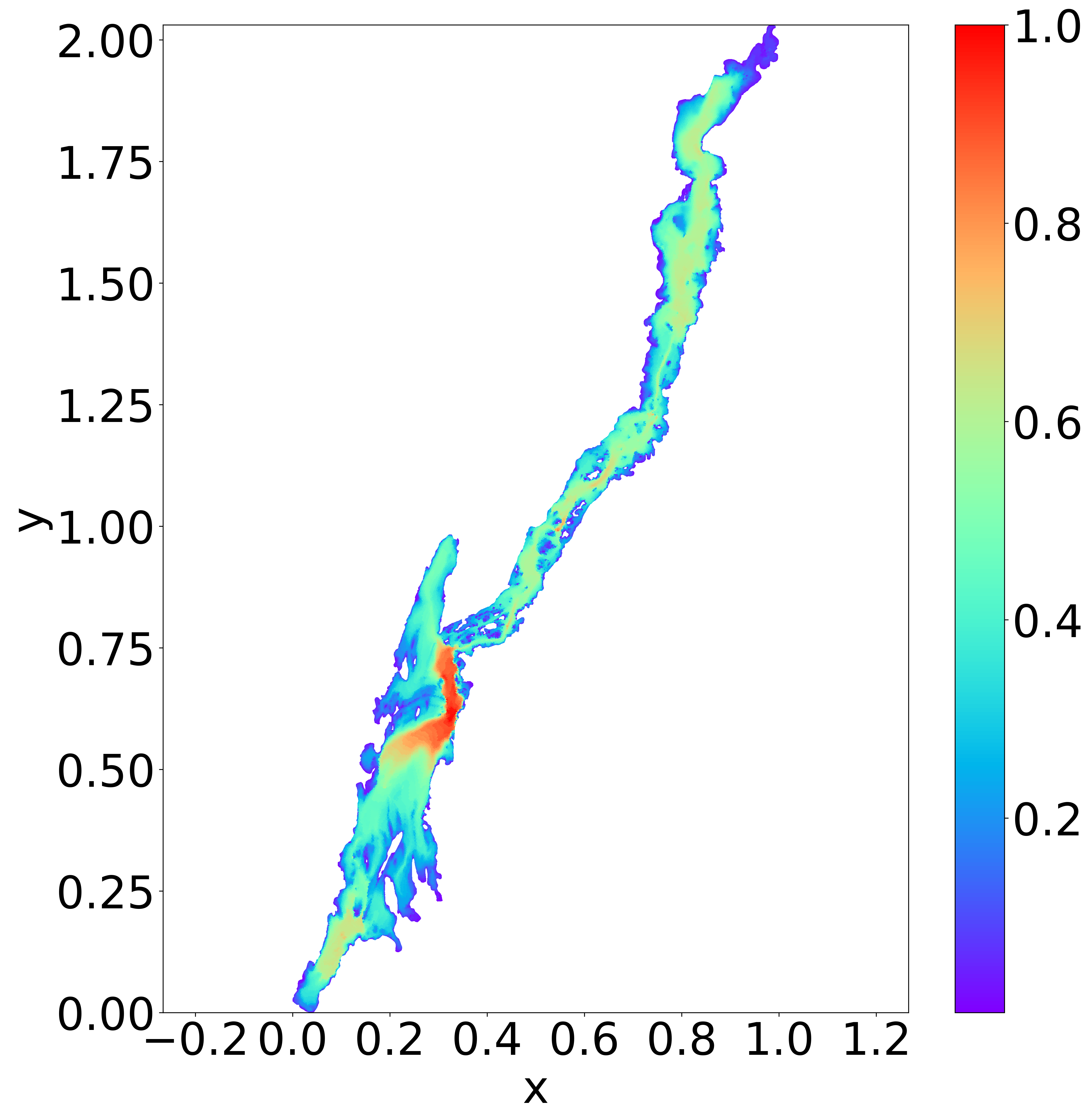} &
\includegraphics[scale=.080]{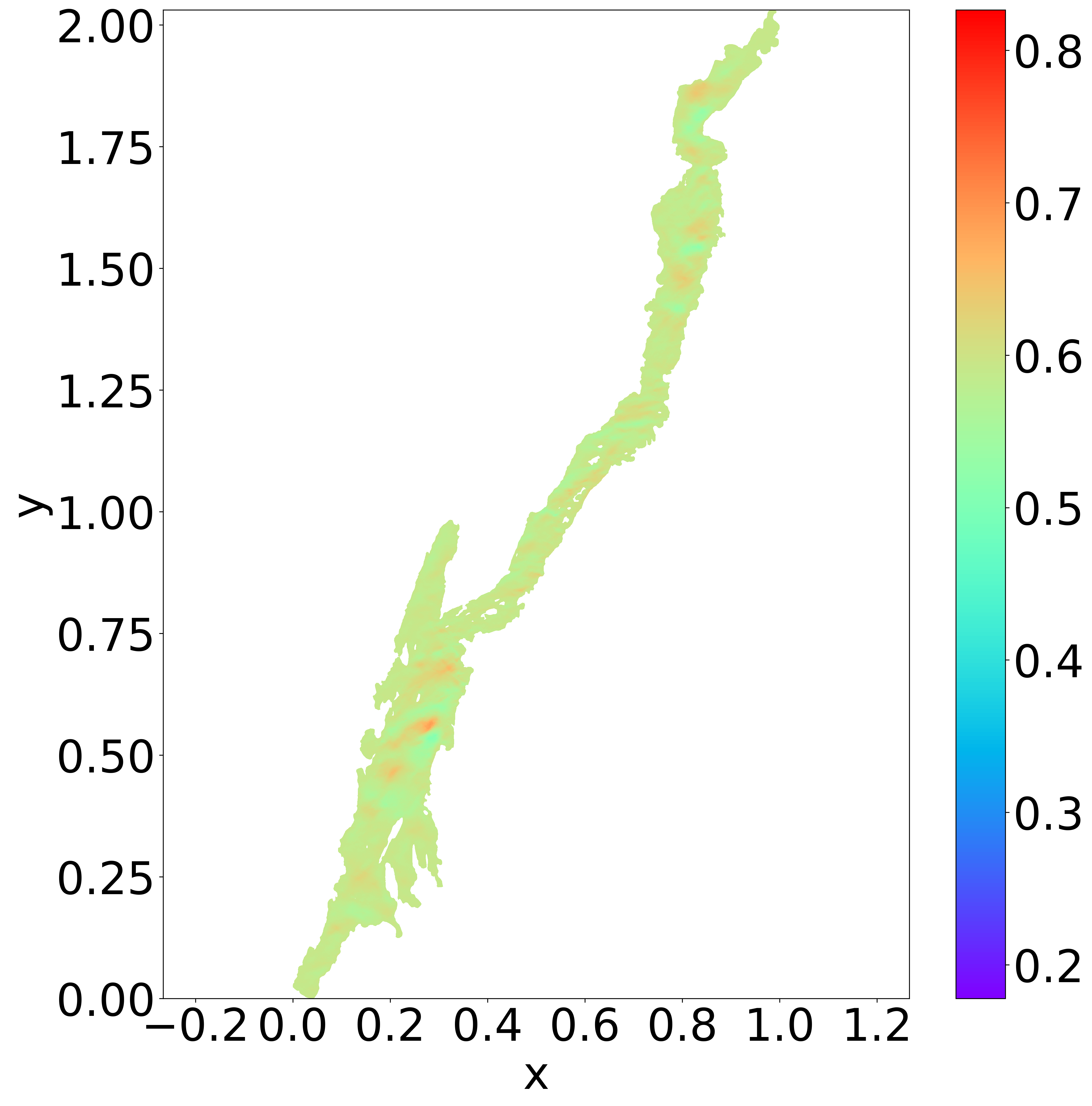} &
\includegraphics[scale=.080]{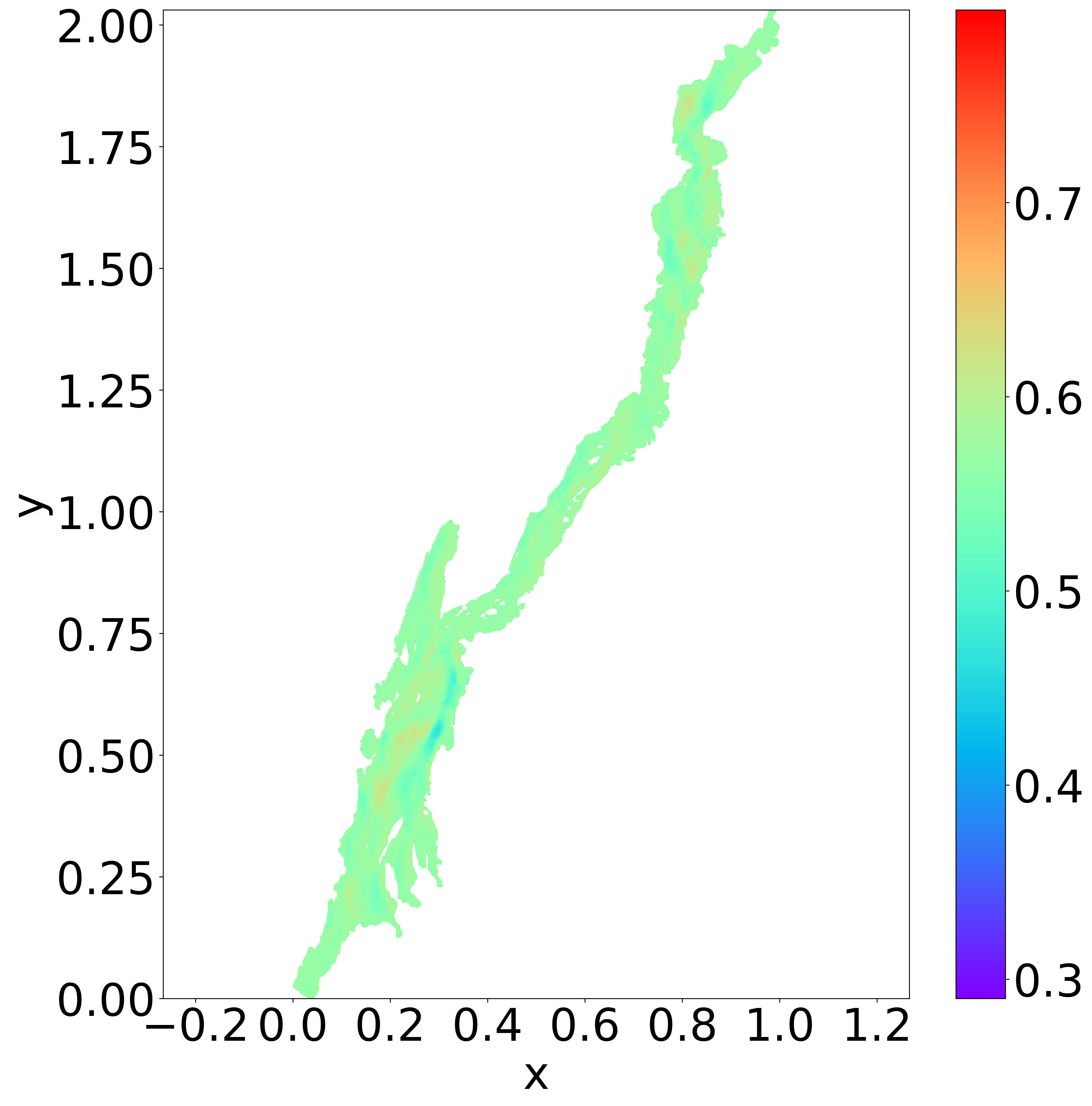}\\
\multicolumn{4}{c}{\textbf{Predicted}}
\end{tabular}
\caption{Full order model solution vs. FCNN ROM approximation for density, temperature and velocity components.}
\label{fcnn_comparison}
\end{figure}

\vspace{-0.5cm}
When the time coefficients are multiplied by the spatial modes according to equation (\ref{Q_approx}), the FCNN ROM produces the desired 36 hour hydrodynamic forecast for Lake George. Figure \ref{fcnn_comparison} shows the comparison between the predicted physical state of Lake George at the end of the 36 hours forecast, and its "true" physical state from SUNTANS. In these plots, the physical variables (temperature, density and velocity) have been normalized (their values are restricted to the range $[0,1]$). 

\begin{figure}[h!]
\vspace{-0.5cm}
\centering
\begin{tabular}{cc}
\includegraphics[scale=0.35]{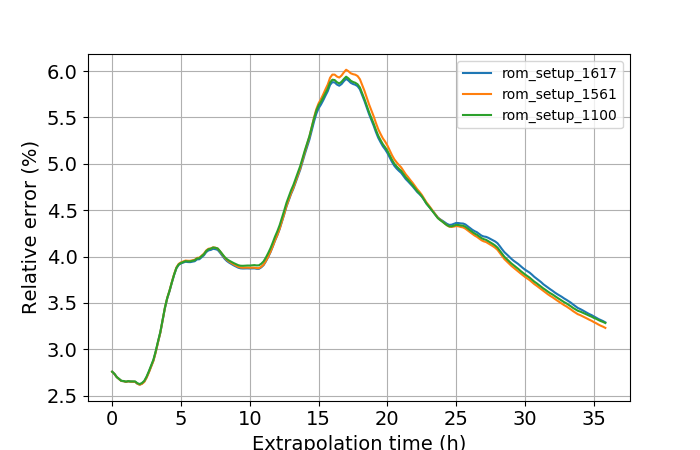} &
\includegraphics[scale=0.35]{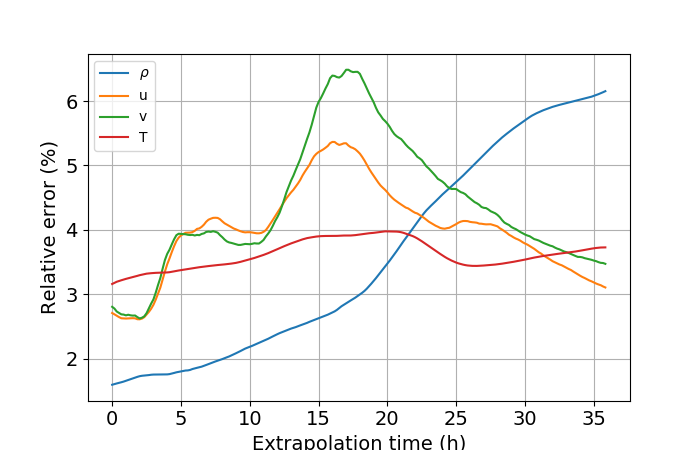} \\
a) Relative error for the 3 best & b) Relative error for each \\
hyperparameter settings & variable: $\rho, u, v, T$
\end{tabular}
\caption{a) Best performances based on the average $\frac{u+v}{2}$ error; b) Best configuration: model 1617.}
\label{fcnn_error}
\end{figure}

It is evident that variables which change slowly with time, such as density and temperature, are better approximated by the FCNN ROM than water velocity. The water velocity is acutely sensitive to surface winds which can change rapidly with time. The error plots in Figure \ref{fcnn_error}b shows the impact of a strong wind event between 12 and 23 hours that significantly increases the error of the water velocity components, degrading the performance of the whole model. Such high frequency fluctuations are difficult for the neural network to capture because the correlation matrix $\mathbf{C}$ takes into consideration all physical variables at once.

Figure \ref{fcnn_error}a shows the 3 best FCNN ROMs, ranked according to the relative error of the averaged water velocity components ($\frac{u+v}{2}$). This choice was made considering evidence of the difficulty in approximating fast-changing variables. The hyperparameter optimization strategy generates a set of candidates labeled with  arbitrary  numerical  indices  such  as  1617, 1561  and  1100  (cf. Fig. \ref{fcnn_error}) to identify each individual model that was trained. The water velocity components were expected to influence the choice of the best ROMs because they are harder to approximate, as noted previously. Figure \ref{fcnn_error}b shows the $L^{2}$-norm relative error of each variable for the best FCNN ROM setup.

\vspace{-0.5cm}
The relative error of all variables is kept below $7\%$ for the duration of the 36 hour forecast, which is quite remarkable. At the end of the forecast, only the density relative error is above the target of $5\%$, although a visual comparison between the predicted and true density distributions shows very small differences. The error levels achieved are even more impressive if we consider that SUNTANS uses an internal $3$ second time step to reach its solution, while the surrogate model has an effective timestep of $10$ minutes (the frequency of data output from SUNTANS and WRF). We should also highlight that deep neural networks performed much better than shallow ones, none of which were selected in the top three FCNN ROM setups.

Table \ref{opt_hyperparam_fcnn} shows the optimum hyperparameters selection for FCNN ROM number 1617.

\begin{table}[h!]
\vspace{-0.5cm}
\caption{FCNN ROM best hyperparameters configuration.}
\label{opt_hyperparam_fcnn}
\centering
\begin{tabular}{ccccc}
\hline
Hidden layers & Neurons/layer & Dropout/layer &  Learning rate & Adam iter. \\
\hline
8 & $\lbrace83, 65, 67, 145, 59, 103, 81, 129\rbrace$ & 0.3 & 1e-05 & 5040 \\
\hline
\end{tabular}
\end{table}

\vspace{-0.3cm}
For the sake of completeness, we also investigated the ability of the FCNN ROM to provide accurate predictions of flow velocities at specific touristic regions at Lake George. We identified two places of interest using lat-long coordinates: Million Dollar beach in the south and City of Bolton in the middle west. Circular areas with $1.0 $km of radius centered at those points delimit the regions of interest where we evaluated the $36 h$ time averaged error of the best surrogate model (labeled 1617) to predict eastward and northward flow velocities. 

Table \ref{t_avg_error_fcnn} shows the time averaged error of the flow velocity components for the two specific regions of interest.

\begin{table}[h!]
\vspace{-0.3cm}
\caption{Time averaged percentage error of flow velocities at two regions of interest.}
\label{t_avg_error_fcnn}
\centering
\begin{tabular}{ccc}
\hline
Placement & $u$ error (\%) & $v$ error (\%)\\
\hline
Million Dollar beach & 0.6 & 0.5 \\
City of Bolton & 3.0 & 2.4 \\
\hline
\end{tabular}
\end{table}

\vspace{-0.7cm}

\subsubsection{LSTM ROM for hydrodynamics}
The LSTM learning process used an asymmetrical sliding window with 200 minutes input and 10 minutes output. As before, hyperparameter optimization was performed using a very simple strategy of randomly generating different neural network architectures. Such a strategy was essential to create a competitive model compared with the FCNN ROM.

\begin{figure}[h!]
\centering
\begin{tabular}{ccc}
\includegraphics[scale=0.21]{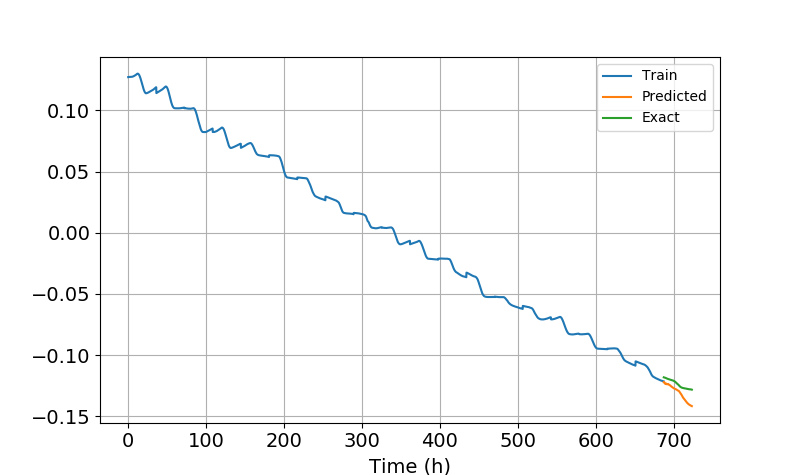} &
\hspace{-0.4cm}
\includegraphics[scale=0.21]{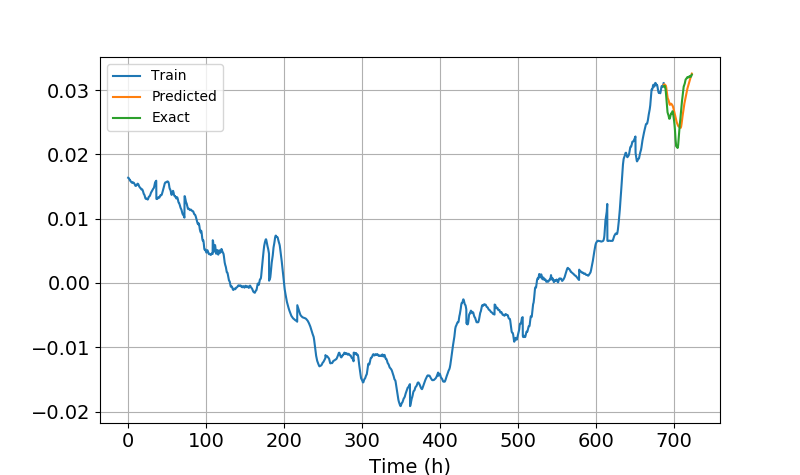} &
\hspace{-0.4cm}
\includegraphics[scale=0.21]{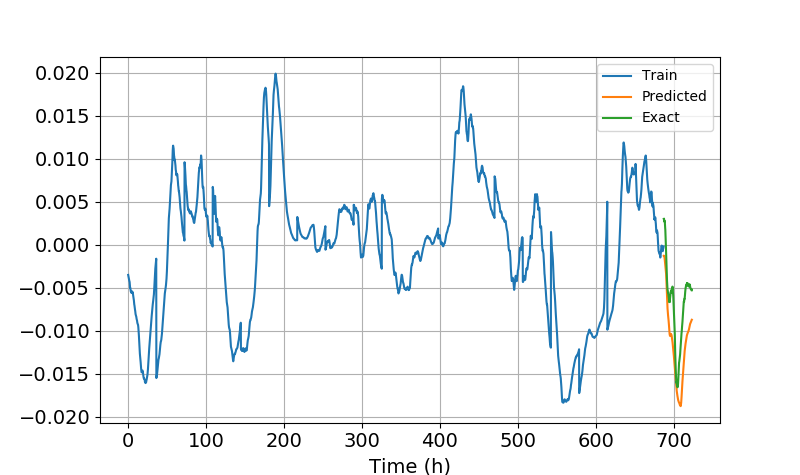} \\
\includegraphics[scale=0.21]{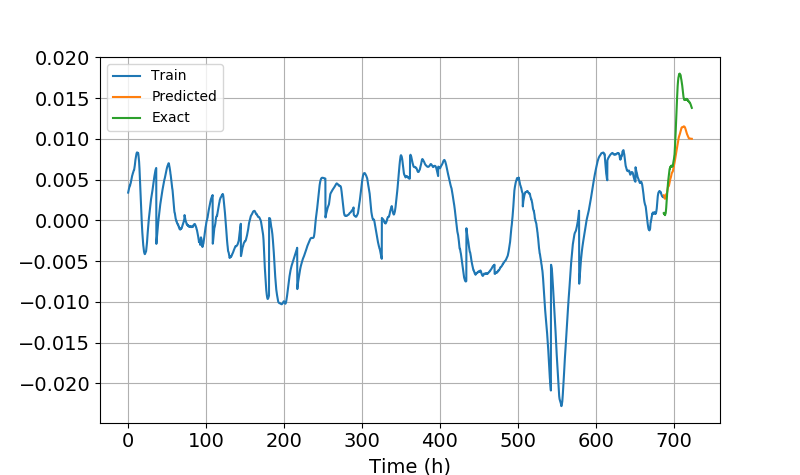} &
\hspace{-0.4cm}
\includegraphics[scale=0.21]{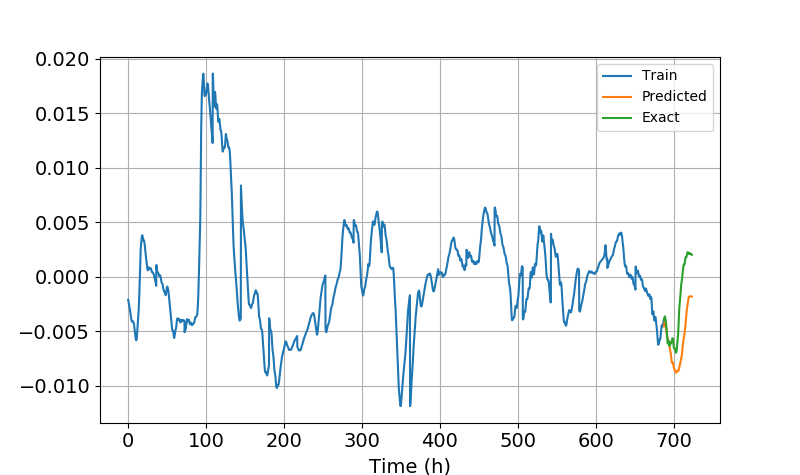} & \\
\end{tabular}
\caption{\small{Temporal coefficients of the LSTM ROM corresponding to the 5 spatial modes. From left to right, first row: $\mathbf{a}_{1}(t), \mathbf{a}_{2}(t)$ and  $\mathbf{a}_{3}(t)$; second row: $\mathbf{a}_{4}(t)$ and $\mathbf{a}_{5}(t)$.}}
\label{lstm_time_coeff}
\end{figure}

Figure \ref{lstm_time_coeff} shows the time series of the 5 temporal coefficients for each spatial model of the LSTM. The spatial modes are the same as those used by the FCNN ROM and have already been depicted in the previous subsection. Remarkably, the LSTM ROM can adequately predict the time series associated with the temporal coefficients without needing to (directly) compute the time derivatives of the temporal coefficients, nor integrate the time derivative approximation with an RK scheme.

\begin{figure}[h!]
\centering
\begin{tabular}{cccc}
\hspace{-0.2cm}
\textbf{Density} $\rho$ & \textbf{Temperature} $T$ & \textbf{East velocity} $u$ & \textbf{North velocity} $v$\\
\hspace{-0.2cm}
\includegraphics[scale=.080]{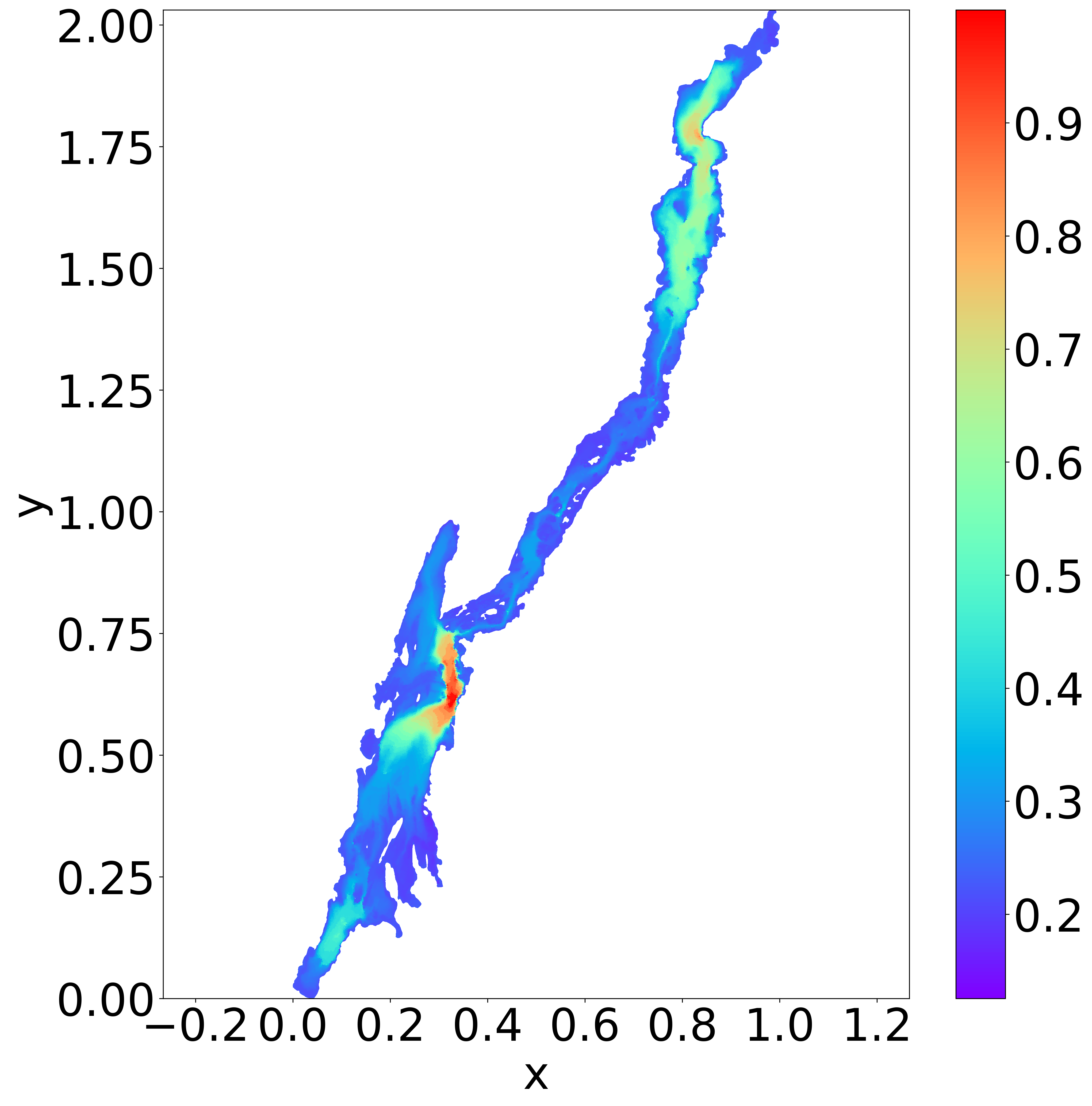} &
\includegraphics[scale=.080]{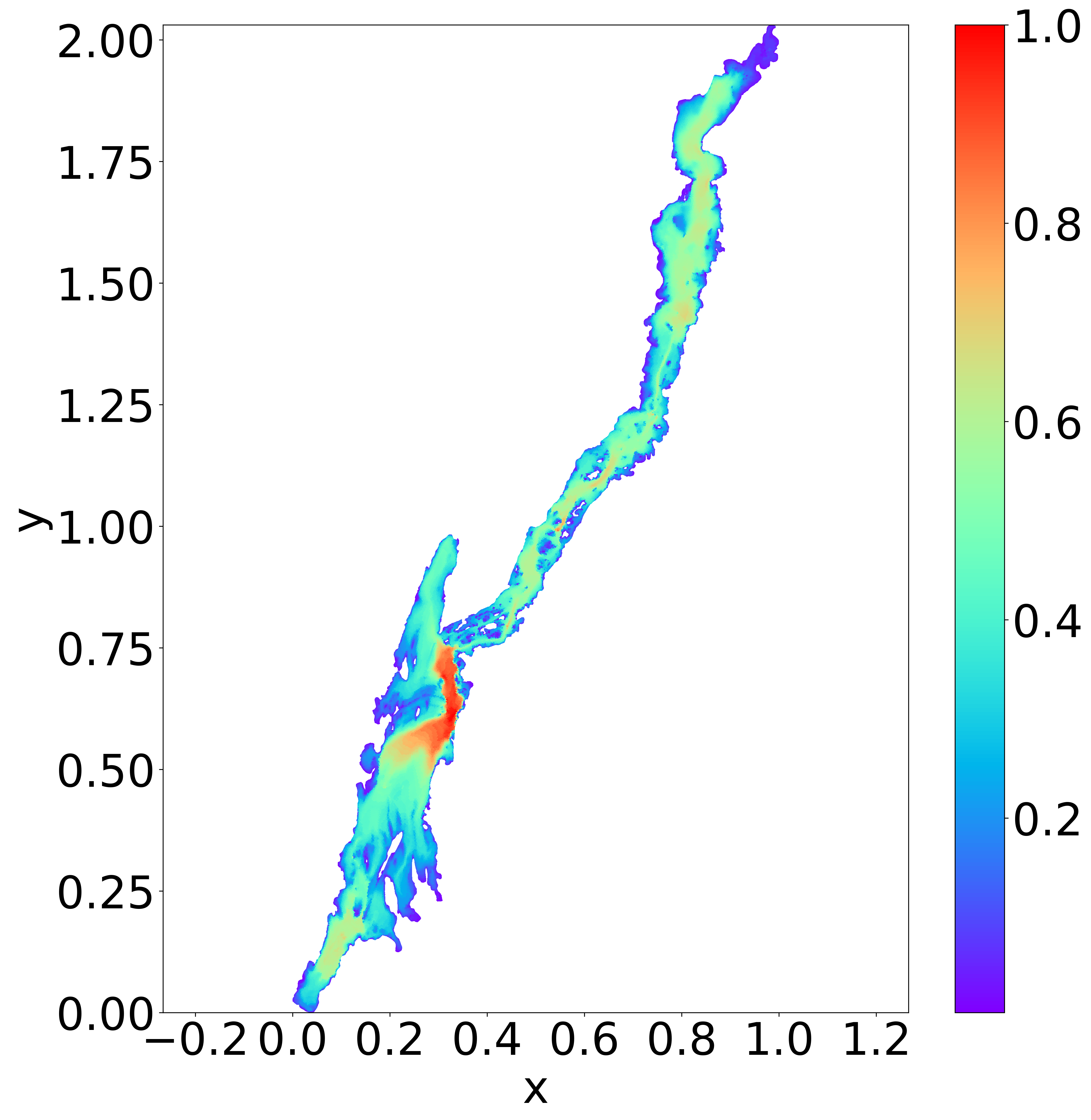} &
\includegraphics[scale=.080]{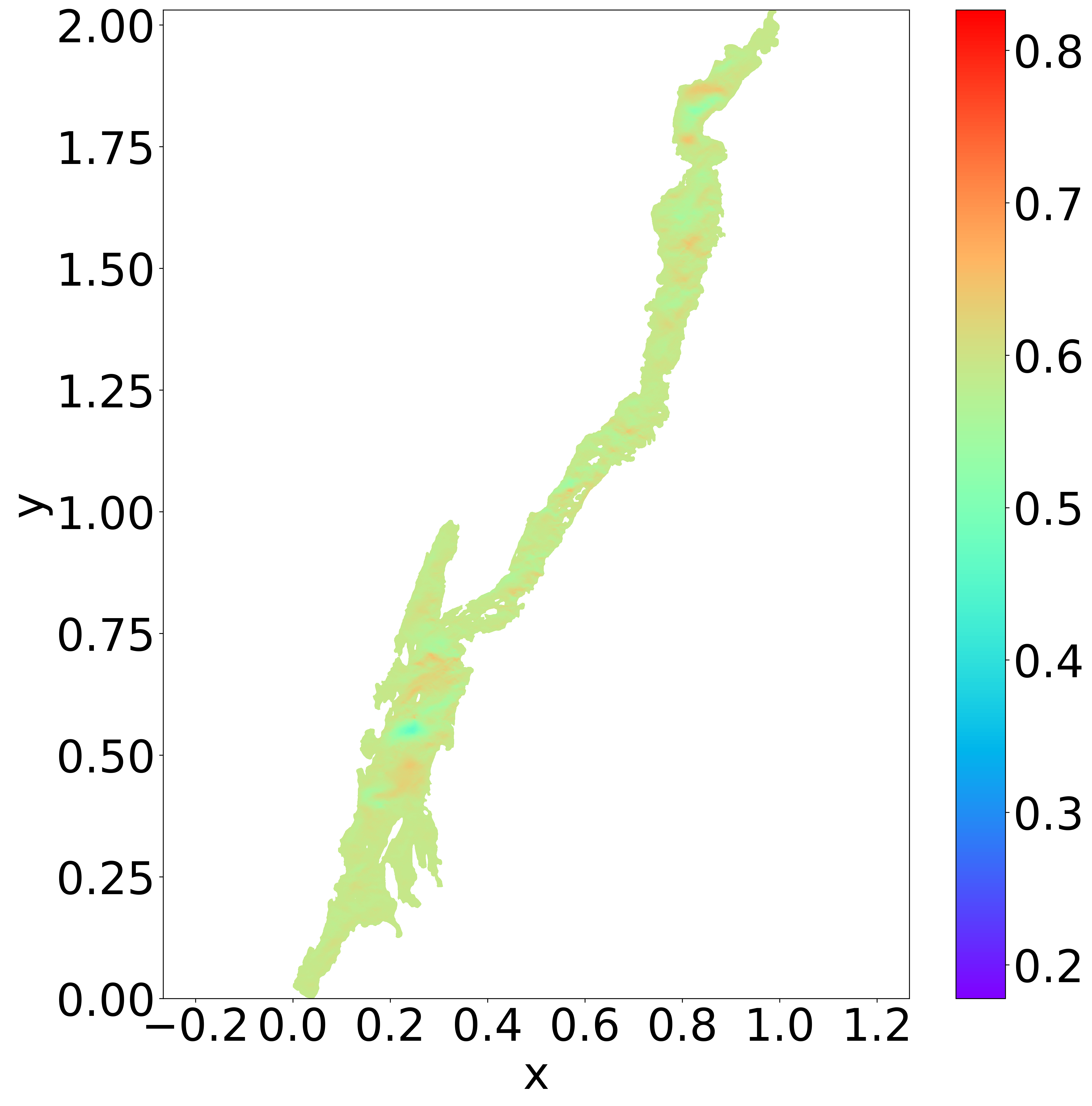} &
\includegraphics[scale=.080]{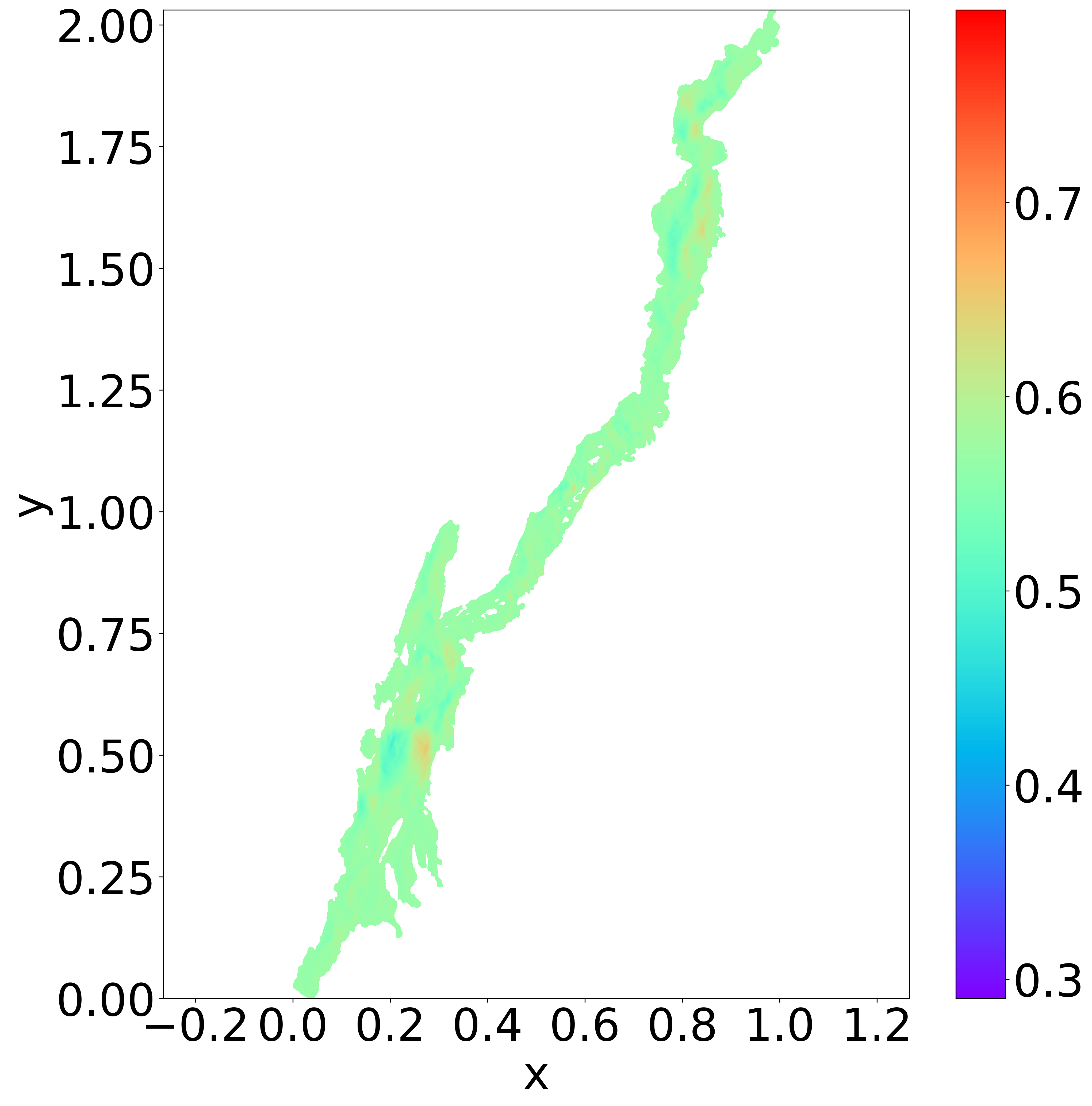}\\
\multicolumn{4}{c}{\textbf{True}} \\
\hspace{-0.2cm}
\includegraphics[scale=.080]{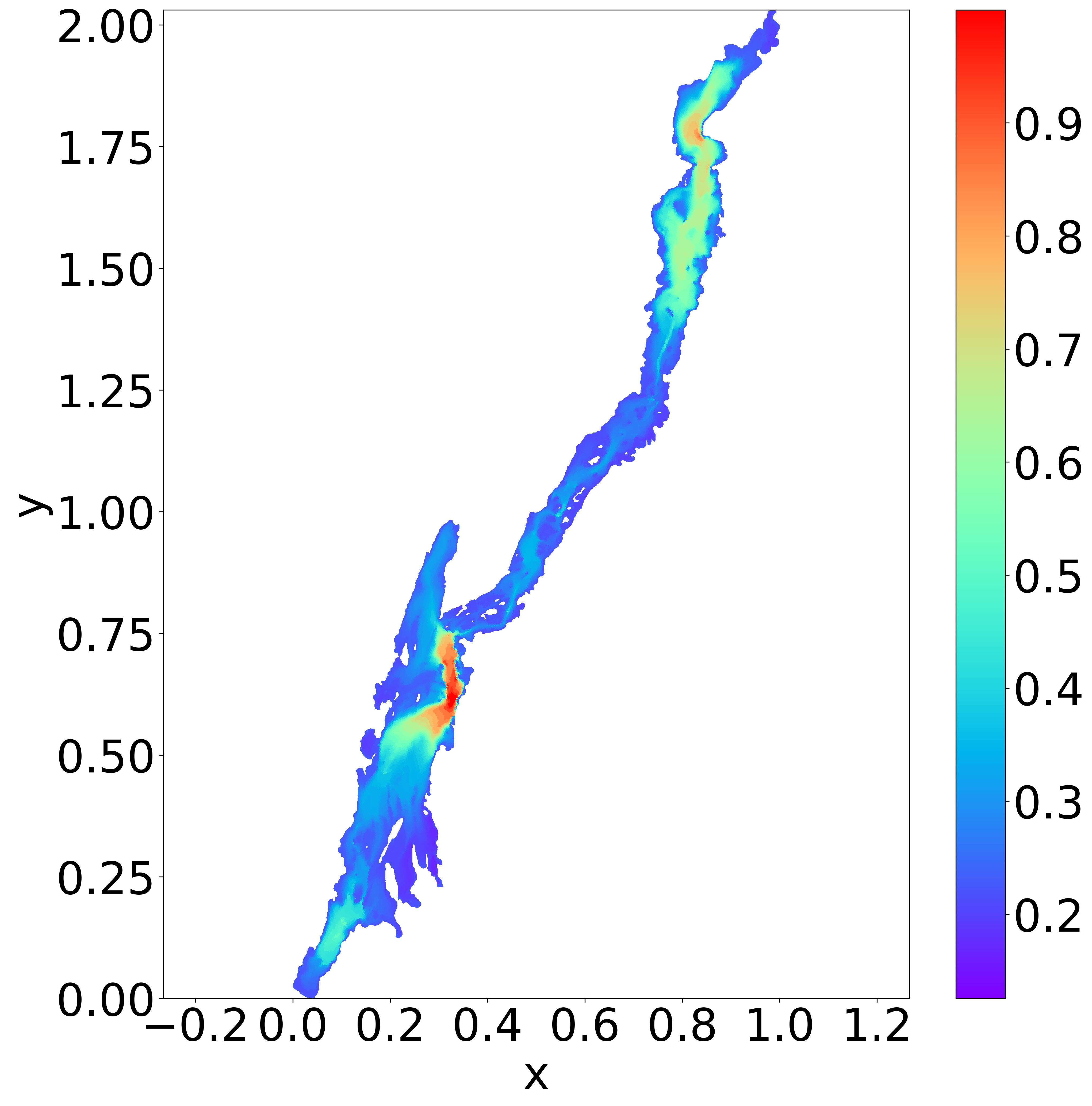} &
\includegraphics[scale=.080]{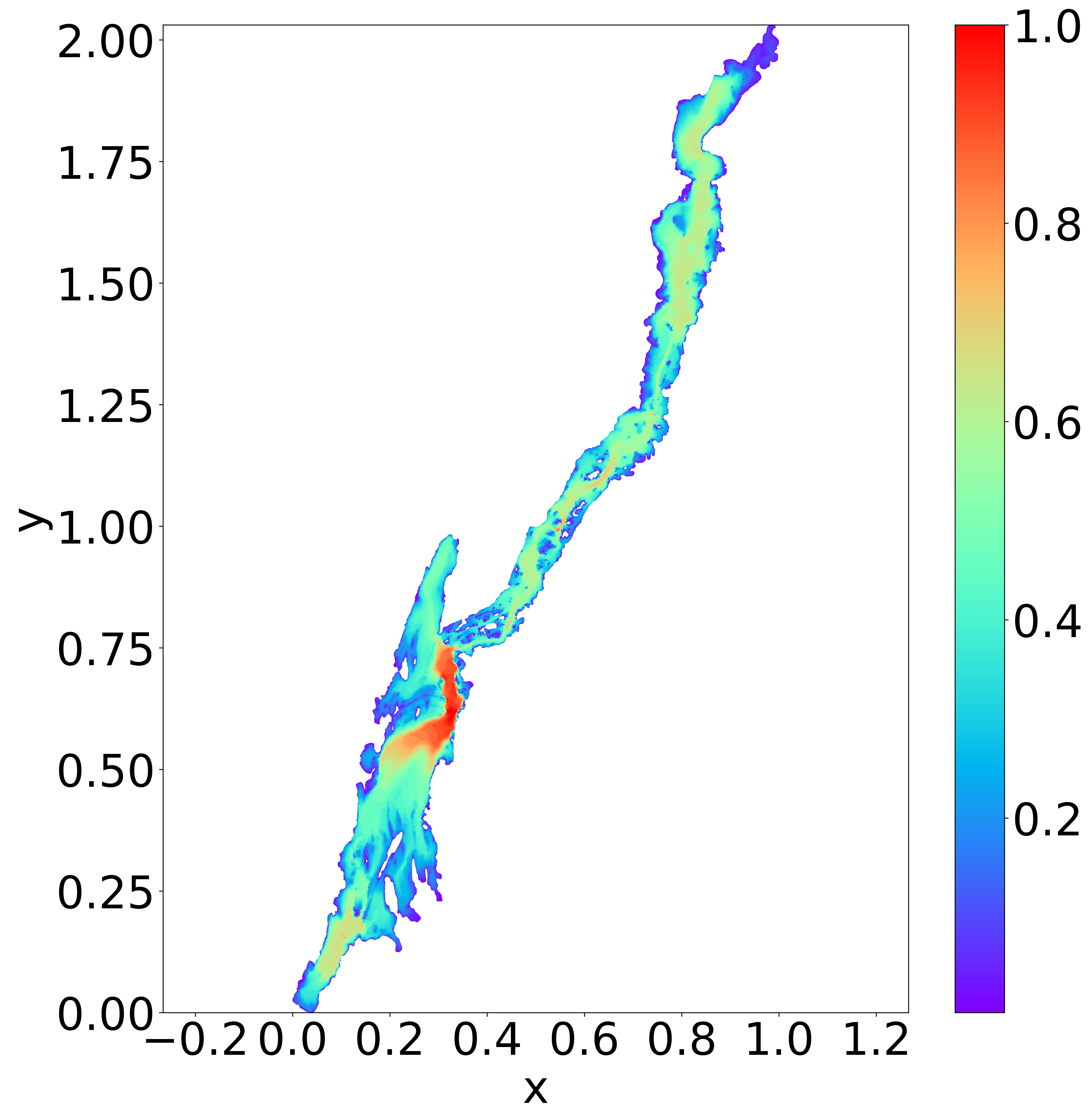} &
\includegraphics[scale=.080]{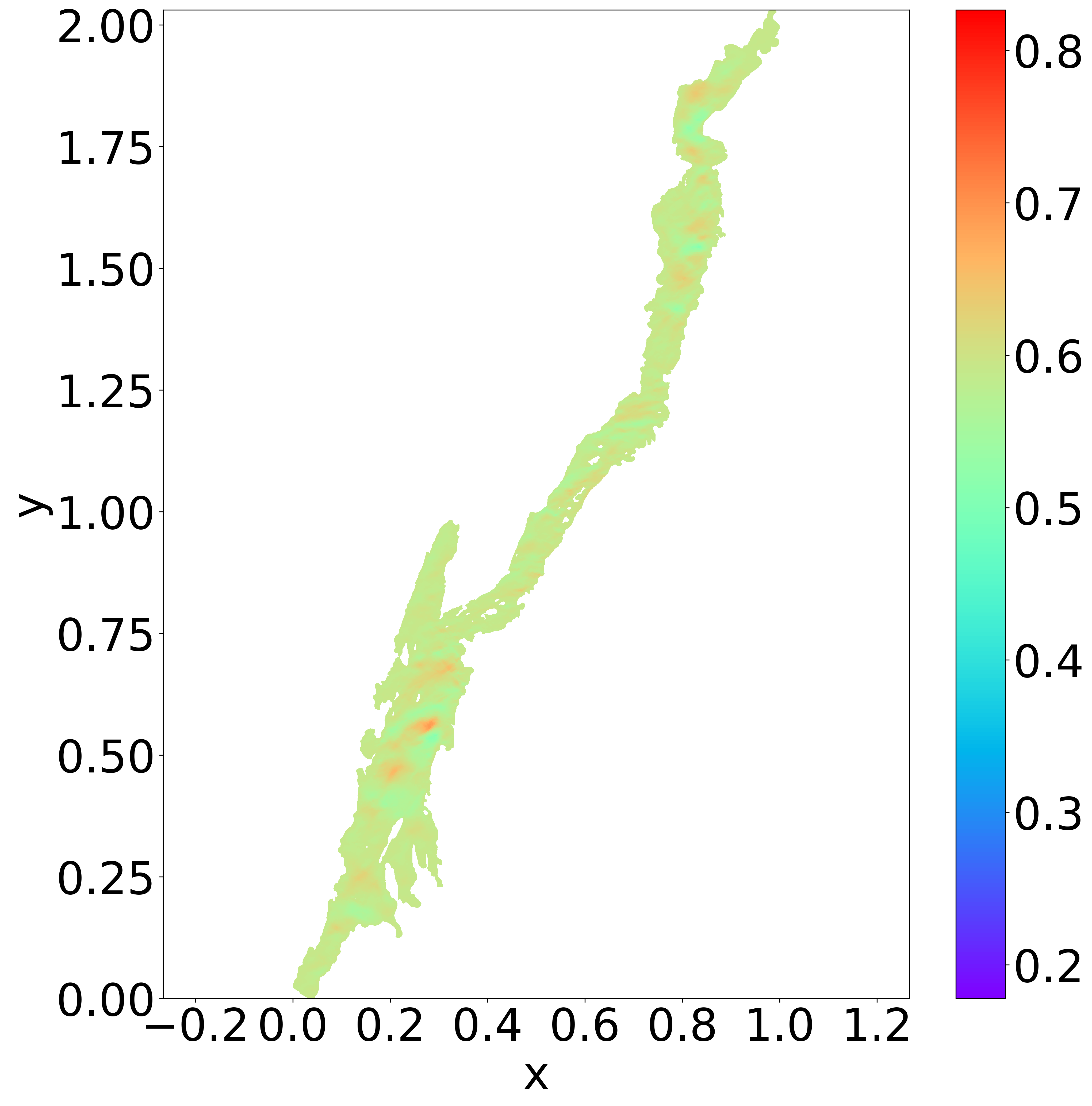} &
\includegraphics[scale=.080]{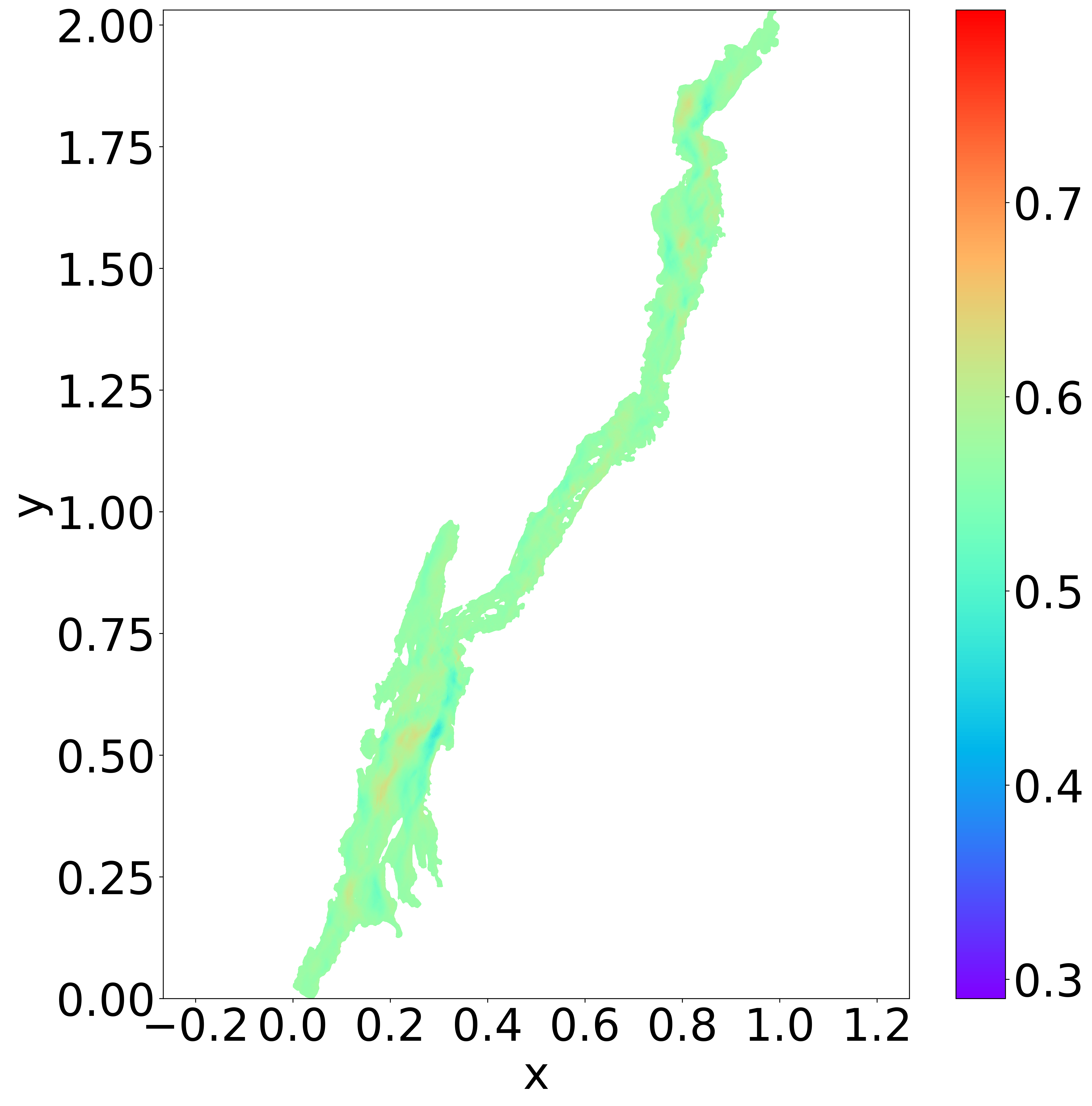}\\
\multicolumn{4}{c}{\textbf{Predicted}}
\end{tabular}
\caption{Full order model solution vs. LSTM ROM approximation for density, temperature and velocity components.}
\label{lstm_comparison}
\end{figure}

Figure \ref{lstm_comparison} shows the comparison of the predicted physical state of Lake George with the assumed true physical state. As with the FCNN ROM, the depth-averaged temperature and density variables are better approximated than velocity due to their large inertial trends. Only the main patterns of the velocity components can be described by the LSTM ROM. However, as with the FCNN ROM, there is a monotonic increasing relative error of the density variable. This is in part because the model performance ranking was based on the relative error of the average velocity components. The increasing error of the density variable was not observed when the performance criterion was changed to include other metrics, such as the error based on the average of all physical variables.

\begin{figure}[h!]
\vspace{-0.8cm}
\centering
\begin{tabular}{cc}
\includegraphics[scale=0.35]{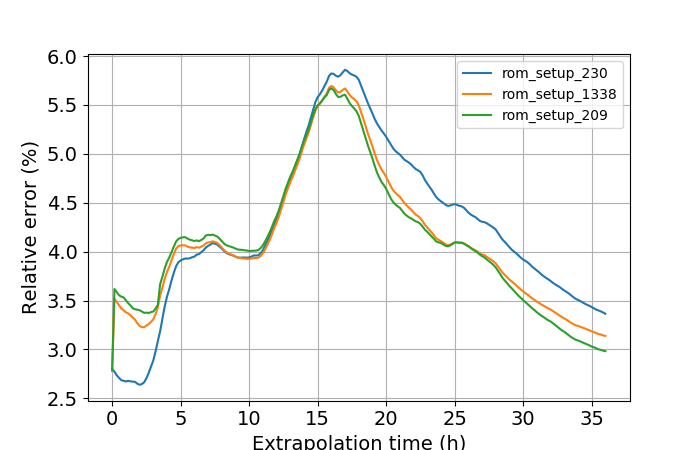} &
\includegraphics[scale=0.35]{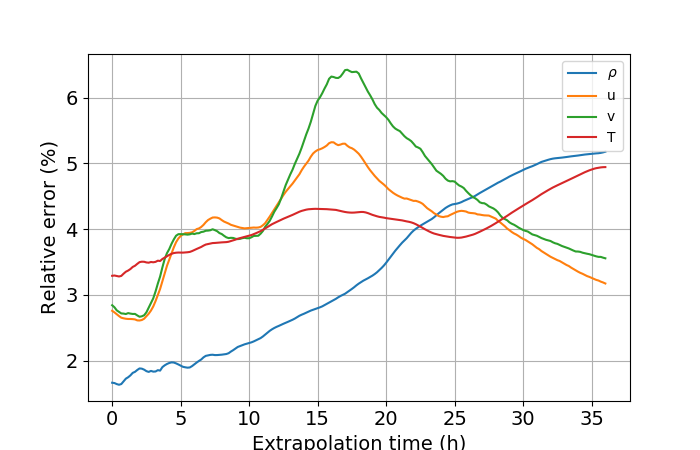} \\
a) Relative error for the 3 best & b) Relative error for each \\
hyperparameter settings & variable: $\rho, u, v, T$
\end{tabular}
\caption{a) Best performances based on the average $\frac{u+v}{2}$ error; b) Best configuration: model 230.}
\label{lstm_error}
\end{figure}

\vspace{-0.5cm}
From the error plots of the 3 best performing LSTM ROMs (Fig. \ref{lstm_error}), the relative error remains below $6\%$ (and below $7\%$ for each variable) for the duration of the forecast, which again is a remarkable result. As with the FCNN ROM, we observe a similar increase in the relative error from 12 to 23 hours when the surface wind strengthened. At the end of the 36 hour forecast, only the density relative error is slightly above the target of $5\%$, demonstrating the strong ability of the LSTM architecture to deal with time-evolving sequences. Table \ref{opt_hyperparam_lstm} shows the optimum hyperparameters selection for LSTM ROM number 230.

However, such improved results comes with a price: the LSTM ROM needs $4.8$ times more degrees of freedom (i.e weights and biases) than the FCNN ROM to perform equivalently\footnote{The LSTM minimization algorithm finds $264,690$ optimal weights and biases while FCNN requires only $55,101$.}.

\begin{table}[h!]
\vspace{-1.2cm}
\caption{LSTM ROM best hyperparameters configuration.}
\label{opt_hyperparam_lstm}
\centering
\begin{tabular}{ccccc}
\hline
Hidden layers & LSTM cells per layer & Dropout &  Learning rate & Adam iterations \\
\hline
3 & $\lbrace133,92,129\rbrace$ & $\lbrace0.2,0.3,0.3\rbrace$ & 0.01 & 1000 \\
\hline
\end{tabular}
\end{table}

\vspace{-0.8cm}

\section{Conclusions}  
In this work, we have developed two ANN-based ROMs, namely an FCNN ROM and an LSTM ROM, which are capable of reproducing a hydrodynamic forecast for a medium-sized freshwater lake with reasonable accuracy (a relative error of less than $6\%$ over the entire prediction range) and low computational cost. Both ANN-based ROMs reduced the dimensionality of the original data provided by the full-order hydrodynamics model, SUNTANS, to a small number of spatial basis through the POD technique. POD shows great promise for hydrodynamics applications as it drastically reduced the computational cost of the forecasting. 

More specifically, both ANN-based ROMs performed well at predicting the temporal coefficients of the depth-averaged hydrodynamic variables. The FCNN ROM (with a deep neural network configuration) and the LSTM ROM showed similar results for a 36 hour hydrodynamic forecast, trained with the 19 days of forecasts. However, the LSTM ROM used $4.8$ times more learning parameters than the FCNN ROM to perform equivalently.

The FCNN approach was also tested on predicting the time series of the Lorenz attractor model. The validation example with hyperparameter optimization for the chaotic regime showed outstanding results reaching 9 Lyapunov units until failing to represent the system's trends. FCNN also described relatively well the unsteady section of the deterministic case although with a clear bias in all discrete variables for the final steady state pattern.

We applied machine learning techniques to data generated by the RANS simulator, SUNTANS. RANS numerical models are based on assumptions that simplify computation by decreasing the fidelity of the approximation. Training the ANN-based ROMs on large eddy simulation (LES) or DNS numerical models output would provide higher fidelity since they include less assumptions, or none for DNS. In the literature, we see usage of neural networks \cite{Wang2017} trained on DNS and LES outputs to better represent the Reynolds stress in a RANS model. We believe this hybrid approach will improve the prediction of mixing events at the boundaries of lakes and oceans. However, running simulations of realistic geophysical flows with these models is currently prohibitive from a computational viewpoint.

Another important point that deserves attention is the capacity of the surrogate models to approximate fast changing physical quantities. In \cite{lee_calrberg2019} the authors point out that POD-ROMs can produce accurate results only if the problem of interest admits a fast decaying Kolmogorov n-width which corresponds to the class of diffusion-dominant problems. Circulation is mostly a convection-dominant problem with strong variability in velocity components. Another factor influencing the accuracy of ROMs is the chaotic/turbulent behavior.

As a next step, we want to test the ability of nonlinear dimensionality reduction \cite{gonzalez2018,lee_calrberg2019} techniques to describe the circulation features with higher fidelity. 


%
%
%
 \bibliographystyle{splncs04}
 \bibliography{references}
%
%
%
%
%
%
\end{document}